 \documentclass[journal]{IEEEtran}
\usepackage{times,amsmath,epsfig,latexsym,amssymb,psfrag,graphicx}
\usepackage{paralist,mathrsfs}
\usepackage{booktabs}
\usepackage{verbatim}
\usepackage{color}
 \usepackage{setspace}
\usepackage{epstopdf}
\usepackage{subcaption}
\usepackage{mathbbol}
\usepackage[ruled,vlined]{algorithm2e}
\usepackage{citesort}
 \usepackage{url}
\usepackage{pifont}
\usepackage[edges]{forest}
\usepackage{enumitem}
 
\usepackage{supertabular}

\usepackage{tikz}
\usepackage{lipsum,adjustbox}
\usetikzlibrary{mindmap,calc,positioning}

 \definecolor{Lightgray}{RGB}{235,235,235}
\usepackage{colortbl}
\usepackage[para]{threeparttable}

\usetikzlibrary{arrows.meta}
\tikzset{decision/.style = {diamond, draw, fill=black!20, text width=4.5em, text badly centered, node distance=2.8cm, inner sep=0pt},
	block/.style    = {rectangle, draw, fill=blue!5, text width=5em, text centered, rounded corners, minimum height=4em},	block2/.style    = {rectangle, draw, dashed,  fill=blue!20, text width=5em, text centered, rounded corners, minimum height=4em},
	line/.style     = {draw, -latex', very thick},
	cloud/.style    = {draw, ellipse,fill=red!20, node distance=2.8cm, minimum height=2em}
	}
	
\usepackage{tikz}
\tikzset{
	basic/.style  = {draw, text width=2cm, drop shadow, font=\sffamily, rectangle},
	root/.style   = {basic, rounded corners, thin, align=center, fill=red!30},
	level-2/.style = {basic, rounded corners=6pt, thin,align=center, fill=blue!20, text width=3cm},
	level-3/.style = {basic, thin, align=center, fill=orange!20, text width=2.4cm}
}

\usetikzlibrary{trees,positioning,shapes,shadows,arrows}
\usepackage[nopostdot,style=super,nonumberlist,toc]{glossaries}
\makeglossaries
\renewcommand*{\CustomAcronymFields}{%
  name={\the\glsshorttok},% name is abbreviated form
  description={\the\glslongtok},% description is long form
  first={\noexpand\emph{\the\glslongtok}\space(\the\glsshorttok)},%
  firstplural={\noexpand\emph{\the\glslongtok\noexpand\acrpluralsuffix}\space(\the\glsshorttok)},%
  text={\the\glsshorttok},%
  plural={\the\glsshorttok\noexpand\acrpluralsuffix}%
}

\newcommand{\secref}[1]{Sec.~\ref{#1}}

\newcommand{\figref}[1]{Fig.~\ref{#1}}

\newcommand{\tabref}[1]{Table~\ref{#1}}

\newcommand*\rot{\rotatebox{90}}  
\newcommand*\OK{\ding{51}}

\newcommand\Tstrut{\rule{0pt}{2.6ex}}         % = `top' strut for tables
\newcommand\Bstrut{\rule[-0.9ex]{0pt}{0pt}}   % = `bottom' strut

\begin{document}
\title{Cellular, Wide-Area, and  Non-Terrestrial IoT: A Survey on 5G Advances and the Road Towards 6G}

\author{\IEEEauthorblockN{Mojtaba Vaezi,   \IEEEmembership{Senior Member, IEEE}, 
		Amin Azari, \IEEEmembership{Member, IEEE},
%		\IEEEmembership{Member, IEEE},  
Saeed R. Khosravirad, \IEEEmembership{Member, IEEE}, Mahyar Shirvanimoghaddam, \IEEEmembership{Senior Member, IEEE},
	 M. Mahdi Azari, \IEEEmembership{Member, IEEE}, \\  Danai Chasaki, \IEEEmembership{Senior Member, IEEE}, and Petar Popovski,  \textit{Fellow, IEEE}}
	%\\ 	 \IEEEmembership{(Invited Paper)}
	
%	\author{\IEEEauthorblockN{co-authors}}
}

\maketitle

%% for glossaries go to http://texblog.org/2014/01/15/glossary-and-list-of-acronyms-with-latex/
%\makenoidxglossaries

\newacronym{csi}{CSI}{channel state information}
\newacronym{cqi}{CQI}{channel quality indicator}
\newacronym{ack}{ACK}{acknowledgement}
\newacronym{arq}{ARQ}{automatic repeat request}
\newacronym{awgn}{AWGN}{additive white Gaussian noise}
\newacronym{cc}{CC}{chase combining}
\newacronym{mmw}{mmWave}{millimeter-wave}
\newacronym{dp}{DP}{dynamic programming}
\newacronym{fec}{FEC}{forward error correction}
\newacronym{harq}{HARQ}{hybrid automatic repeat request}
\newacronym{hspa}{HSPA}{high speed packet access}
\newacronym{iid}{i.i.d.}{independent and identically distributed}
\newacronym{ir}{IR}{incremental redundancy}
\newacronym{lte}{LTE}{Long Term Evolution}
\newacronym{mdp}{MDP}{markov decision process}
\newacronym{mrc}{MRC}{maximal-ratio combining}
\newacronym{nack}{NAK}{negative acknowledgement}
\newacronym{pdf}{pdf}{probability density function}
\newacronym{wimax}{WiMax}{worldwide interoperability for microwave access}
\newacronym{3gpp}{3GPP}{3rd generation partnership project}
\newacronym{ofdm}{OFDM}{orthogonal frequency-division multiplexing}
\newacronym{ofdma}{OFDMA}{orthogonal frequency-division multiple access}
\newacronym{wlan}{WLAN}{wireless local area network}
\newacronym{mmse}{MMSE}{minimum mean-square-error}
\newacronym{gsm}{GSM}{global system for mobile communications}
\newacronym{edge}{EDGE}{enhanced data \gls{gsm} environment}
\newacronym{stbc}{STBC}{space-time block code}
\newacronym{amc}{AMC}{adaptive modulation and coding}
\newacronym{sinr}{SINR}{signal-to-interference-plus-noise ratio}
\newacronym{mi}{MI}{mutual information}
\newacronym{acmi}{ACMI}{accumulated mutual information}
\newacronym{nacmi}{NACMI}{normalized ACMI}
\newacronym{cdi}{CDI}{channel distribution information}
\newacronym{latr}{LATR}{long-term average transmission rate}
\newacronym{rtr}{RTR}{round transmission rate}
\newacronym{pomdp}{POMDP}{Partially Observable Markov Decision Process}
\newacronym{fd}{FD}{full-duplex}
\newacronym{hd}{HD}{half-duplex}
\newacronym{td}{TD}{Time Division}
\newacronym{tdma}{TDMA}{time-division multiple access}
\newacronym{mac}{MAC}{media access control}
\newacronym{uwb}{UWB}{ultra wideband}
\newacronym{ieee}{IEEE}{institute of electrical and electronics engineers}
\newacronym{dB}{dB}{decibel}
\newacronym{cdf}{cdf}{cumulative density function}
\newacronym{ccdf}{ccdf}{complementary cumulative density function}
\newacronym{min}{Min.}{minimum}
\newacronym{med}{Med.}{median}
\newacronym{avg}{Avg.}{average}
\newacronym{ul}{UL}{uplink}
\newacronym{app}{APP}{a-posteriori probability}
\newacronym{logmap}{LogMAP}{log maximum a-posteriori}
\newacronym{llr}{LLR}{log-likelihood ratio}
\newacronym{ue}{UE}{user equipment}
\newacronym{qos}{QoS}{quality of service}
\newacronym{4g}{4G}{fourth-generation}
\newacronym{6g}{6G}{sixth-generation}
\newacronym{tti}{TTI}{transmission time interval}
\newacronym{rrm}{RRM}{radio resource management}
\newacronym{mmib}{MMIB}{mean mutual information per bit}
\newacronym{dsi}{DSI}{decoder state information}
\newacronym{tb}{TB}{transport block}
\newacronym{tbs}{TBS}{transport block size}
\newacronym{cb}{CB}{code block}
\newacronym{cbg}{CBG}{code block group}
\newacronym{cbs}{CBS}{code block size}
\newacronym{prb}{PRB}{physical resource block}
\newacronym{rb}{RB}{resource block}
\newacronym{bler}{BLER}{block error rate}
\newacronym{blep}{BLEP}{block error probability}
\newacronym{crc}{CRC}{cyclic redundancy check}
\newacronym{tdd}{TDD}{time division duplex}
\newacronym{fdd}{FDD}{frequency division duplex}
\newacronym{embb}{eMBB}{enhanced mobile broadband}
\newacronym{mcc}{MCC}{mission critical communication}
\newacronym{mmc}{MMC}{massive machine communication}
\newacronym{mtc}{MTC}{machine type of communication}
\newacronym{mmtc}{mMTC}{massive machine type of communication}
\newacronym{umtc}{uMTC}{ultra-reliable \gls{mtc}}
\newacronym{urllc}{URLLC}{ultra-reliable low-latency communications}
\newacronym{rtt}{RTT}{round trip time}
\newacronym{rs}{RS}{reference symbols}
\newacronym{kpi}{KPI}{key performance indicator}
\newacronym{kpis}{KPIs}{key performance indicators}
\newacronym{tx}{Tx}{transmitter node}
\newacronym{rx}{Rx}{receiver node}
\newacronym{cran}{C-RAN}{centralized radio access network}
\newacronym{rru}{RRU}{remote radio unit}
\newacronym{bbu}{BBU}{baseband unit}
\newacronym{fhd}{FHD}{fronthaul delay}
\newacronym{cch}{CCH}{control channel}
\newacronym{saw}{SAW}{stop-and-wait}
\newacronym{qci}{QCI}{\gls{qos} class identifier}
\newacronym{gbr}{GBR}{guaranteed bit rate}
\newacronym{mbr}{MBR}{maximum bit rate}
\newacronym{ngbr}{non-GBR}{non-\gls{gbr}}
\newacronym{arp}{ARP}{allocation and retention priority}
\newacronym{effcr}{ECR}{effective coding rate}
%\newacronym{cbs}{CBS}{code block size}
\newacronym{mcs}{MCS}{modulation and coding scheme}
\newacronym{eva}{EVA}{extended vehicular A}
\newacronym{epa}{EPA}{extended pedestrian A}
\newacronym{etu}{ETU}{extended typical urban}
\newacronym{re}{RE}{resource element}
\newacronym{reS}{REs}{resource elements}
\newacronym{nr}{NR}{new radio}
\newacronym{qpsk}{QPSK}{quadrature phase shift keying}
\newacronym{qam}{QAM}{quadrature amplitude modulation}
\newacronym{siso}{SISO}{single-input and single-output}	
\newacronym{miso}{MISO}{multiple-input single-output}
\newacronym{mimo}{MIMO}{multiple-input multiple-output}
\newacronym{bs}{BS}{base station}
\newacronym{phy}{PHY}{physical layer}
\newacronym{rlc}{RLC}{radio link control}
\newacronym{bcfsaw}{BCF-SAW}{BCF-SAW}
\newacronym{bcf}{BCF}{backwards composite feedback}
\newacronym{bac}{BAC}{binary asymmetric channel}
\newacronym{bsc}{BSC}{binary symmetric channel}
\newacronym{dtx}{DTX}{discontinued transmission}
\newacronym{bpsk}{BPSK}{binary phase shift keying}
\newacronym{bep}{BEP}{bit error probability}
\newacronym{ndi}{NDI}{new data indicator}
\newacronym{dci}{DCI}{downlink control information}
\newacronym{csit}{CSIT}{channel state information at the transmitter}
\newacronym{lt}{LT}{loudest talker}
\newacronym{ct}{CT}{cooperative transmission}
\newacronym{bps}{bps}{bits per second}
\newacronym{bpcu}{bpcu}{bits per channel use}
\newacronym{los}{LOS}{line-of-sight}
\newacronym{nlos}{NLOS}{non-line-of-sight}
\newacronym{regsaw}{Reg-SAW}{Regular SAW}
\newacronym{Lrep}{$L$-Rep-ACK}{Increased feedback repetition order}
\newacronym{Lack}{$L$-ACK-SAW}{$L$ required ACK per packet}
\newacronym{Asym}{Asym-SAW}{Asymmetric feedback detection for SAW}
\newacronym{bretx}{Blind-ReTx}{Blind retransmission}
%\newacronym{df}{DF}{decode-and-forward}
\newacronym{af}{AF}{amplify-and-forward}
\newacronym{ap}{AP}{access point}
\newacronym{icn}{ICN}{industrial control network}
\newacronym{comp}{CoMP}{coordinated multi-point}
\newacronym{rhs}{RHS}{right hand side}
\newacronym{lhs}{LHS}{left hand side}
%\newacronym{ap}{AP}{access point}
%\newacronym{comp}{CoMP}{Coordinated Multipoint}
\newacronym{sumiso}{SU-MISO}{single-user multiple-input-single-output}
\newacronym{ibl}{IBL}{infinite block length}
\newacronym{fbl}{FBL}{finite block length}
%\newacronym{icn}{ICN}{industrial control network}
\newacronym{lan}{LAN}{local area network}
\newacronym{wsn}{WSN}{wireless sensor network}
\newacronym{rt}{RT}{real-time}
\newacronym{tdm}{TDM}{time division multeplxing}
\newacronym{isi}{ISI}{inter-symbol interference}
\newacronym{nist}{NIST}{National Institute of Standards and Technology}
\newacronym{cbrs}{CBRS}{Citizens Broadband Radio Service}
%\newacronym{los}{LOS}{line-of-sight}
%\newacronym{nlos}{NLOS}{non-line-of-sight}
\newacronym{itu}{ITU}{International Telecommunications Union}
\newacronym{mmwave}{mmWave}{millimeter-wave}
\newacronym{nsr}{NSR}{noise-to-signal ratio}
\newacronym{das}{DAS}{distributed antenna system}
%\newacronym{pdf}{PDF}{probability density function}		
\newacronym{pmf}{PMF}{probability mass function}
\newacronym{srs}{SRS}{sounding reference signal}
\newacronym{dmrs}{DMRS}{demodulation reference signal}
\newacronym{psd}{PSD}{power spectral density}
\newacronym{rf}{RF}{radio frequency}
\newacronym{id}{ID}{identifier}
\newacronym{df}{DF}{decode-and-forward}
\newacronym{iiot}{IIoT}{industrial IoT}
\newacronym{icsi}{I-CSI}{imperfect CSI}
\newacronym{pcsi}{P-CSI}{perfect CSI}
\newacronym{andcoop}{ANDCoop}{adaptive network-device cooperation}
\newacronym{dmt}{DMT}{diversity-multiplexing tradeoff}
\newacronym{scs}{SCS}{sub-carrier spacing}
\newacronym{trp}{TRP}{transmission reception point}
\newacronym{hpbw}{HPBW}{half-power beamwidth}
\newacronym{ss}{SS}{spread-spectrum}
\newacronym{adc}{ADC}{analog-to-digital converter}
\newacronym{scfde}{SC-FDE}{single-carrier frequency domain equalization}
\newacronym{d2d}{D2D}{device-to-device}
\newacronym{prose}{ProSe}{proximity service}
\newacronym{v2x}{V2X}{vehicle-to-everything}
\newacronym{v2v}{V2V}{vehicle-to-vehicle}
\newacronym{cdd}{CDD}{cyclic delay diversity}
\newacronym{ldpc}{LDPC}{low-density parity-check}
\newacronym{e2e}{E2E}{end-to-end}
\newacronym{InF}{InF}{indoor factory}
\newacronym{scm}{SCM}{spatial channel modeling}
\newacronym{cu}{CU}{central unit}
\newacronym{du}{DU}{distributed unit}
\newacronym{sl}{SL}{side-link}
\newacronym{im}{IM}{interference measurement}
\newacronym{cp}{CP}{cyclic prefix}
\newacronym{sir}{SIR}{signal to interference ratio}
\newacronym{dvbt}{DVB-T}{digital video broadcasting-terrestrial}
\newacronym{sa}{SA}{service availability}
\newacronym{sfn}{SFN}{single-frequency network}
\newacronym{sf}{SF}{spreading factor}
\newacronym{twc}{TWC}{tactile wireless control}
\newacronym{per}{PER}{packet error rate}
\newacronym{cper}{CPER}{consecutive packet error rate}
\newacronym{aoi}{AoI}{age of information}
\newacronym{uhd}{UHD}{ultra-high definition}
\newacronym{npn}{NPN}{non-public networks}

\newacronym{1g}{1G}{first-generation (cellular networks)}
\newacronym{2g}{2G}{second-generation}
\newacronym{3g}{3G}{third-generation}
\newacronym{3d}{3D}{three-dimensional}
\newacronym{3gpp}{3GPP}{third-generation  partnership project}
\newacronym{4g}{4G}{fourth-generation}
\newacronym{5g}{5G}{fifth-generation}
\newacronym{6g}{6G}{sixth-generation (cellular networks)}
\newacronym{acma}{ACMA}{asynchronous coded multiple access}
\newacronym{adc}{ADC}{analog-to-digital converter}
\newacronym{ae}{AE}{autoencoder}
\newacronym{afc}{AFC}{analog fountain code}
\newacronym{ai}{AI}{artificial intelligence}
\newacronym{aka}{AKA}{authentication and key agreement}
\newacronym{ann}{ANN}{artificial neural network}
\newacronym{ap}{AP}{access point}
\newacronym{b5g}{B5G}{beyond 5G}
\newacronym{bc}{BC}{broadcast channel}
\newacronym{ble}{BLE}{Bluetooth low-energy}
\newacronym{bler}{BLER}{block error rate}
\newacronym{bp}{BP}{belief propagation}
\newacronym{bs}{BS}{base station}
\newacronym{bw}{BW}{bandwidth}
\newacronym{cagr}{CAGR}{compound annual growth rate}
\newacronym{cdd}{CDD}{cyclic delay diversity}
\newacronym{cdma}{CDMA}{code division multiple access}
\newacronym{cnn}{CNN}{convolutional neural network}
\newacronym{cqi}{CQI}{channel quality indicator}
\newacronym{crc}{CRC}{cyclic redundancy check}
\newacronym{csi}{CSI}{channel state information}
\newacronym{css}{CSS}{chirp spread spectrum}
\newacronym{d2d}{D2D}{device-to-device}
\newacronym{ddos}{DDoS}{distributed denial-of-service}
\newacronym{dl}{DL}{deep learning}
\newacronym{dln}{DL}{downlink}
\newacronym{dnn}{DNN}{deep neural network}
\newacronym{dns}{DNS}{domain name system}
\newacronym{drl}{DRL}{deep reinforcement learning}
\newacronym{embb}{eMBB}{enhanced mobile broadband}
\newacronym{fdma}{FDMA}{frequency division multiple access}
\newacronym{fl}{FL}{federated learning}
\newacronym{gan}{GAN}{generative adversarial network}
\newacronym{gfa}{GFA}{grant-free access}
\newacronym{gmsk}{GMSK}{Gaussian frequency-shift keying}
\newacronym{gnb}{gNB}{gNodeB}
\newacronym{gps}{GPS}{global positioning system}
\newacronym{gsm}{GSM}{global system for mobile communications}
\newacronym{harq}{HARQ}{hybrid automatic repeat request}
\newacronym{https}{HTTPS}{hypertext transfer protocol secure}
\newacronym{idma}{IDMA}{interleave division multiple access }
\newacronym{ids}{IDS}{intrusion detection system}
\newacronym{igma}{IGMA}{interleave-grid multiple access }
\newacronym{iiot}{IIoT}{industrial IoT}
\newacronym{imt}{IMT}{international mobile telecommunications}
\newacronym{iot}{IoT}{Internet of things}
\newacronym{isd}{ISD}{inter-site distance}
\newacronym{ism}{ISM}{industrial safety medical}
\newacronym{itu}{ITU}{international telecommunications union}
\newacronym{lcrs}{LCRS}{low code rate spreading}
\newacronym{ldpc}{LDPC}{low density parity check}
\newacronym{lds}{LDS}{low-density signature }
\newacronym{leo}{LEO}{low earth orbit}
\newacronym{llr}{LLR}{log-likelihood ratio}
\newacronym{lora}{LoRa}{long range}
\newacronym{lorawan}{LoRaWAN}{LoRa wide area network}
\newacronym{los}{LoS}{line-of-sight}
\newacronym{lpwa}{LPWA}{low-power wide-area}
\newacronym{lssa}{LSSA}{low code rate and signature based shared access}
\newacronym{lte}{LTE}{Long Term Evolution}
\newacronym{lte-a}{LTE-A}{LTE-Advanced}
\newacronym{lte-m}{LTE-M}{LTE-MTC}
\newacronym{m2m}{M2M}{machine-to-machine}
\newacronym{mac}{MAC}{ media access control}
\newacronym{ma-mec}{MA-MEC}{multi-access mobile edge computing}
\newacronym{miso}{MISO}{multiple-input and single-output}
\newacronym{ml}{ML}{machine learning}
\newacronym{mmtc}{mMTC}{massive machine-type communications}
\newacronym{mmwave}{mmWave}{millimeter-wave}
\newacronym{mpa}{MPA}{message passing algorithm}
\newacronym{musa}{MUSA}{multi-user shared access}
\newacronym{nbiot}{NB-IoT}{narrowband IoT}
\newacronym{noca}{NOCA}{non-orthogonal coded access}
\newacronym{noma}{NOMA}{non-orthogonal multiple access}
\newacronym{npbch}{NPBCH}{narrowband physical broadcast channel}
\newacronym{npdcch}{NPDCCH}{narrowband physical downlink control channel}
\newacronym{npdsch}{NPDSCH}{narrowband physical downlink shared channel}
\newacronym{nprach}{NPRACH}{narrowband physical random access channel}
\newacronym{npss}{NPSS}{narrowband primary synchronization signal}
\newacronym{npusch}{NPUSCH}{narrowband physical uplink shared channel}
\newacronym{nr}{NR}{new radio}
\newacronym{nrs}{NRS}{narrowband reference signal}
\newacronym{ntn}{NTN}{non-terrestrial network}
\newacronym{ofdm}{OFDM}{orthogonal frequency-division multiplexing}
\newacronym{ofdma}{OFDMA}{orthogonal frequency-division multiple access}
\newacronym{osd}{OSD}{ordered statistics decoder}
\newacronym{pdcch}{PDCCH}{physical downlink control channel}
\newacronym{pdma}{PDMA}{pattern division multiple access}
\newacronym{phy}{PHY}{physical layer}
\newacronym{pkc}{PKC}{public key cryptography}
\newacronym{pucch}{PUCCH}{physical uplink control channel}
\newacronym{puf}{PUF}{physically unclonable function}
\newacronym{pusch}{PUSCH}{physical uplink shared channel}
\newacronym{qos}{QoS}{quality of service}
\newacronym{rach}{RACH}{random access channel}
\newacronym{rach}{RACH}{random access}
\newacronym{rb}{RB}{resource block}
\newacronym{rbm}{RBM}{restricted Boltzmann machine }
\newacronym{rdma}{RDMA}{repetition division multiple access }
\newacronym{rsma}{RSMA}{resource spread multiple access}
\newacronym{relu}{ReLU}{rectified linear units}
\newacronym{rf}{RF}{radio frequency}
\newacronym{rfid}{RFID}{radio-frequency identification}
\newacronym{rl}{RL}{reinforcement learning}
\newacronym{rnn}{RNN}{recurrent neural network}
\newacronym{rs}{RS}{rate-splitting}
\newacronym{rss}{RSS}{received signal strength}
\newacronym{sc}{SC}{successive cancellation}
\newacronym{sc-fdma}{SC-FDMA}{single-carrier FDMA}
\newacronym{scma}{SCMA}{sparse code multiple access }
\newacronym{sc-sic}{SC-SIC}{superposition coding-successive interference cancellation}
\newacronym{sdn}{SDN}{software defined networking}
\newacronym{sic}{SIC}{successive interference cancellation}
\newacronym{siso}{SISO}{single-input and single-output}
\newacronym{snr}{SNR}{signal-to-noise ratio}
\newacronym{sql}{SQL}{structured query language}
\newacronym{sram}{SRAM}{static random-access memory}
\newacronym{svd}{SVD}{singular value decomposition}
\newacronym{svm}{SVM}{support vector machine}
\newacronym{tbcc}{TBCC}{tail-biting convolutional code}
\newacronym{tdma}{TDMA}{time division multiple access}
\newacronym{thz}{THz}{terahertz}
\newacronym{tls}{TLS}{transport layer security}
\newacronym{tls}{TLS}{transport layer security}
\newacronym{trp}{TRP}{transmission reception points}
\newacronym{trp}{TRP}{transmission reception point }
\newacronym{tti}{TTI}{transmission time interval}
\newacronym{uav}{UAV}{unmanned aerial vehicle}
\newacronym{ue}{UE}{user equipment}
\newacronym{ul}{UL}{uplink}
\newacronym{urllc}{URLLC}{ultra-reliable low-latency communications}
\newacronym{uwb}{UWB}{ultra wideband}
\newacronym{vlc}{VLC}{visible  light communication}
\newacronym{wpt}{WPT}{wireless power transfer}
\newacronym{wsn}{WSN}{wireless sensor network}
\newacronym{dpc}{DPC}{dirty paper coding}
\newacronym{re}{RS}{rate splitting}
\newacronym{edrx}{eDRX}{extended DRX}
\newacronym{psm}{PSM}{power saving mode}
\newacronym{drx}{DRX}{discontinues reception}
\newacronym{bch}{BCH}{Bose-Chaudhuri-Hocquenghem}
\newacronym{dft}{DFT}{discrete Fourier transform}
        
%\newacronym{}{}{}
%\newglossaryentry{fast fading}{
%  name = fast-fading,
%  description={ridiiiiiiifast-fading}
%}

\begin{abstract}
	The next wave of wireless technologies is proliferating in
connecting things among themselves as well as to humans.
In the era of the Internet  of  things  (IoT), 
billions of sensors, machines, vehicles, drones, and robots will be connected, making the world around us smarter. The IoT will encompass devices that must wirelessly communicate a diverse set of data gathered from the environment for myriad new applications. The ultimate goal is to extract insights from this data and develop solutions that  improve quality of life and generate new revenue. 	 
Providing large-scale, long-lasting, reliable, and near real-time connectivity is the major challenge in enabling a smart connected world.  This paper provides a comprehensive survey on existing and emerging communication solutions for serving IoT applications in the context of cellular,  wide-area, as well as  non-terrestrial networks. Specifically, wireless technology enhancements for providing IoT access in  fifth-generation (5G) and beyond cellular networks, and communication networks over the unlicensed spectrum are presented. 
%like SigFox and \gls{lora} are presented.
Aligned with the main key performance indicators of 5G and beyond 5G networks, we investigate solutions and standards that enable energy efficiency, reliability, low latency, and scalability (connection density) of current and future IoT networks. 
The solutions include grant-free access and channel coding for short-packet communications, non-orthogonal multiple access, and on-device intelligence. 
Further, a vision of new paradigm shifts in communication networks in the 2030s is provided, and the integration of the associated new technologies like artificial intelligence, non-terrestrial networks, and new spectra is elaborated. In particular, the potential of using emerging deep learning and federated learning techniques for enhancing the efficiency and security of IoT  communication are discussed,
and their promises and challenges are introduced.  
Finally, future research directions toward beyond 5G IoT networks are pointed out.
\end{abstract}

\begin{IEEEkeywords}
IoT, IIoT, 5G, 6G, SigFox, LoRa, LTE-M, NB-IoT, security, reliability, survival time, service availability,  energy-efficiency, blockchain, SDN,  non-terrestrial, satellite, UAV, 3D, NOMA, random access, grant-free access, turbo code, LDPC, polar, deep leaning, federated learning.  \end{IEEEkeywords}
 
%\tableofcontents

\section{Introduction}\label{intro1}
{\let\thefootnote\relax\footnotetext{
% 		%Manuscript received January 16, 2017; revised May 5, 2017, and accepted July 16, 2017.
% 		%%Date of publication August 16, 2017; date of current version August 16, 2017.
% 		%This research was supported in part by the U. S. National Science
% 		%Foundation under Grant CMMI-1435778, and  in part by a Canadian NSERC fellowship.
 			M. Vaezi is with the Department of Electrical  and Computer Engineering, Villanova University, Villanova, PA, USA (e-mail: mvaezi@villanova.edu). 
		
 			A. Azari is with the Ericsson Research, Ericsson AB, Sweden (e-mail: amin.azari@ericsson.com).
			
			S. R. Khosravirad is with the Nokia Bell Labs, Murray Hill, NJ, USA (e-mail: saeed.khosravirad@nokia-bell-labs.com).
			
 			M. Shirvanimoghaddam is with the School of Electrical and Information Engineering, The University of Sydney, Australia (e-mail: mahyar.shm@sydney.edu.au).
 			
 			M. M. Azari is with Interdisciplinary Centre for Security, Reliability and Trust (SnT), University of Luxembourg, Luxembourg (e-mail: mohammadmahdi.azari@uni.lu)
			
			D. Chasaki is with the Department of ECE, Villanova University, Villanova, PA, USA (e-mail: danai.chasaki@villanova.edu).
			
		 Petar Popovski is with the Department of Electronic Systems, Aalborg University, Denmark (e-mail: petarp@es.aau.dk).
			
		%	H. V. Poor is with the Department of Electrical Engineering, Princeton University, Princeton, NJ 08544 USA (e-mail: poor@princeton.edu).
			}}

With over 20 billion connected devices in 2020, the \gls{iot} is transforming the business and consumer world in an unforeseen manner  and is driving a new industrial revolution. 
The term \gls{iot} was coined by Kevin Ashton at MIT's Auto-ID Center in 1999 \cite{IoTCisco} to define a network that not only connects people but also `things or objects' around them.
By  2009 there were more things or objects connected to the Internet than people \cite{IoTCisco}. 
Today, the \gls{iot} is a huge network of things---from simple sensors to smartphones, wearables, autonomous cars, and drones---which are connected together and are used to collect  and analyze  data to perform  actions. 
 Although  the forecasts by different analysts and consulting agencies  for \gls{iot}   market volume vary a lot, these all agree that \gls{iot} market volume is immense  and is growing at a significant rate.
According to IoT Analytics \cite{analytics2021}, by 2025 there will be more than 30 billion active \gls{iot} devices---smartphones are not included in this statistic. 
Business Insider forecasts that there will be more than 64 billion \gls{iot} devices by 2025 \cite{5GAmerica}.
The \gls{iot} market is expected to reach 1,854.76 billion US
dollars in 2028 compared to 381.3 billion US dollars in 2021  at a compound annual growth rate \gls{cagr} of 25.4\% \cite{BI}.

The \gls{iot} is about \textit{networks}, \textit{devices}, and \textit{data}.  It embodies the communication and interaction  between objects and embedded devices from
 diverse environments to collect  data, process  it,   and leverage resulting information and knowledge to make the world around us better via smarter and/or automated decision-making.
To progress  toward this goal, \gls{iot} devices, data, and networks are intertwined.  

\gls{iot} devices and their associated hardware are very diverse and include a wide range:  low-cost   sensing modules (e.g., temperature,  light, 
  pressure, humidity,  proximity  sensors) used in smart homes, industry, agriculture, logistics,  retails, etc.; wearable electronic devices worn on different parts of the body (e.g., glasses, helmets, collars, watches, wristbands, socks, shoes, etc.);  more powerful devices such as automation actuators, robot arms, drones, driver-less cars, and so on; as well as 
 desktops, tablets, and cellphones which are integral parts of \gls{iot} today.  
The diversity of devices and associated applications results in heterogeneous data
in various senses. 
 
% Examples of applications are 
%\begin{itemize}
%	\item []
%\begin{itemize}
%\item  Smart home (temperature and light control; optimized use of energy, water, gas) 
%\item Smart city (smart parking, lighting, garbage collection) 
%\item Industry (quality control,asset utilization, machinery management)
%\item Logistics, supply chain, and retail (product tracking, warehouse management, door delivery)
%\item Automotive industry (driver-less cars, car diagnostics, optimized traffic flow)
%\item Agriculture (water management, crop care,  soil analysis, drones for field monitoring)
%\item Healthcare (remote surgery, optimized patient care)
%\item Smart grids (smart metering, network management)
%\item Environment (forest fire detection, weather forecasting)
%\end{itemize}\end{itemize}
%

The data generated by \gls{iot} devices is large-scale streaming data that is heterogeneous, highly noisy, and highly correlated in space and time.  \gls{iot} data is classified according to 6Vs: \textit{volume} (large quantity), \textit{velocity} (high rate of data production), \textit{variety} (different types of data, e.g., text, audio, video, sensory data),  \textit{variability} (constantly changing), \textit{veracity} (level of confidence or trust in the data), and \textit{value} (valuable information and business insights out of it) \cite{mohammadi2018deep}. 
To extract valuable information and insights, such diverse data  is typically transmitted to the \textit{cloud} or fusion centers via wireless networks.

\begin{figure}[!t]
	\centering
	\includegraphics[width=\columnwidth]{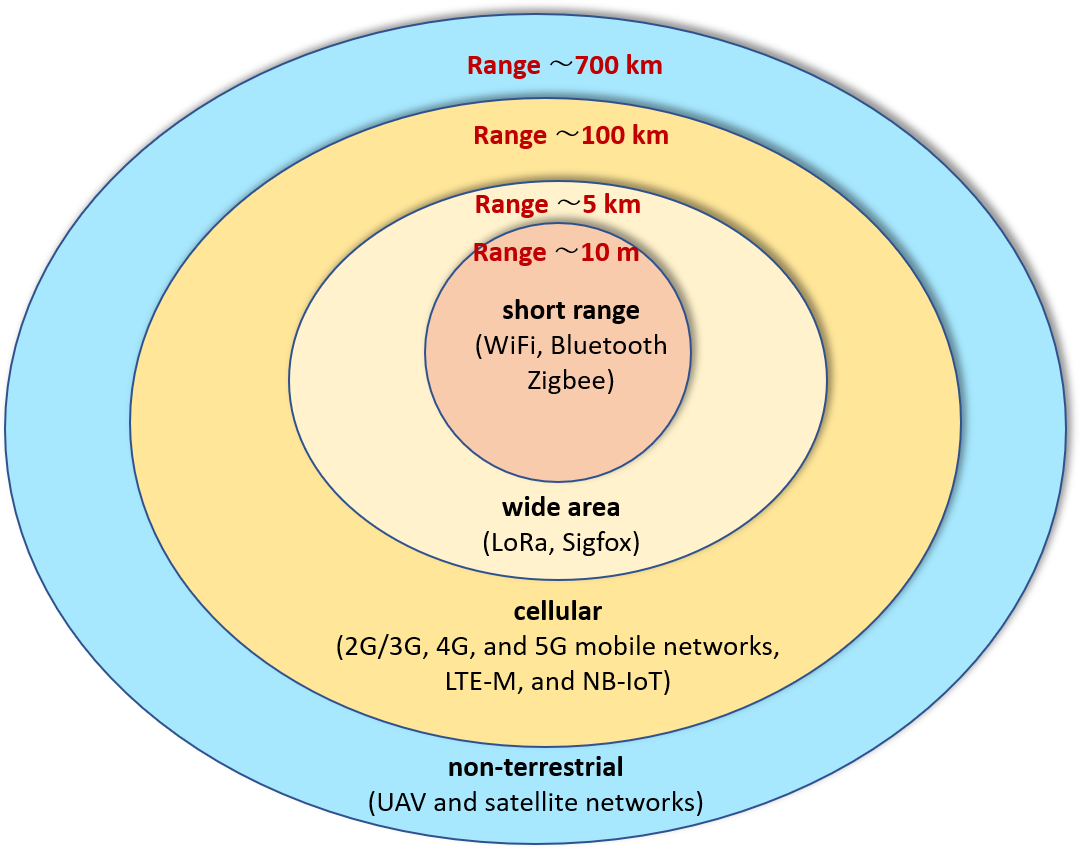}
	\caption{Different types of \gls{iot} networks in terms of range (non-terrestrial, cellular, wide-area, and short-range), their approximate coverage range, and main representative protocols.}\label{fig:3nets}
\end{figure}

Wireless access networks are crucial for the widespread adoption of \gls{iot} devices. 
\gls{iot} technologies have been extensively studied  and classified in recent
years.  A  widely accepted classification is in terms of coverage range  which divides \gls{iot} networks into \textit{cellular}, \textit{wide-area}, and \textit{short-range} \gls{iot} \cite{ericsson}. 
\textit{\Glspl{ntn}} represent a new category that provides extremely wide-area and \gls{3d} coverage.\footnote{The terms \glspl{ntn} and \gls{3d} networks are used interchangeably in this paper and refer to both \gls{uav}-based
and satellite-based systems.} \glspl{ntn} can be seen as a generalization of cellular and wide-area networks.
The approximate range of networks, along with some well-known examples of them, are illustrated in Fig.~\ref{fig:3nets}.
Cellular and  wide-area \gls{iot} are the fastest-growing types.  Ericsson mobility report projects \glspl{cagr}
of 23\% and 22\%  for these \gls{iot} types,   between 2020 and 2026 \cite{ericsson}. This will challenge the incumbent and emerging cellular networks in various senses.   

This survey is mainly about cellular,  wide-area, and  non-terrestrial  \gls{iot}, their applications and challenges. To better understand the challenges and requirements, we first elaborate on  existing access networks and main \gls{iot} applications  in the following.

\subsection{Application Types}

\begin{figure*}[ht]
	\centering
	\includegraphics[width=6in]{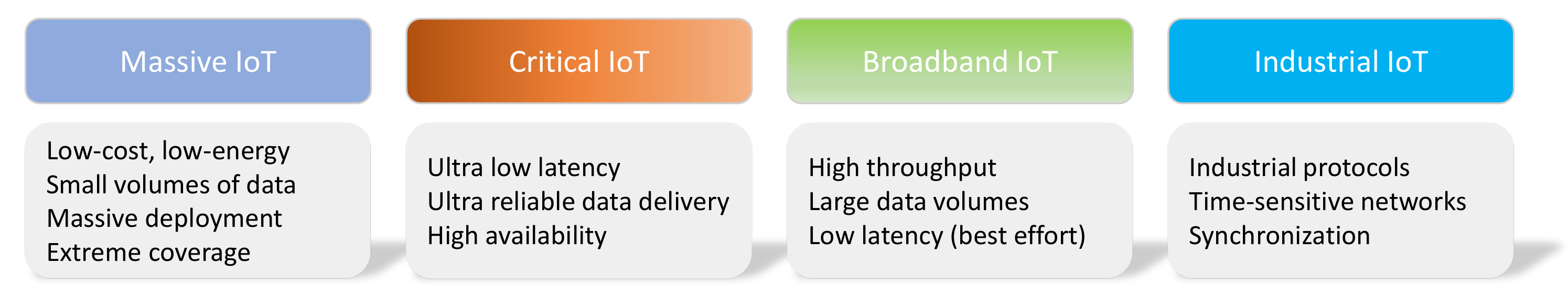}
	\caption{Different \gls{iot} segments and their characteristics \cite{ericsson2}.} \label{fig:IoTtypes}
\end{figure*}

In terms of applications (required data rate, latency, etc.), we can have two very different types of \gls{iot}: \textit{massive \gls{iot}} and \textit{critical \gls{iot}}. There are also two other important forms of \gls{iot}, namely, \textit{broadband} and \textit{industrial \gls{iot}}, as shown in Fig.~\ref{fig:IoTtypes} \cite{ericsson2}. These four  \gls{iot} segments are briefly described in the following.  

\begin{tikzpicture}
\node {IoT types}
[edge from parent fork right,grow=right, level distance=30mm, sibling distance=7mm]
child {node {Broadband IoT}}
child {node {Industrial IoT}}
child {node {Critical IoT}}
child {node {Massive IoT}
	%	child {node {short-range}}	
	%	child {node {wide-area}}	
	%	child {node {cellular}}
};
\end{tikzpicture}

\textit{Massive \gls{iot}}, a.k.a. \gls{mmtc}, describes applications with data collection purpose where a massive number of  low-power endpoints (e.g., low-cost sensors) continuously and infrequently (mostly sporadically) transmit small volume of data to the cloud or fusion centers. 
Examples of massive \gls{iot} are the sensors used to read the temperature, pressure, light, etc., in smart homes/buildings, smart meters, and smart agriculture.
In this case, transmitted data has a low volume but the number of \gls{iot} devices is very large.
Massive \gls{iot} devices are usually expected to work for up to 10 years without a change of battery. 
\gls{lte} networks already support massive \gls{iot} via \gls{nbiot}/\gls{lte-m}.

\textit{Critical \gls{iot}}, in contrast, involves fewer endpoints  handling large volumes of data. \gls{iot} devices in this category go beyond just data collection and represent high-bandwidth and low-latency applications, technically known as  \gls{urllc}.
Industrial control, robotic machines and autonomous vehicles are examples of critical \gls{iot}, all of which require  real-time data transmission.
These two types of \gls{iot} applications have very different requirements, as shown in Fig.~\ref{fig:IoTtypes}.

\emph{\Gls{iiot}} connects people, data, and machines  in the world of manufacturing and its virtual \emph{digital twin}.   The \gls{iiot} is a key enabler of 
\textit{Industry~4.0} \cite{lasi2014industry}  which refers to the fourth phase in the industrial revolution--the previous three being mechanical production (1784), mass production (1870), and then automation via electronics/computers (1970) \cite{Industry}.
Industry~4.0 is a digital transformation that focuses heavily on exploiting  inter-connectivity, automation, and real-time data.
\gls{iiot}, \gls{ai}, big data, cloud infrastructure, and cybersecurity are the key technologies working together to realize this transformation \cite{deloitte}.
The data created by the \gls{iiot} can be leveraged into optimizing efficiencies across the manufacturing operation from supply to distribution and  after-sale services. The most noteworthy benefits manufacturers can expect from Industry~4.0 are optimized processes, higher asset utilization, greater labor productivity, better insight into the entire supply chain and production process, and after-sale services. 

% \begin{tikzpicture}
% \node {Industry~4.0 technologies}
% [edge from parent fork right,grow=right, level distance=45mm, sibling distance=7mm]
% child {node {Cybersecurity}}
% child {node {Cloud infrastructure}}
% child {node {Big data/Analytics}}
% child {node {Artificial intelligence}}
% %child {node {Automation}}
% child {node {Industrial IoT}
% };
% \end{tikzpicture}	

\textit{Broadband \gls{iot}}
brings \textit{mobile broadband} connectivity to \gls{iot} devices, providing lower latency and higher throughput than the \gls{mmtc}. Typical applications are advanced wearables, drones, augmented reality, and virtual reality  devices and sensors that require greater capabilities than \gls{lte-m} or \gls{nbiot} can provide \cite{ericsson2}. \gls{lte}-capable smartwatches and \gls{lte} networks already supporting broadband \gls{iot}. \gls{lte}-connected drones two examples of such device categories.
 \Gls{5g} cellular networks will boost broadband \gls{iot} performance.

\subsection{Access Networks}

A reliable connection between devices, sensors, and  \gls{iot} platform is a key for any successful \gls{iot} project.		
\gls{iot} wireless networks must support a wide variety of connected devices and applications, as described earlier. Such a broad diversity makes it difficult, if not impossible, to have one technology that fits every situation. 	
Choosing the right network requires
consideration of various factors from coverage, power consumption, battery size, and  deployment cost. Hence, we have to make a trade-off somewhere by choosing any of these wireless technologies.

\begin {figure}[!tbp]
\small
\centering
\begin{tikzpicture}[edge from parent fork down]
\tikzstyle{every node}=[fill=red!30,rounded corners]
\tikzstyle{edge from parent}=[red,-o,thick,draw]
\tikzstyle{every sibling}=[fill=blue!90,rounded corners]
\tikzstyle{level 1}=[sibling distance=30mm]
\tikzstyle{level 2}=[sibling distance=15mm]
\tikzstyle{level 3}=[sibling distance=5mm]
\node {IoT Networks}
child {node {Cellular}
	child {node {LTE-M}}
	child {node { NB-IoT}}
}
child {node {WAN/MAN}
	child {node {Sigfox}} 
	child {node {LoRaWAN}}
	%	}
	%	child {node {Mesh}
	%		child {node {Zigbee}}
}
child {node {LAN/PAN}
	child {node {WiFi}}
	child {node {Bluetooth}}	
	%	child {node {BLE}}
};
\end{tikzpicture}
\caption{\gls{iot} access network types  some representative wireless protocols within each category.}
\label{fig:IoTnets}
\end{figure}
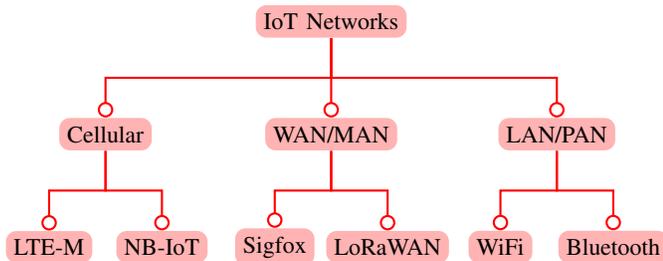

\gls{iot} wireless connectivity and technologies can be classified in several ways, e.g., in terms of coverage range (short, medium, long, or \gls{3d}), spectrum (licensed or unlicensed), data rates (low or high), technologies (cellular or non-cellular), power requirements (low or high), standard (\gls{3gpp}  or non-\gls{3gpp}), cost (low or high), etc.
A  common classification is to  have cellular \gls{iot}, wide-area \gls{iot}, and short-range \gls{iot} \cite{ericsson}, which respectively have long, medium, and short ranges. Within each category, there may be several wireless protocols as depicted in Fig.~\ref{fig:IoTnets}.
In particular, the cellular \gls{iot} standards can be divided as

 	\begin{forest}
 	for tree={
 		align=left,
 		edge = {draw, semithick, -stealth},
 		anchor = west,
 		font = \small\sffamily\linespread{.84}\selectfont,
 		forked edge,          % for forked edge
 		grow = east,
 		s sep = 0mm,    % sibling distance
 		l sep = 8mm,    % level distance
 		fork sep = 4mm,    % distance from parent to branching point
 		tier/.option=level
 	}
 	[Cellular \gls{iot} \\ solutions 
 	[\gls{5g}]  [\gls{nbiot}] [\gls{lte-m}] [4G]  [2G/3G] ] 	 
 	]
 \end{forest}

% \begin{tikzpicture}
% \node {Cellular IoT }
% [edge from parent fork right,grow=right, level distance= 30mm, sibling distance=6mm]
% child {node {\gls{5g}}}
% child {node {\gls{nbiot}}}
% child {node {\gls{lte-m}}}
% child {node {4G}}
% child {node {2G/3G}
% 	%	child {node {short-range}}	
% 	%	child {node {wide-area}}	
% 	%	child {node {cellular}}
% };
% \end{tikzpicture} 

%
%\begin{forest}
%	for tree={
%		align=left,
%		edge = {draw, semithick, -stealth},
%		anchor = west,
%		font = \small\sffamily\linespread{.84}\selectfont,
%		forked edge,          % for forked edge
%		grow = east,
%		s sep = 0mm,    % sibling distance
%		l sep = 8mm,    % level distance
%		fork sep = 4mm,    % distance from parent to branching point
%		tier/.option=level
%	}
%	[IoT \\ solutions \\ categories
%	[standard [non 3GPP][3GPP] ]
%	[power [high][low] ] 
%	[data rate [high][low] ]  [spectrum [unlicensed][licensed] ] 	[ range    [long][medium][short]] 	 
%	]
%\end{forest}
%

In general, \gls{iot} devices may access any of the above cellular networks including the legacy \gls{2g} and \gls{3g} cellular networks. 
%, as well as the incumbent \gls{4g} and \gls{5g} networks. 
These networks still play an important role and accommodate many \gls{iot} devices. Specifically, nearly one billion \gls{iot} devices are still connected to the legacy \gls{2g}/\gls{3g} networks (see \cite[Fig.~14]{ericsson}) and, according to   \cite{elnashar2019iot}, this number will increase. Also, an exponentially increasing number of broadband \gls{iot} devices are being connected to the incumbent \gls{4g} networks. However, the above solutions are not designed for massive \gls{iot} devices, which are low
power modules and need extended coverage.

 To handle such use cases, new solutions like \gls{nbiot} and \gls{lte-m} have been introduced in \gls{lte} networks.
\gls{nbiot} was standardized in the \gls{3gpp}  \gls{lte} Release 13 in 2016 to provide
wide-area connectivity for \gls{mmtc} \cite{kanj2020tutorial}.  
In \gls{lte} Release 14, \gls{nbiot}
was further developed to deliver an enhanced user
experience \cite{hoglund2017overview}.\footnote{The main  difference between \gls{nbiot} and
	legacy \gls{lte} is in the uplink design.  In \gls{nbiot} the
	subcarrier spacing can be 3.75 kHz as well as  the 15 kHz used in legacy
	\gls{lte} systems \cite{wang2017primer}.}  
\gls{nbiot} supports major \gls{mmtc} requirements, such as a long battery lifetime and a massive number of devices in a cell, and extends coverage  beyond existing cellular technologies.

Unlike \gls{4g}   networks which were primarily designed with MMB applications in mind, and \gls{iot} solutions like \gls{nbiot} came in later,  
\gls{5g} wireless access is intended not just to evolve mobile broadband but to be a key \gls{iot} enabler. Specifically,	\gls{5g} wireless access has been developed with
three broad  use cases  in mind \cite{Zaidi20175g}: 
\begin{itemize}
	\item [] 
	\begin{itemize} \item \gls{embb}
		\item massive machine-type
		communications (mMTC)
		\item ultra-reliable low-latency communications (URLLC)
\end{itemize}\end{itemize}

%
%In terms of spectrum, \gls{iot} access technology could be based on \textit{licensed} or \textit{unlicensed} bands. As the name implies, technologies in the licensed spectrum,  require a license to use a frequency channel.  These are basically cellular technologies and include \gls{2g}, 3G, \gls{lte},  \gls{nbiot}, \gls{lte-m}, and \gls{5g}.
%In general, solutions based on cellular networks provide a large range (coverage area) and a higher level of reliability but at relatively high cost \cite{de20195g}. 
%On the other hand, solutions based on unlicensed spectrum (e.g., WiFi, Sigfox, \gls{lora}, and  ZigBee) have lower cost but provide limited or no \gls{qos} as they do not have dedicated spectrum.

 \begin{table*} \small 
	\caption {\gls{5g} key performance indicators}
	\label{tab:5GKPI} 
	\centering
	\begin{tabular}{@{} cl*{4}c }
		\rowcolor{blue!20} \cellcolor{white}
		%		& & \multicolumn{14}{c}{Topic} \\[2ex]
		& \Tstrut KPI &  Requirement & Category \\
		\cmidrule{2-4}
		\rowcolor{black!15} \cellcolor{white}
		&Peak data rate		&  20 Gbps (downlink),
	  10 Gbps (uplink) & eMMB  \\
		&User  experienced data rate 	& 100 Mbps (downlink),
	50 Mbps (uplink)& eMMB   \\
			\rowcolor{black!15} \cellcolor{white}
		&Spectral efficiency		& 30 bit/s/Hz (downlink),
	  15 bit/s/Hz (uplink) & eMMB  \\
			&Latency (user plane)		&   4 ms for \gls{embb}, 
		1 ms for \gls{urllc} & eMMB, \gls{urllc}     \\
			\rowcolor{black!15} \cellcolor{white}
		&Reliability		&  $1-10^{-5}$  &  \gls{urllc}   \\
		&Energy efficiency	& Qualitative  & eMMB   \\
		\rowcolor{black!15} \cellcolor{white}
		&Connection density			&  $10^6$   devices/Km$^2$&  \gls{mmtc}   \\
			&Mobility			& up to 500 Km/h&  eMMB   \\
		\cmidrule[1pt]{2-4}
	\end{tabular}
\end{table*}

			\begin{figure}[!ht]
			\centering
			\includegraphics[width=3in]{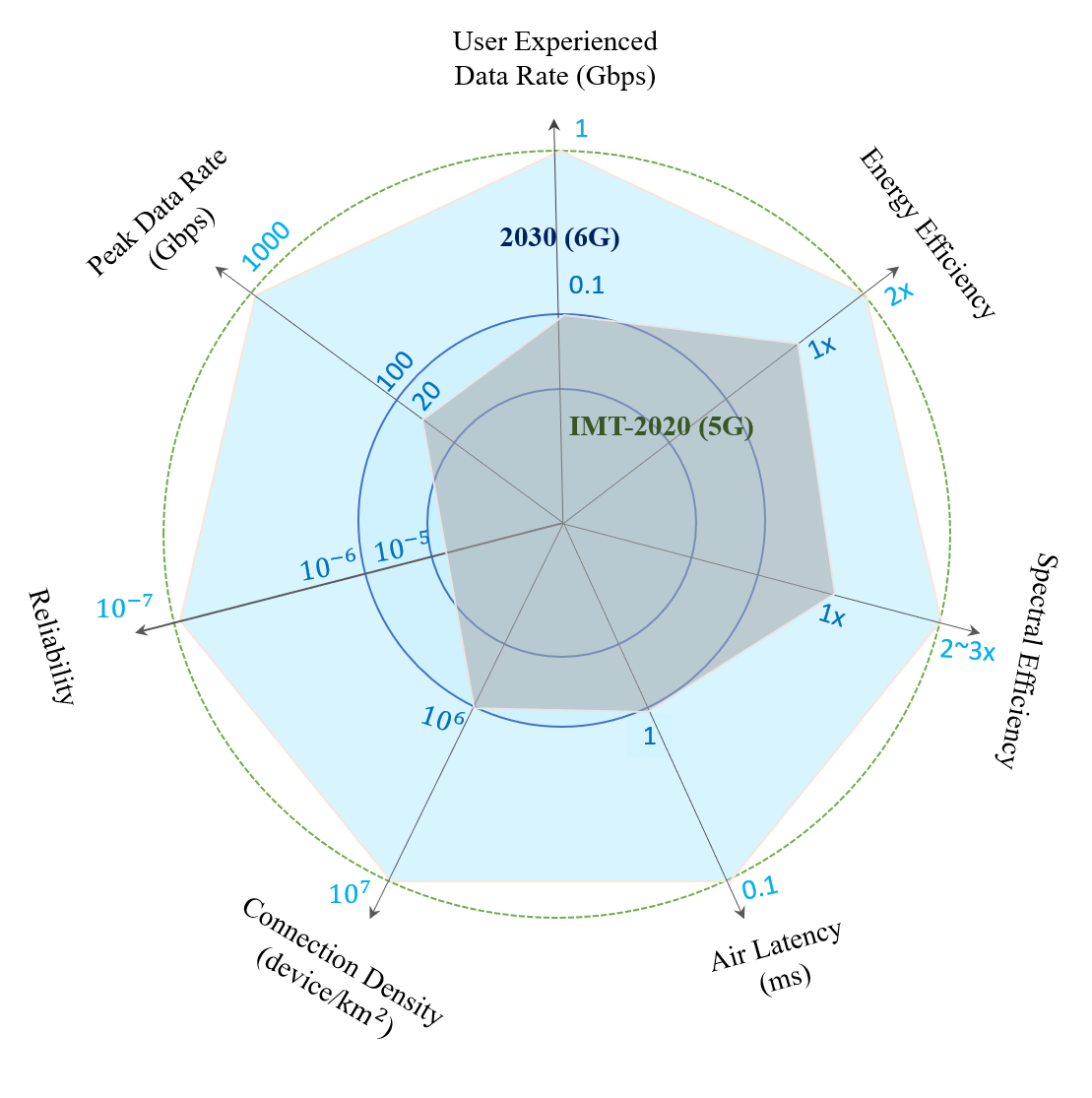}
			\caption{Expected key performance specifications and requirements in the 2030s  versus \gls{5g} (IMT-2020)} \label{fig:6G}
		\end{figure}

The \gls{iot} segments in Fig.~\ref{fig:IoTtypes} can be related to these uses cases as follows: \gls{mmtc} represents massive,  \gls{embb} represents broadband \gls{iot}, and  \gls{urllc} may represent both  critical and industrial \gls{iot}.
Besides, there are various \gls{kpi} requirements for \gls{5g} cellular networks as listed in Table~\ref{tab:5GKPI} \cite{ITUR2017}. Each KPI is related to one or more use cases. The KPIs on \gls{mmtc} and \gls{urllc} are specifically  related to \gls{iot}.
It is worth mentioning that, despite challenging requirements, only a subset of Industry~4.0 use cases are addressable
by current \gls{5g} KPIs, and  \gls{6g} still has much work to do.

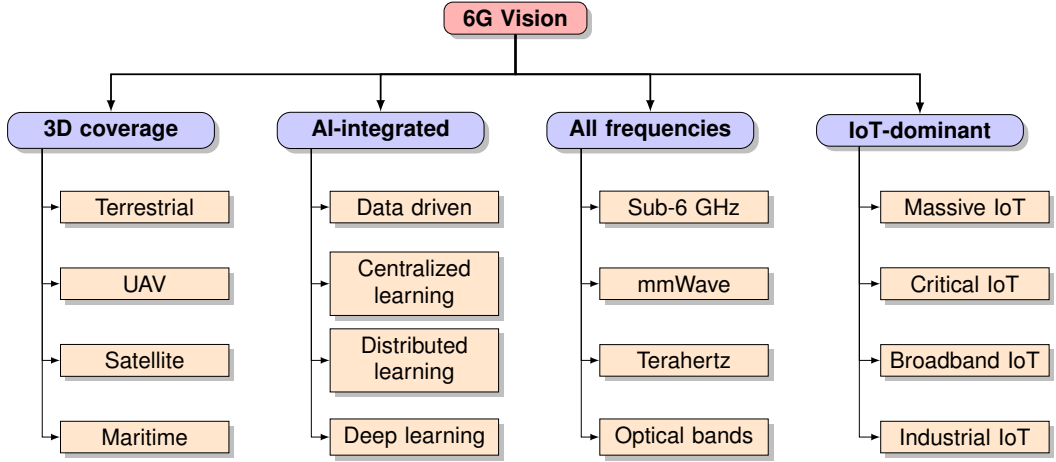
\begin{figure*}
\centering
\scalebox{.85}{
	\begin{tikzpicture}[
	level 1/.style={sibling distance=12em, level distance=5em},
	{edge from parent fork down},
	edge from parent/.style={->,solid,black,thick,draw}, 
	edge from parent path={(\tikzparentnode.south) -- (\tikzchildnode.north)},
	>=latex, node distance=1.2cm, edge from parent fork down]
	% define root %
	\node[root] {\textbf{\gls{6g} Vision}}	
	% Level-1 children %
	child {node[level-2] (c1) {\textbf{3D coverage}}}
	child {node[level-2] (c2) {\textbf{\gls{ai}-integrated}}}
	child {node[level-2] (c3) {\textbf{All frequencies}}}
	child {node[level-2] (c4) {\textbf{\gls{iot}-dominant}}};	
	% For Category 1 children %
	\begin{scope}[every node/.style={level-3}]
	\node [below of = c1, xshift=15pt] (c11) {Terrestrial};
	\node [below of = c11] (c12) {UAV};
	\node [below of = c12] (c13) {Satellite};		
	\node [below of = c13] (c14) {Maritime};	
	% For Category 2 children %
	\node [below of = c2, xshift=15pt] (c21) {Data driven};
	\node [below of = c21] (c22) {Centralized learning};
	\node [below of = c22] (c23) {Distributed learning};
	\node [below of = c23] (c24) {Deep learning};	
	% For Category 3 children %
	\node [below of = c3, xshift=15pt] (c31) {Sub-6 GHz};
	\node [below of = c31] (c32) {mmWave};
	\node [below of = c32] (c33) {Terahertz};
	\node [below of = c33] (c34) {Optical bands};
	%\node [below of = c34] (c35) {item 3-5};	
	% For Category 4 children %
	\node [below of = c4, xshift=20pt] (c41) {Massive IoT};
	\node [below of = c41] (c42) {Critical IoT};
	\node [below of = c42] (c43) {Broadband IoT};
	\node [below of = c43] (c44) {Industrial IoT};
	\end{scope}	
	% Draw arrows from level-1 to it's children %
	\foreach \value in {1,2,3,4}
	\draw[->] (c1.195) |- (c1\value.west);	
	\foreach \value in {1,...,4}
	\draw[->] (c2.195) |- (c2\value.west);	
	\foreach \value in {1,...,4}
	\draw[->] (c3.195) |- (c3\value.west);	
	\foreach \value in {1,2,3,4}
	\draw[->] (c4.195) |- (c4\value.west);
	\end{tikzpicture}
}
\vspace{.5cm}
\caption{A vision of new paradigm shifts in \gls{6g} communication networks.}
\label{fig:6Gvision}
\end{figure*}

	\subsection{\gls{6g} Vision and Requirements}

	Although industry and standardization bodies are still working around \gls{5g} mobile technology, and \gls{5g} commercialization is still in its initial stages, many researchers in academia and industry have already started asking what is next. 
	\gls{imt}\footnote{\gls{imt} is the generic term used by the \gls{itu}
		community to designate  mobile broadband systems.
		\gls{imt} standards are specifications and requirements for 
		mobile broadband service that fulfill established technical parameters such as peak data rate, latency, reliability,  spectrum efficiency, and so forth.
	}  plans to complete its \gls{6g} vision by 2023 and to develop technical requirements for \gls{6g} via industry standards organizations, such as the \gls{3gpp}.

	\subsubsection{Technical requirements}		
		Broadly speaking, the goal of \gls{6g} wireless networks is to move \gls{5g} forward  another order of magnitude on various technical parameters like throughput, number of connections, latency, and reliability.  \gls{6g} is also expected to make spectral and energy efficiency even greater than \gls{5g}, to allow  new connected services such as network-connected drones,  tactile Internet, real-time augmented reality glasses, etc., and to support mobility exceeding 1000 km/h for flying objects (e.g., airplanes) \cite{you2021towards,6gera}.
			
	Figure~\ref{fig:6G} illustrates the aspired requirement of  wireless networks in the 2030s, or loosely speaking \gls{6g} networks, based on the  \gls{itu}. The requirements are also compared with those of \gls{5g} networks, i.e.,  \gls{imt}-2020. 
	Other aspects of networks in the 2030s, such as   architecture, new use cases, evaluation methodology, are under study \cite{Network2030}.

		\subsubsection{Vision}		\gls{6g} is envisioned to bring new paradigm shifts to wireless access networks in terms of coverage, spectrum, applications, and security \cite{you2021towards}. 
	\begin{itemize}
		\item \textbf{3D coverage:} In the 2030s,  \glspl{ntn} such as \glspl{uav} \cite{shin2021uav,antennagain} and satellite  communication networks will complement  the  current terrestrial. This will result in a global 3D communication network overage in the ground, sea, and space.  
		\item \textbf{\gls{ai}-integrated:} \gls{ai} and machine learning have the potential to disrupt the design of future
		wireless networks. It has already penetrated network technologies and 
		will be ubiquitous  across all domains (core, access, edge, device), will impact the design of \gls{6g} networks,  and  will proliferate new smart applications. Particularly, big data technologies combined with deep learning will improve network management, automation, and efficiency. 
		\item \textbf{All frequencies:}		
		While \gls{5g} is based on the \gls{rf} bands, i.e., sub-6 GHz and \gls{mmwave} bands (spectrum between 30 GHz and 300 GHz), new frequency bands like  terahertz band (0.1–10 \gls{thz}) \cite{tripathi2021millimeter}
		and optical frequency band \cite{haas2020introduction,chowdhury2018comparative} will be exploited in \gls{6g}. 
		Moving into higher frequency bands  is a key trend for achieving
		1000 Gbps peak data rates 
		and for driving more values from the spectrum.
		
			\begin{tikzpicture}
		\node {6G bands}
		[edge from parent fork right,grow=right, level distance=30mm, sibling distance=8mm]
		child {node {Optical bands}}
		child {node {THz bands }}
		child {node {RF bands (5G)}
						child {node {mmWave}}	
						child {node {sub-6 GHz}}	
		};
		
		\end{tikzpicture} 
% 		\begin{itemize}
% 			\item \textit{sub-6 GHz bands} have been the main frequency in the \gls{3g}, \gls{4g}, and \gls{5g} for its wide coverage range (due to smaller path loss).
% 			\item \textit{mmWave bands} (spectrum between 30 GHz and 300 GHz) have been largely investigated to boost the sub-6 GHz Gbps level target data rate to about 10 Gbps in \gls{5g} and entail very different propagation characteristics and hardware requirements. 	
% 			\item \textit{THz band} (0.1–10 \gls{thz}) is envisioned as a key technology to satisfy the demand for ultra-high peak data rates (1000 Gbps) of \gls{6g} and is  a new frontier in   communications research \cite{tripathi2021millimeter}. \gls{thz} can enable users to ‘see-through’ materials and structures so it can be used for surveillance and security screening. 
% 			\item \textit{Optical bands}
% 			have more than three orders of magnitude larger spectral resources than the radio frequency spectrum (3 kHz -- 300~GHz) \cite{haas2020introduction,chowdhury2018comparative}. Optical wireless communication  technologies such as visible light communication,  wireless
% 			networking with light or light fidelity (LiFi),  and free-space optics. These technologies are  very timely for \gls{6g} cellular communications due to their wide bandwidth, low latency, high security,  and high energy efficiency. 
% 		\end{itemize} 
	Channel propagation
	characteristics, use cases, and required hardware for these bands are very different.
		\item \textbf{\gls{iot} dominant:} By 2030, cellular networks  will be \gls{iot} dominant. Ericsson projects that there will be 5.9 billion cellular \gls{iot} connections in 2026, up from 1.7 billion in 2020,  and  expressing a 23\% compound annual growth rate \cite{ericsson}. All four types of \gls{iot}  (massive, critical, industrial, and broadband) are growing exponentially. These devices are more and more going to connect to non-terrestrial, cellular, and wide area networks.

	\end{itemize}

 \subsection{Survey Scope and Related Works}

Beginning the digital transformation initiated by \gls{iot}, there are a  few trends focus across academia and industry:
\begin{enumerate}
		\item \gls{iot} creates true global connectivity.	
		\item \gls{ai} is managing \gls{iot} devices and decision-making.
		\item The   `edge' is becoming closer than ever before.
	\item \gls{iot} security issues are  rising continuously.

\end{enumerate} 

The high data rates, high reliability, global coverage, and low latency of \gls{5g} and beyond connections \cite{lin20215g}, coupled with cutting-edge \gls{ai} technologies, such as deep learning and federated learning, will be indispensable toward making the reality of the above trend.
On the other hand, there are several serious concerns  in the areas of privacy and security. Further, the continual rise of \gls{iot} security issues besides global connectivity makes \gls{iot} security threats and privacy issues more alarming and necessitates  novel solutions in this area. 

 \subsubsection{Survey scope} 
  This survey covers existing and emerging solutions related to  all three big players, i.e., \gls{5g} and beyond access, integration of \gls{ai} to \gls{iot}, and  security issues. We review a wide range of technologies that enable \gls{iot} applications in cellular,  wide-area, and  non-terrestrial  networks. Particularly we discuss the access networks over licensed and unlicensed spectrum. 
  For cellular \gls{iot}, we cover the advances made to satisfy most of the \gls{5g} KPIs, including reliability, energy efficiency, massive connectivity (grant-free and non-orthogonal access), and   latency.  We also investigate the main \gls{iot} security/privacy threats and solutions.
 In addition, we elaborate on several emerging technologies and solutions that are expected to be a part of \gls{iot} in the era of \gls{6g}. Specifically, we overview various deep learning architectures and federated learning  and explore the \gls{iot} applications that can
 benefit from these algorithms. Further, we discuss the remaining challenges and open issues that need to be addressed in the future. 
This survey does not cover
 traditional machine learning algorithms for \gls{iot}. It also
 does not go into the details of the \gls{iot} standards and infrastructure.

 \subsubsection{Related works} 
To the best of our knowledge, there is not a paper surveying
\gls{iot} advances related to all \gls{5g} \glspl{kpi} in the literature.
There is also no paper about \gls{iot} solutions and architectures beyond  \gls{5g} and their relation to those in \gls{5g}. 
 There are, however,  a number of \gls{iot} surveys and tutorials related to certain topics covered in this survey. These surveys and their primary focus are listed in Table~\ref{tab:related-survey}. Our work is much more comprehensive and covers almost all \gls{5g} KPIs, as well as emerging \gls{ai} and security solutions. 
 
 \begin{table*} 
 	\caption {Existing \gls{iot} surveys and tutorials and their primary focus}
 	\label{tab:related-survey} 
 	\centering
 	\begin{tabular}{@{} cr*{14}c }
 		\rowcolor{blue!20} \cellcolor{white}
 		%		& & \multicolumn{14}{c}{Topic} \\[2ex]
 		& Year & Reference & Main Topic & \rot{Reliability/Latency} &\rot{Energy Efficiency} & \rot{Spectral Efficiency}&  \rot{\shortstack[l]{Non-Terrestrial\\  Networks}}  &  \rot{Network Slicing} &\rot{\shortstack[l]{Grant-Free Access \\ NOMA}}  
 		& \rot{Access Networks} &   \rot{Security/Privacy} 
 		& \rot{Deep Learning} & \rot{Federated Learning} & \rot{Positioning} &  \rot{6G Vinson} \\
 		\cmidrule{2-16}
 	
 		&2021	& This work & Cellular, wide-area, and \gls{3d} IoT   &\OK &\OK  & & \OK &  &\OK &\OK &\OK  &\OK & \OK &  &\OK \\
 		\rowcolor{black!15} \cellcolor{white}
  		&2021	&\cite{nsr}&Network slicing for IoT    &  &   & &  &  \OK &  &  &   &  &   &  &  \\
  		&2021	&\cite{posr}& Positioning for IoT    &  &   & &  &   &  &  &   &  &   & \OK &  \\
 		\rowcolor{black!15} \cellcolor{white}		
 		&2021	& \cite{khalil2021deep} & Deep learning in IIoT &    &   &   &   &   &   &   &   &\OK   &   &   &   \\
 		&2021	& \cite{s21} & \gls{6g} vision for IoT&    &   &   &   &   &   &   &   &   & \OK  &   &\OK   \\
 		\rowcolor{black!15} \cellcolor{white}
 		&2020	& \cite{stoyanova2020survey} & IoT  forensics &   &   &   &   &   &   &   &\OK   &   &   &   &   \\
 		&2020	& \cite{elbayoumi2020noma} & NOMA-assisted MTC   &    &   &   &   &   &\OK   &   &   &   &   &   &   \\
 		\rowcolor{black!15} \cellcolor{white}
 		& 2020	& \cite{shahab2020grant}  &    Grant-free NOMA for IoT &  & &&&&\OK&& & &  & &\\
 		& 2020	& \cite{lei2020deep}  &    DRL for autonomous  IoT &  & &&&&&& &\OK &  & &\\
 		\rowcolor{black!15} \cellcolor{white}
 		 &2020	& \cite{maraqa2020survey} & Power-domain NOMA &   &   &   &   &   & \OK   &   &   &   &   &   &   \\
 		& 2020	& \cite{hussain2020machine}  &     ML in IoT security  &  & &&&&&&\OK &\OK &  & & \\	
 		\rowcolor{black!15} \cellcolor{white}
 		& 2020	& \cite{kanj2020tutorial}  &    	Narrowband IoT PHY/MAC &  & &&&&&\OK& & &  & &\\
 		& 2020	& \cite{s20}  &    	Access control in IoT &  & &&&&&&\OK & &  & &\\
 		\rowcolor{black!15} \cellcolor{white}
 			& 2020	& \cite{s20l}  &   Enabling technologies  for IoT &  &\OK &&&&&\OK& & &  & &\\
 		 & 2019	& \cite{de20195g}  &    \gls{5g} waveform for  IoT &  & &&&&&\OK& & &  & &\\
 		 \rowcolor{black!15} \cellcolor{white}
 		 & 2019	&  \cite{chettri2019comprehensive}  &    Enabling technologies IoT-\gls{5g}  &  & &&&&&\OK& & &  & &\\
 		 % & 2019	&  \cite{zhang2019deepTut}  &    DL for mobile networks  &  & &&&&&& & \OK&  & &\\
 		 & 2019	&  \cite{s19g}  &   Energy-efficient IoT  &  &\OK &&&&&& & &  & &\\
 		\rowcolor{black!15} \cellcolor{white}
 		 & 2018	& \cite{stellios2018survey}  &  IoT-enabled cyberattacks  &  & &&&&&& \OK& &  & &\\
 		 			
 		 & 2018	& \cite{mohammadi2018deep} & Deep learning for IoT   &    &   & &&&&&& \OK& &  & \\
 	%	&2017	& \cite{lin2017survey} & Enabling technologies   &    &   &   &   &   &   &   &   &   &   &   &   \\
 	  \rowcolor{black!15} \cellcolor{white}
 		 & 2017	& \cite{akpakwu2017survey}  &     Communication technologies  &  & &&&&&\OK& & &  & & \\
 		
 		 & 2017	& \cite{xu2017narrowband}  &    	Narrowband IoT   &  & &&&& &\OK& & &  & &\\
 		   \rowcolor{black!15} \cellcolor{white}
 		 	 & 2017	& \cite{s17au}  &    	IoT security    &  & &&&& &&\OK & &  & &\\
 		 & 2016	& \cite{sua} & Aerial IoT   &   &   & &\OK&&&&& & &  & \\
 		  		   \rowcolor{black!15} \cellcolor{white}
 		 & 2015	& \cite{al2015internet} & Enabling technologies   &\OK    &   & &&&&\OK&\OK& & &  & \\
 			\cmidrule[1pt]{2-16}
 	\end{tabular}
 \end{table*}

   \subsection{Contributions} 
   
   %This paper is intended for IoT researchers 
  
\begin{figure*}
 	\begin{forest}
 		for tree={
 			align=left,
 			edge = {draw, semithick, -stealth},
 			anchor = west,
 			font = \small\sffamily\linespread{.84}\selectfont,
 			forked edge,          % for forked edge
 			grow = east,
 			s sep = 0mm,    % sibling distance
 			l sep = 8mm,    % level distance
 			fork sep = 4mm,    % distance from parent to branching point
 			tier/.option=level
 		}
 		[Papers \\Structure
 		[{\color{blue}Section X:}   Conclusions  
 		]
 	%%%%%%%%%%%%%%%%%%%%%%%%%%%%%%%%%%
 		 		[{\color{blue}Section IX:}  Remaining Challenges  and Future Directions 
 		]
 		%%%%%%%%%%%%%%%%%%%%%%%%%%%%%%%%%%
% 		[{\color{blue}Section VI:  \\Remaining Challenges \\ and Future Directions}  
% 		[Future Directions]
% 		[NOMA for IoT [ Wake-up Receivers][Grant-Free Access]]
% 		[Limited Battery Lifetime]
% 		]
 		 %%%%%%%%%%%%%%%%%%%%%%%%%%%%%%%%%%
 		 		[{\color{blue}Section VIII:} \\ 
 		Non-Terrestrial \\ IoT Networks  [Satellite-Integrated  Networks] [\gls{uav}-Integrated  Networks [Energy  and Information Transfer] [Localization] [Data Collection]]]
 		 		%%%%%%%%%%%%%%%%%%%%%%%%%%%%%%%%%%
 		 [ {\color{blue}Section VII: }  \\ Deep  \\Learning \\ for  IoT
 	    [IoT-Friendly DL [Network  Compression] [Hardware Acceleration]]  [Federated Learning]  [Deep Reinforcement   Learning] [Autoencoder]  [Deep Learning Models] 	[Neural Networks] 
 		]
 		%%%%%%%%%%%%%%%%%%%%%%%%%%%%%%%%%%
 			[{\color{blue}Section VI: \\IoT \\Security}  
 		[ SDN and Blockchains [Privacy and Trust]  [SDN] ]
 		 [Security Attacks  [ Network] [ Software] [Physical]]]
 		%%%%%%%%%%%%%%%%%%%%%%%%%%%%%%%%%%
 		[{\color{blue}Section V: \\Massive \\Connectivity}  
 		[ Random Access [Unsourced]  [Grant-free ]  [Grant-based ]
 		] [ Multiple Access [Uplink NOMA] [Downlink NOMA]]]
 		%%%%%%%%%%%%%%%%%%%%%%%%%%%%%%%%%%
 		[{\color{blue}Section IV: \\Channel Code \\ Design}  
 		[Decoder Design][ \gls{csi}-free Codes][State-of-the-Art]]
 		 	%%%%%%%%%%%%%%%%%%%%%%%%%%%%%%%%%
 	%	[{\color{blue}Section IV: \\ Enabling  \\ Solutions}  
 	%	[mmWave IoT ]
 	%	[ \gls{d2d} Relaying  [Relaying 	for Reliability] [Relaying for Coverage]]
 	%	]
 		%%%%%%%%%%%%%%%%%%%%%%%%%%%%%%%%%
 		[{\color{blue}Section III: \\ Enabling \\ Access \\ Solutions}  
 	  [Domain of Future Access Solutions] [Unlicensed  Spectrum]	[Licensed  Spectrum] 
 		]
 		%%%%%%%%%%%%%%%%%%%%%%%%%%%%%%%%%
 		[{\color{blue}Section II:}  \\ Key \\Performance \\ Indicators 
 		 	[Data Rate] 	[Connection density]  [Energy Efficiency] [Age of Information]	[Survival  Time and  Service Availability] 		[Latency and  Reliability] 
 		]
 		%%%%%%%%%%%%%%%%%%%%%%%%%%%%%%%%%
 		[{\color{blue}Section I:}   \\ Introduction
 		[Contributions] [Survey Scope and Related Works][\gls{6g} Vision  [Technical  Requirements] [\gls{6g} Vision ]] 
 		[Access Networks] [Application Types]
 		]
 		]
 	\end{forest}
 \caption{Structure of the paper.} \label{fig:str}
 \end{figure*}
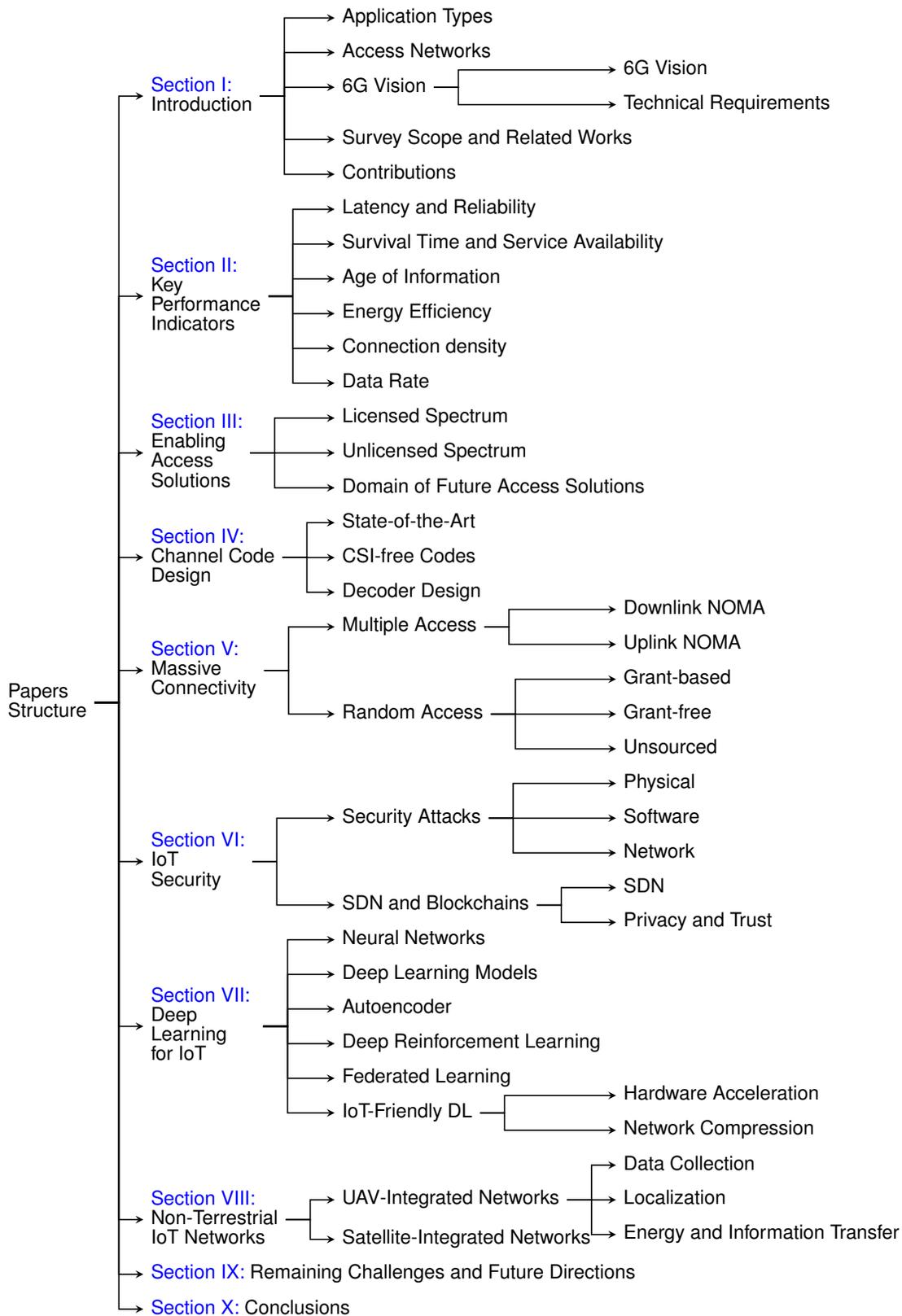

   Figure~\ref{fig:str} depicts
the structure of the paper.
 The contributions of this paper
   can be summarized as follows:  
   
 In Section~\ref{sec:kpis}, a comprehensive overview of the \glspl{kpi} and service requirements for \gls{iot} is presented. The various envisioned \gls{iot} use cases have a diverse set of service requirements, often in deep contrast with one another. Namely, we address the importance of latency and reliability for mission-critical \gls{iot}, survival time and service availability for cyber-physical and industrial systems, \gls{aoi} for sensor networks, battery lifetime and energy efficiency, connection density of \gls{iot} devices in the envisioned deployments, and finally, the required data rates and spectral efficiency. We address the   outlined requirements for  \gls{5g} as well as   the  visions  for  cyberphysical  and \gls{iiot}  applications while keeping an open eye on the visions  for  future  generations  of  cellular and wide-area \gls{iot} technologies, particularly with respect to \gls{6g} developments.
 
 In Section~\ref{Sec:enablingtech},  major IoT connectivity providers have been introduced and compared. Namely, we have categorized the existing technologies into solutions over the licensed and unlicensed spectrum. Then, for each category, major technologies in the market have been introduced. Finally, a comparative study of IoT technologies, including (i) communications characteristics, e.g. type of radio access and frequency; (ii) performance indicators, e.g. link budget and sensitivity; and (iii) major limitations, e.g. data rate and number of packets per day,   has been presented. 
 %The enabling access technologies for \gls{iot} are discussed in \secref{enablingaccesstech}. Particularly, we 
 We further address the significant but challenging potentials of \gls{mmwave} access technologies for future high-rate \gls{iot} access. Moreover, we discuss relaying, especially in form of \gls{d2d} relaying, as an enabling scheme for extending coverage, improving spectral efficiency, and enhancing reliability for \gls{iot} systems. The role of cooperation between networks and devices is discussed as a necessary leap to enable efficient, low-energy, and yet dependable communication service. The vision is to move away from rigid network architectures where all devices communicate directly with a network access point and enable a fluid architecture where devices and the network can form mesh-type sub-networks in an on-demand manner.

Section~\ref{sec:channelcode} provides an overview of channel coding techniques for \gls{5g} and future wireless communication technologies for \gls{iot}. We first review the well-known channel coding techniques that have been considered for cellular systems and highlight their pros and cons for \gls{iot} applications. We then explain new requirements for channel coding and decoding mainly imposed by new \gls{iot} scenarios. This includes low-capacity communications and \gls{csi}-free communications. We then review some rate-adaptive coding techniques which can be effectively applied in these scenarios. We finally highlight the main requirements of the decoding techniques for resource-constraint devices.

Section~\ref{sec:NOMA} discusses massive connectivity solutions in \gls{5g} and beyond where solutions based on multiple access, random access, and new frequencies are considered. In multiple access, we discuss how new, non-orthogonal multiple access (NOMA) schemes in the uplink and downlink can increase the number of connected devices. We describe the state-of-the-art in power-domain NOMA as well as various code-domain NOMA schemes. 
Random access could be grant-based, grant-free, or unsourced. We argue that the last two are more suitable for massive IoT whereas the first one could be used for broadband IoT applications.  
Further, we describe how new frequencies (e.g., \gls{mmwave} and \gls{thz}) are useful for broadband and critical IoT and can be used to accommodate such applications.

Section \ref{sec:security} provides an overview of the most common security threats in cellular,  wide-area, and  non-terrestrial  IoT deployments. We categorize possible attacks in terms of their attack medium: physical, software, or network attacks, and then discuss how emerging networking and database technologies can help address some of the security threats in IoT. Finally, going forward, we stress the need for better device authentication since traditional certificates/key exchange algorithms do no scale well. Better hardware-based authentication and key generation schemes, that can not be tampered with, are available in the FPGA/hardware trust research community and can be directly applied to future cellular,  wide-area, and  non-terrestrial  networking devices.

Section~\ref{sec:AI} covers deep learning, distributed learning, and federated learning for IoT. We being with a  tutorial on various supervised and unsupervised deep learning models, their advantages, and limitations for IoT uses. We then overview recent advances made using the above learning methods in 
various aspects of the IoT, including physical layer, network layer, and application layer. We also discuss
characteristics and needs of the IoT-friendly deep learning.
Particularly, we describe recent literature on network compression and hardware acceleration.

% Section \ref{sec:NTN} describes how \glspl{ntn}  are capable of providing a larger and global coverage which are less vulnerable to disasters. Such an inherent feature of \glspl{ntn} along with the advances in the more cost-effective small satellites/UAVs and their corresponding industries are leading the world towards new types of IoT services through space/air. 
% %Motivated by this, in Section \ref{sec:NTN} we overview SoA regarding various aspects of NTNs integration into IoT networks, use cases, challenges and existing solutions.
Section \ref{sec:NTN} provides an overview of the role of \glspl{ntn} in IoT, the corresponding challenges, and opportunities. 
Section \ref{sec:future} sheds light on the major remaining challenges of IoT communications, including the battery lifetime of IoT devices. Furthermore, several promising solutions for enabling further energy efficiency in communications have been explored. Among these solutions, great emphasis has been put on grant-free access for uplink-oriented communication and wake-up receiver for downlink-oriented communication, as two unexplored areas in cellular-based IoT communications, with great potential in saving energy for IoT devices.

A complete list of abbreviations is given in the Appendix. %~\ref{sec:appendix}.
%   The remainder of this paper is organized as follows. 

%--------------------------------------------------------
 
\section{Key Performance Indicators}\label{sec:kpis}

The envisioned requirements for \gls{iot} are as diverse as the plethora of use cases that are associated with it. In this section, we provide a  comprehensive overview of the  requirements and \glspl{kpi} that are outlined by technical expertise and industry visions. To this end, we build upon the outlines  proposed in \cite{imt2020} as part of the \gls{5g} requirements, in \cite{5gacia2019} and \cite{3gpp22104} as the visions for cyber-physical control applications, and, in \cite{6gera} as the visions for  future generations of cellular telecommunications technologies.

%---------------------------
\begin{figure*}[t]
\begin{center}
\includegraphics[width=\textwidth,keepaspectratio]{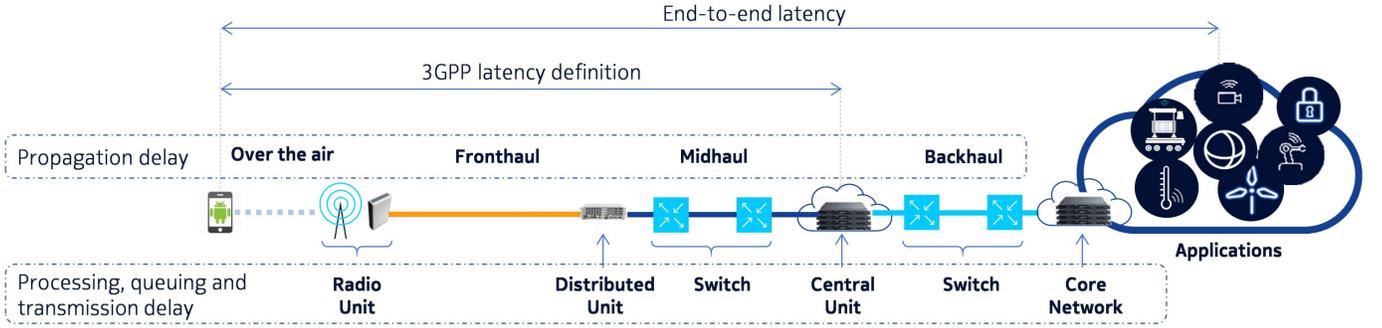}
\caption{Breakdown of the various elements of delay in E2E communication service latency.}
\label{Fig:latencybreakdown}
\end{center}
\end{figure*}
%----------------

\subsection{Latency and Reliability}\label{kpi:latencyreliability}

The initial momentum for low-latency wireless communication in the \gls{5g} era targeted  mission-critical communication. The  \gls{5g} vision aimed to achieve  1 ms  latency for small size packets (see \gls{3gpp} definition illustrated in \figref{Fig:latencybreakdown}), used in applications such as remote surgery and machine control. Such applications also required high reliability of packet delivery. Therefore, in \gls{3gpp} Release 15, the main target of air interface, \gls{phy} and \gls{mac} layer advancements was to enable   99.999$\%$ reliability within a 1 ms latency for 32 bytes packet size~\cite{pedersenflexible}. This initiated a category of use cases, commonly addressed under the umbrella term \gls{urllc}. New access techniques including flexible \gls{tti} size and mini-slots~\cite{7842312},  grant-free  channel access and preemptive scheduling (also know as punctured scheduling)~\cite{8287951} were deployed for low latency, while multi-connectivity with multiple access points~\cite{Rebal:2018TCOM}, enhanced \gls{harq} techniques~\cite{pedersen2017harqmag}, and carriers and packet duplication~\cite{mahmood2019resource} were introduced for reliability. The applications for critical communications have since expanded significantly, currently including use cases from energy management and smart  grid control, health care and telesurgery, entertainment and gaming, and mobility and traffic control. Some of the new applications require lower latency, higher reliability and in some cases, significantly high data rates. In \tabref{tab:QoS_latency}, a few of those use cases are depicted according to \cite{3gpp22263} based on performance requirements of professional low-latency audio and video transport services. The terminology in \tabref{tab:QoS_latency} are defined  as follows:

\paragraph{Transfer interval}  the  time difference between two consecutive transfers of packets from the application to the wireless communications system.

\paragraph{End-to-end latency}  the time that takes to transfer a given piece of information from a source to a destination, measured at the communication interface, from the moment it is transmitted by the source to the moment it is successfully received at the destination.

As illustrated in \figref{Fig:latencybreakdown},  \gls{e2e} latency can generally consist of several delay components, including but not limited to, the propagation delay over the air and over x-haul links, as well as the processing and queuing delay in switches and processing units. Among those, processing delays and x-haul propagation delay are critical in the experienced round-trip delay~\cite{khosravirad2016overview}. For applications that sit far away from the user, the physical distance between the two ends of the communication link is a strict limiting factor on how small the \gls{e2e} latency can get. For instance, the telesurgery entry in \tabref{tab:QoS}, denoted by motion control and haptic feedback, an upper bound of 20 ms is envisioned for \gls{e2e} latency due to the fact that the communicating points may be sitting kilometers away from each other. On the other hand, some applications require sub-milliseconds \gls{e2e} latency as listed in \tabref{tab:QoS} and \tabref{tab:QoS_latency}. For those applications, it is important to bypass as many delay components as possible of the ones  depicted in \figref{Fig:latencybreakdown}. For instance, for motion control applications the controller and the actuators are usually located in close proximity of each other, thus allowing the communication link to be established by the central unit of the \gls{gnb}, without the need to go through the core network. In such situation, the processing and routing of the traffic are controlled by the \gls{gnb} in the \emph{edge} cloud, as opposed to the core cloud. With proper design of the air interface, and ensuring a smooth queuing process at the \gls{gnb}, this will enable \gls{e2e} latency of sub-milliseconds in future cellular technologies.

Incidentally, the use case requirements are widely different in terms of reliability and latency, ranging from $10^{-4}$ to $10^{-10}$ \gls{per}, only in the presented examples. Furthermore, the differences in traffic types call for various types of radio access techniques, including stationary point-to-point, mobile device-to-device, and multi-cast and broadcast access.

\begin{table*}[t]
\scriptsize
%\footnotesize
\centering
\caption{Performance requirements of professional low-latency  audio and video transport services~\cite{3gpp22263}}.
\label{tab:QoS_latency}
\begin{tabular}{|m{0.1\textwidth}|p{0.05\textwidth}|p{0.045\textwidth}|p{0.04\textwidth}|p{0.06\textwidth}|p{0.06\textwidth}|p{0.1\textwidth}|p{0.2\textwidth}|}
\hline 
\Tstrut \textbf{Packet  error rate}  & \textbf{E2E \;\;latency} &  \textbf{Transfer interval} & \textbf{Number of UEs} & \textbf{Uplink data rate} 	& \textbf{Downlink data rate}  & \textbf{Service area} & \textbf{Application}   \\  \hline \hline
$10^{-6}$ \Tstrut &
750 $\mu$s &
250 $\mu$s &
30 &
5 Mbit/s & 
NA &
10 m $\times$ 10 m
& Audio studio (periodic deterministic traffic) \\[.5ex] \hline
$10^{-6}$ \Tstrut &
750 $\mu$s &
250 $\mu$s &
10 &
NA &
3 Mbit/s & 
30 m $\times$ 30 m
& Audio and video production (periodic deterministic traffic) \\ [1ex] \hline
$10^{-6}$ \Tstrut&
750 $\mu$s &
250 $\mu$s &
200 &
500 kbit/s & 
NA &
500 m $\times$ 500 m
& Music festival (periodic deterministic traffic) \\ [.5ex]\hline
$10^{-10}$  \Tstrut uplink $10^{-7}$ downlink  &
400 ms &
--- &
1 &
12 Gbit/s & 
20 Mbit/s &
500 m $\times$ 500 m
& 
Uncompressed \gls{uhd} video \\ \hline
$10^{-8}$  \Tstrut uplink $10^{-7}$ downlink  &
3 ms &
--- &
10 &
100 Mbit/s & 
20 Mbit/s &
1 km$^2$
& 
non-public networks  radio Camera \gls{uhd}\\ \hline
$10^{-4}$   \Tstrut &
7 ms &
--- &
50000 &
--- & 
200 kbit/s &
1500 m $\times$ 1500 m & 
Integrated multi-cast audience services (periodic deterministic traffic)\\ \hline
\end{tabular}
\vspace{-10pt}
\end{table*}

\begin{figure*}[t]
\begin{center}
\includegraphics[width=.6\textwidth,keepaspectratio]{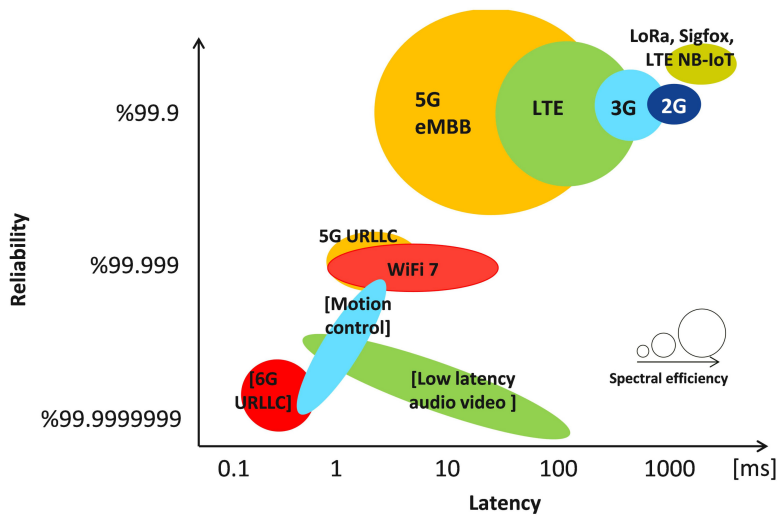}
\caption{Latency and reliability capability of different technologies. Envisioned use cases and technologies are noted in brackets.}
\label{Fig:latencyvsreliability}
\end{center}
\end{figure*}

For future cellular,  wide-area, and  non-terrestrial  technologies in the \gls{6g} era,   lower radio latencies of the order of 100 $\mu$s at Gbits/s data rates are required to enable extreme networking solutions~\cite{6gera,Bennis:2020arxiv}. This will for instance enable replacing traditional industrial wireline connectivity solutions such as Sercos or EtherCAT. The  requirement on reliability for some applications, such as in \tabref{tab:QoS_latency}, is significantly more challenging as compared to \gls{5g} \gls{urllc}. For others, such as cyber-physical applications, reliability is  based on the downtime of the control application which is represented by the number of consecutive packet errors (for requirements concerning those applications, see \secref{kpi:survival}).  \figref{Fig:latencyvsreliability} compares the latency and reliability performance of different wireless technologies addressed in this paper. The numbers in this figure must be taken as approximated based on the reported measurements and envisioned capabilities of different technologies in the literature. For the future generation wireless network, some extreme use cases are anticipated in \figref{Fig:latencyvsreliability} according to \gls{6g} \gls{urllc}~\cite{6gera}, WiFi 7 promises to support low latency and ultra-reliable services~\cite{adame2019time}, and the use cases in \tabref{tab:QoS_latency} and \tabref{tab:QoS}.

\subsection{Survival Time and Service Availability}\label{kpi:survival}

In the context of  cyber-physical and industrial  systems, dependable and real-time wireless communications are crucial to enable mobility. The wireless transformation in future industrial automation will reduce the bulk and cost of installation while enabling a highly flexible and dynamically re-configurable  environment. This vision is shared by various other use cases too, including,   smart grid, mobility and traffic, health care, entertainment, and gaming. The challenge is for the wireless communication system to  be as dependable as a wired connection, i.e., at extremely high reliability while guaranteeing every time/everywhere service.  Those qualities are realized in form of \gls{qos} characteristics introduced by \gls{3gpp} in the context of wireless cyber-physical systems, using the  new notions of survival time and service availability~\cite{3gpp22104,38.824}. \tabref{tab:QoS} shows a few example applications from~\cite{3gpp22104}. The terminology in \tabref{tab:QoS} is defined  according to the interaction between the application and the wireless communications system, as follows:

\paragraph{Survival time}  the time that an application consuming a communication service may continue without an anticipated message. The anticipation is due to the periodic pattern of the traffic.

\paragraph{Communication service availability}  the  percentage value of the amount of time the    \gls{e2e} communication service \gls{qos} requirements---including \gls{e2e} latency and survival time requirements---are satisfied for the application.

In light of these definitions, some rather interesting observations can be made from the examples provided in \tabref{tab:QoS}. The duration of the survival time depends on the  application. In fact, for  motion control systems  responsible of controlling moving parts of machines, e.g., printing  or packaging machines, the survival time is comparable to the transfer interval. This is also true for motion control with haptic feedback, e.g., in the case of robotic-aided telesurgery where the feedback provides tactile guidance for the patient body model. Mainly, the high-precision of operation that is vital in those use cases,  translates into a tolerable duration of lost packets comparable to the transfer interval. On the other hand, in some applications with less stringent operation accuracy, such as remote control of a harbor crane, the survival time can be as large as six consecutive transfer intervals~\cite{Kokkoniemi2019}. The immediate impact of introducing survival time as a \gls{qos} requirements of the wireless link is that burst of consecutive errors, longer than that of the survival time duration must be avoided. For instance, with survival time equivalent to one transfer interval, the system can tolerate one packet failure, however failure in delivering two consecutive packets results in the control application to fail. 

Moreover,  there  is no direct  relation between latency and survival time, where, depending  on the use case and service area, those requirements may vary. For motion control applications within small areas, low \gls{e2e} latency is achievable (physical distance allowing for it) and desirable as seen in the examples in~\tabref{tab:QoS}. For example, the haptic-enabled controller node  and the actuators of a motion control network can be located in the same room and connected to the same \gls{5g} base station, resulting in a short \gls{e2e} link.
As the distance between nodes increases (e.g., in case of a nationwide mission-critical network as depicted in \tabref{tab:QoS}), the \gls{e2e} latency  naturally increases too.

\begin{table*}[t]
\scriptsize
\centering
\caption{\gls{qos} characteristics for cyber-physical and medical applications~\cite{3gpp22104,3gpp22263}.
\label{tab:QoS}}
\begin{tabular}{|p{0.08\textwidth}|p{0.05\textwidth}|p{0.05\textwidth}|p{0.05\textwidth}|p{0.05\textwidth}|p{0.06\textwidth}|p{0.07\textwidth}|p{0.2\textwidth}|}
\hline 
\Tstrut \textbf{Commun. service availability} & \textbf{E2E \;\;latency} & \textbf{Message size} [byte] & \textbf{Transfer interval} & \textbf{Survival time} & \textbf{Number of devices} & \textbf{Service area} & \textbf{Application}  \\ \hline \hline
 \Tstrut 99.999 - 99.99999 &
%$10^{-7}-10^{-5}$ &
$<$ TI
& 50 &
500 $\mu$s & 500 $\mu$s & $\leq$ 20 & 50 m $\times$ 10 m $\times$ 10 m
& Motion control \\ \hline
 \Tstrut 99.9999 - 99.999999 &
%$10^{-8}-10^{-6}$ &
$<$ TI
& N/A &
$\leq$ 1 ms &
3 $\times$ TI 
& 2--5 & 100 m $\times$ 30 m $\times$ 10 m
&  Wired-2-wireless 100 Mbps link replacement    \\
\hline
 \Tstrut 99.999999 &
%$ 10^{-8}$ &
$<2$ ms & 250 to 2000 &
1 ms & 1 ms & 1 & room
&  Motion control and haptic feedback\\
\hline
 \Tstrut 99.9999 &
%$ 10^{-6}$ &
$<20$ ms & 250 to 2000 &
1 ms & 1 ms & $<$ 2 per 1000 $\text{km}^2$ & national
&   Motion control and haptic feedback \\ 
\hline
 \Tstrut 99.999999 &
%$ 10^{-8}$ &
$<2$ ms & 50 &
2 ms & 2 ms & $>$ 2 
&  100 $\text{m}^2$ &    Mobile Operation Panel: Haptic feedback data stream
\\ \hline
 \Tstrut $>$99.99999 &
%$ 10^{-8}$ &
$<1$ ms & 1500 to 9000 &
--- & $~$8 ms &  1 
&  100 $\text{m}^2$ &    Imaging/video traffic for medical applications
\\ \hline
\end{tabular}
\vspace{-10pt}
\end{table*}

The introduction of survival time will result in a shift in design paradigm from typical link reliability---i.e., minimizing \gls{per}---to service availability---i.e., minimizing the chance of a burst of consecutive  errors~\cite{Tirk2106:Optimized}. Recently, analysis of the consecutive packet failure rate and service availability for periodic traffic has  gained attention in the literature. We refer to~\cite{Samarakoon,Demel2020,Gebert2020,Paris2020}, where initial studies have been carried out on characterizing consecutive error rate statistics, both analytically and using  empirical tools. In case of~\cite{Demel2020}, the impact of several scheduling schemes on consecutive packet failure rate is empirically studied, suggesting a conservative link-adaptation for a link in survival mode.

A situation where a communication link has failed to deliver the last packet within the \gls{e2e} latency budget is commonly referred to as \emph{survival mode}. Motivated by the fact that  cyber-physical applications can tolerate   packet failures as long as the duration of consecutive failures is bounded by the survival time  of the application, a proper strategy for survival mode operation is vital. A two-way communication link---which could consist of isochronous traffic between a controller and an actuator, or, a one-way traffic to the actuator followed by acknowledgment feedback---can  inform the scheduler of packet failures~\cite{Tirk2106:Optimized}. Therefore, survival mode can  trigger a more cautious transmission---i.e., with higher reliability---of the immediate next packet.   This further motivates  the network to discriminate between two modes of operation, namely the \emph{normal mode} of operation which takes up most of the network operation time, providing the link with reasonable reliability levels, and the \emph{survival mode} of operation, which expectantly occurs infrequently but  requires a more reliable transmission, e.g., by means of increasing redundancy for the transmission. This analogy is deliberately reminiscent of the special treatment that an ambulance in a survival challenge receives on a congested  highway, as opposed to normal cars.

%---------------------------
%----------------

\subsection{Age of Information}

A closely related \gls{qos} metric to survival time is the notion of \gls{aoi}, which has been studied extensively in the literature, e.g., see~\cite{Kaul2011,AbdelAziz2018,Zhang2020,8486307,8006590},   the survey in~\cite{8187436} and the references within. \gls{aoi} characterizes the freshness of the information at the receiver end of a communication link by measuring the elapsed time since the freshest packet was generated at the transmitter end. It has been mostly studied for \emph{status update} type of traffic, addressing the analysis of average \gls{aoi}. This is particularly important in control applications, where the freshness of sensory information plays a key role in the decision-making of a central controller.

Compared to \emph{latency} which is measured with respect to a fixed point in time when a data packet has been created, \gls{aoi} is measured  with respect to alteration of the state of, or the occurrence of an event in, the application or system that utilizes the communication service~\cite{popovski2021perspective}. 
The notion of  \gls{aoi} can be coupled with survival time and service availability from~\tabref{tab:QoS} by looking at the low percentiles of \gls{aoi} distribution (one in millions and less, corresponding to the residue of time where service is not available)  as follows: communication service is deemed available if the age of the freshest packet at the receiver is less than the sum of the \gls{e2e} latency and survival time budgets.

\subsection{Battery Lifetime and Energy Efficiency}\label{kpi:battery}
 Smart devices are usually battery-driven and a long battery life is crucial for them, especially for devices in remote areas, as there would be a huge amount of maintenance effort if their battery lives are short.  For most reporting IoT applications, packet generation at each device is modeled as a Poisson process~\cite{3g}.
The energy consumption of a device can be then modeled as a semi-regenerative process where the regeneration point is located at the end of each successful data transmission epoch~\cite{nL}.
For a typical device, denote the stored energy in the battery at the reference time as $E_{0}$, static energy consumption per reporting period for data acquisition from environment and processing as $E_{\text{st}}$, and energy consumption in data transmission as $E_{\text{tx}}$. Then, the expected battery lifetime  is derived as $$ L(i)= \frac{{E_{0}}T_i}{{{E_{\text{st}}+E_{\text{tx}}}}}.$$
Modeling of $E_{\text{tx}}$ is closely coupled with the IoT technology, e.g. \gls{nbiot}, and the multiple access scheme used for handling IoT communications. We can extend the  definition and define the battery lifetime of an IoT application.  The lifetime of an IoT application could be defined as the length of time between the  {reference time} and when the service is considered to be  {nonfunctional}.
 {The instant at which the service becomes nonfunctional depends on the level of correlation between gathered data by neighboring devices.}
In critical applications with a sparse deployment of sensors, where losing even one node deteriorates the performance or coverage, the shortest individual battery lifetime may characterize the application lifetime.
On the other hand, in denser deployments and/or when correlation amongst gathered data by neighboring nodes is high, the longest individual battery lifetime, or average individual battery lifetime could be used to characterize the application lifetime.

\subsection{Connection Density} %\CM{Mojtaba}
\label{sec:KPIdensity}

 \gls{5g} is expected to support a connection of up to 1 million devices per square kilometer (km$^2$), ten times higher than that in \gls{4g}. With the explosive growth in the number of IoT devices, the connection density of devices is expected to reach $10^7$ devices/km$^2$ in 2030 (see Fig.~\ref{fig:6Gvision}).
 Designing appropriate multiple access and random access techniques is the key to supporting massive IoT, also referred to as \gls{mmtc}, in cellular networks. 
 It should be highlighted that the characteristics of massive access for
 cellular IoT are very different from those of \gls{embb} and \gls{urllc}, the other two use cases of \gls{5g}. 
In particular, often,  \gls{embb} is for broadband IoT whereas \gls{urllc} is for critical IoT. In the former,  providing high data rates  is important whereas in the latter guaranteeing
 ultra-reliable low-latency communications is the main goal. Representative examples of the two use cases are augmented/virtual reality and autonomous driving, respectively. The main concern of \gls{mmtc} is, however, to ensure wide coverage and low power consumption such that IoT sensors could communicate for a long period of time. More accurately, massive IoT is characterized by    
 \begin{itemize}
\item []\begin{itemize}
	\item sporadic traffic	
	\item small payload
	\item low power consumption
	\item limited capacity
	\item extensive coverage
\end{itemize} 
 \end{itemize} 
With sporadic traffic and small payload, the traffic will be busty, and 
short-packet transmission will be preferred. Further,
idle devices do not access the network in order to save power and extend the battery life. Then, low power standards, like \gls{lte-m} and \gls{nbiot}, are necessary for battery lifetime.  Limited processing and storage capacity   of these
devices implies that sophisticated signal processing cannot be implemented. With the above requirements of massive access, in beyond \gls{5g} cellular IoT,   multiple access is evolving to massive access \cite{chen2020massive} for such devices.

\subsection{Data Rate and Spectral Efficiency}

A few KPIs are related to the data rate in one way or another. These are peak data rate, user experiences peak data rate, and spectral efficiency. \gls{5g} networks are expected to offer peak data rates of up to 20 Gbps  whereas peak data rates for \gls{6g} will be about 1,000 Gbps (see Table~\ref{tab:5GKPI} and Fig.~\ref{fig:6G}). Also, compared to \gls{5g}, user experienced data rate is expected to increase about an order of magnitude in 2030. Further, spectral efficiency must improve two to three times. the technologies facilitating these escalations are mmWave, \gls{thz}, and massive \gls{mimo}. This paper is not meant to discuss solutions related to the data rate in detail. 

			\begin{figure*}[!ht]
			\centering
			\includegraphics[width=6in]{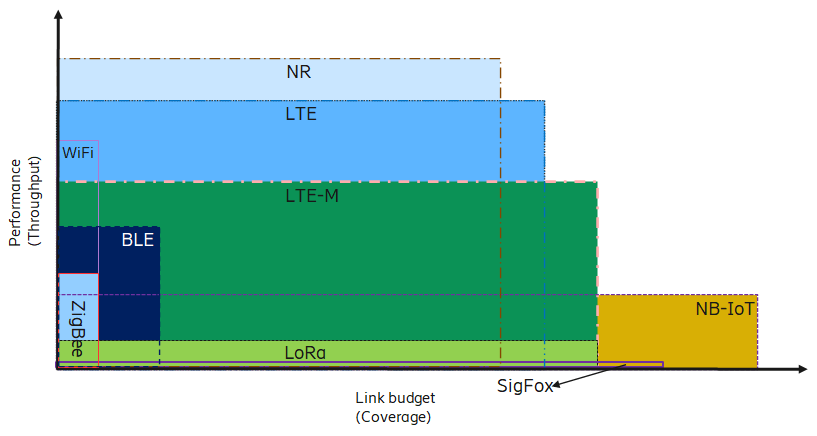}
			\caption{A comparison of throughput and coverage of different connectivity technologies} \label{comp_fig}
		\end{figure*}

\section{Enabling Access Solutions}\label{Sec:enablingtech}
The continuing growth in demand for connecting smart things has encouraged different industries  to investigate evolutionary and revolutionary connectivity solutions for serving IoT communications \cite{w_sony}. The major IoT connectivity solutions can be classified into two classes based on the criteria of data rate and transmission range requirements. Range-wise  for instance, Zigbee like protocols are well-suited for medium-range IoT applications and complement WiFi-like IoT use cases, home automation, smart lighting and energy management just to name few. Additionally, Bluetooth and \gls{ble}, so called Bluetooth Smart are well-optimized for short-range consumer IoT applications owing to their low power dissipation, which are standardized in the Wireless Personal Area Networks, where \gls{ble} mainly targets for fitness and medical wearables, as well as for smart home devices. Another medium-range WiFi derivative for assisting IoT technologies is the WiFi-HaLow (IEEE 802.11ah), which is capable of providing reasonable coverage and less energy dissipation compared to WiFi. However, mainly due to security issues Wifi-HaLow has not taken the deserved attention of the industry. Finally, \gls{rfid} can be exploited as part of the IoT technologies owing to its very small amount of data transmission requirement within a very short distance. \gls{rfid} is mainly used for real-time monitoring of inventory and assets in order to provide efficient stock management and production planning, which is now taking its place for self-checkout, smart mirrors and shelves as emerging IoT applications. After having a brief discussion on short-range IoT technologies without emphasizing on the data rate requirements, we  focus our attention on the long range IoT wireless technologies, which can be further divided into two categories based on their usage of radio spectrum: i) solutions over the licensed spectrum like \gls{lte} category M and \gls{nbiot} \cite{ciot}; and ii) solutions over the unlicensed spectrum like SigFox and \gls{lorawan} \cite{mag_all}. In the following, we investigate these categories in details. 

 \subsection{IoT Communications over Licensed Spectrum}
 Cellular networks  are expected to play a significant role in large-scale deployment of IoT solutions because of their ubiquitous coverage and roaming capability \cite{sulyman2017expanding,phda}. Following
the early-stage predictions of a huge expected increase in the IoT traffic by Cisco \cite{forecast2019cisco}, network providers have started devising solutions for adopting IoT traffic over their networks in order to find new revenue streams \cite{ratasuk2015overview}. Generally speaking, the IoT traffic over cellular networks is  served in three different ways: 
\begin{itemize}
\item it is served along with the legacy traffic, i.e. it passes the same procedures as non-IoT traffic;
\item it is served along with the legacy traffic but using new protocols aiming at enabling lo power/cost IoT connectivity; 
\item it is served over a dedicated pool of
radio resources using IoT-specific algorithms and protocols. 
\end{itemize}
The former, i.e. serving IoT and non-IoT traffic streams using the same resources and procedures, is usually the case in \gls{2g} networks, and it out of our scope here. The other two cases are investigated in the sequel.

\subsubsection{{IoT/non-IoT traffic serving over the same resources using dedicated communications protocols}}
\label{SENH}
A legacy connectivity procedure in \gls{lte}/\gls{lte-a}/LTE-Pro  networks consists of different steps,  including synchronization, connection establishment, authentication, scheduled data transmission, confirming successfulness of data transmission, and connection termination, as described in details in \cite{sched}. The \gls{rach} of \gls{lte} networks is the typical way for IoT devices to start connection with the \gls{bs}, and attach  to the network for following data transmissions. Over the  \gls{rach}, each device  chooses a preamble randomly from a set of broadcasted preambles for access establishment with the \gls{bs}. Since the total number of available preambles at each time is limited, the connection over \gls{rach} introduces collisions and energy wastage, especially  when  a multitude of of devices try simultaneously to get access to the network, e.g. after a power outage in a building or alarm \cite{sched}.
Once a device successfully passes the \gls{rach} procedure and becomes connected,  it can send scheduling requests to the \gls{bs} over the \gls{pucch}. Now, the \gls{bs} performs scheduling and responds to the device by sending  the scheduling grantsover the   \gls{pdcch}. Then, a granted device can  send its data over the \gls{pusch} resources to the \gls{bs}. 
In this procedure, even while there is no communications between the device and the BS, device needs to be ready for listening to the PDCCH for potential downlink data, which is heavily battery consuming. Then, one  solution for energy saving is to let device get into sleeping mode for a certain period of time and wakes up again for checking if there is there is any data queued at the network side,  and getting into the sleeping mode again after communications (or lack of queue data). This kind of periodic transfer between sleep and wake up modes is called discontinuous transmission (DRX). DRX provides energy saving at user device by introducing short and long sleep periods during. For sure, configuration of these sleep/wake up periods to the traffic pattern plays a crucial role in energy efficiency of IoT communications. However, the existing connection establishment, scheduling,  scheduled data transmission, and DRX protocols in the legacy networks have been neither designed nor optimized for  IoT communications. Hence, while they perform well for a limited number of long-lasting streaming/call sessions, for constrained  IoT devices  with a massive number of  short-length  packets (communication sessions) these procedures could be the bottlenecks. Thus, congestion in serving IoT communications,  including radio network congestion and signaling network congestion, is likely to occur \cite{cusm}.  Towards isolating the impact of  IoT communications from the legacy communications, the use of dedicated resources for IoT communications  has been proposed in \gls{lte}-Rel 13 onwards, to be discussed in the following.

\subsubsection{{IoT traffic serving over dedicated resources and communications protocols (\gls{lte-m} and \gls{nbiot})}}\label{SNBT}
From \gls{lte}-Rel 12, the evolution of \gls{lte} for serving low-cost  IoT communications over dedicated resources has been started,  and it has been continued in Release 13 \cite{nok1}. In Release 13,  \gls{lte-m} and \gls{nbiot} systems have been introduced. These systems provide IoT connectivity over dedicated 1.4 MHz and 200 KHz   bandwidths, respectively \cite{nok1}. In the following, we focus on the \gls{nbiot}, as the leading IoT connectivity solution from the \gls{3gpp} for low-cost low-power IoT devices.

 \gls{nbiot}  represents a big step towards accommodation of low-cost low-power IoT devices over cellular networks \cite{nbiot}. Communications in \gls{nbiot} takes place in a very narrow bandwidth, i.e.  200~KHz. This results  which results in more than 20 dB extra  link budget compared to the standard \gls{lte-a} systems \cite{nbiot}. Thanks to this link budget, \gls{nbiot} enables  smart devices deployed in remote areas, e.g. rural areas, to directly communicate with the base stations, as compared to the gateway/relay-based solutions. In the above, we discussed that the legacy communication protocols and logical channels have been  designed and optimized for data/bandwidth-hungry applications, e.g.  20 MHz in \gls{lte-a}. Then, \gls{nbiot} introduces  five novel narrowband physical  channels, including: \gls{nprach}; \gls{npusch}; \gls{npdsch}; \gls{npdcch}; and \gls{npbch}   \cite{prim,wp}. Moreover, \gls{nbiot} introduces four new physical signals: \gls{nrs}; \gls{npss},   demodulation reference signal (DMRS), which is sent along with user data on \gls{npusch}; and narrowband secondary synchronization signal (NSSS). In \gls{nbiot}, a detached device which aims at connecting to the network, follows the folloing procedures. It first listens for receiving cell information, e.g. \gls{npss} and NSSS, through which, it makes itself synchronized with the \gls{nbiot} \gls{bs}. Then, it waits for the \gls{nprach} to establish connection over random access resources, for \gls{npdcch} to receive the scheduling response, and finally for \gls{npusch}/\gls{npdsch} to send/receive data to/from the \gls{bs}.

In contrast to the \gls{lte} networks, in \gls{nbiot} IoT devices are allowed to go from the idle mode to the deep sleep mode during which,  device is still registered with the \gls{bs}. Then, upon waking up, device becomes connected and ready for communications by sending/receiving a fewer number of handshaking messages \cite{ciot}. This new functionality enables saving a considerable amount of energy, especially in the connection establishment procedure and resource reservation \cite{nbt}. Moreover, coverage extension has been devised for \gls{nbiot} systems, which is achieved by time-domain repetition of signals. Towards this end,  \gls{bs} assigns each device to a coverage class, from a limited number of defined classes, based on several parameters, including the average experienced device-\gls{bs} path-loss. This is the  coverage class that characterizes the number of replicas per  data packet in communications. For example, based on the specifications in \cite{wp}, each device belonging to group $i$ shall repeat  $2^{i-1}$ times the preamble message transmitted over the \gls{nprach}, where $i\in \{1,\cdots,8\}$.
This repetition, enables  \gls{nbiot} systems  to offer deep coverage to most indoor areas \cite{ciot}.

In LTE Rel.13, 3GPP standardization body introduced extended DRX (eDRX) and power-saving mode (PSM), as two main components for NB-IoT and LTE-M \cite{gsm2018lte,sultania2018energy,sultania2021optimizing}. eDRX enables IoT application developers to dynamically change how long an IoT device should spend in the sleep period. The minimum sleep times for eDRX are 320 milliseconds (ms) for LTE-M and 10.24 seconds for NB-IoT; and the maximum sleep times are 43 minutes LTE-M, and  3 hours for NB-IoT \cite{gsm2018lte,expl}. Considering the fact that in the legacy DRX, the sleep times are typically 1.28 seconds or 2.56 seconds, one observes how eDRX can significantly help IoT devices to save energy. Moreover, while in the DRX scheme, DRX parameters are tuned by the network, the eDRX approach enables application developers to set and change the eDRX configurations, which brings much more flexibility to the applications. Finally, in eDRX,  device can listen for pending data at the network side without the need to establish a full network connection, which also saves significant energy for the device \cite{expl}.  In contrast to eDRX, in PSM device is goes to a very-low power consumption mode in which, there is only an active timer for counting the length of the PSM state \cite{gsm2018lte}. The length of PSM could be as high as 413 days, which brings significant saving for IoT devices that rarely need data transmissions. In the PSM mode, once the device wakes up and sends data, it is reachable over 4 consecutive time windows. In these time windows, network can ask the device to stay awake for a longer period of time, or send data to the device \cite{expl,gsm2018lte}.  

In order to further improve \gls{nbiot} systems,  in \cite{nb_ra,nbiotaa},  preamble design for  improving \gls{rach} of \gls{nbiot} has been investigated. Energy efficient radio resource scheduling of IoT communications  has been investigated in  \cite{nb_sch}.  For \gls{nbiot} deployment in \gls{lte} guard-band, a deployment option which introduces external interference to the IoT communications, coverage extension has been investigated in  \cite{nbi_cov}. In \cite{nbt}, energy consumption  in data transmission  in different coverage regimes, including normal, robust, and extreme cases, has been studied. Then,  in robust and extreme regimes, the extra coverage  by signal repetition in the time domain has been characterized. When compared to the legacy \gls{lte} systems,  energy consumption analysis in \cite{nbt}  illustrates that \gls{nbiot} systems can significantly reduce the consumed energy in connection establishment due to their capability in easier move between sleep and active modes. Regarding the fact that we are still waiting for large-scale deployment of \gls{nbiot} networks and wide access of IoT devices to its services, there is a lack of a comprehensive study on capacity, energy efficiency, and scalability of \gls{nbiot} systems. Especially, when it comes to coexistence of coverage classes with different repetition orders over the same system, the work in \cite{mnbt} has shown that \gls{nbiot} enhances energy efficiency for far-away devices at the cost of increasing  latency for nearby devices, which calls for further investigation.

 \subsection{IoT Networks over Unlicensed Spectrum}\label{SUL}
From \gls{1g} to \gls{4g}, the telecommunications industry has spent a great deal of resources investigating how to realize high-throughput infrastructure for serving data-hungry applications, and IoT solutions have been out of main scope. Along with telecoms' transition from \gls{2g}/\gls{3g}  to \gls{4g}   networks (2010-2020), some \gls{3gpp}-independent companies have started entering the low-poer wide-area market, by providing large-scale IoT connectivity solutions, especially over the unlicensed spectrum, e.g. SigFox and \gls{lorawan}.  While reliability of communications over the unlicensed spectrum is a big challenge due to grant-free access to radio resources, the simplified connectivity in these solutions has significantly reduced the cost per device. It has also capable of increasing  battery lifetime of devices.  Then, they are able to deliver their service at low cost for IoT applications in which, reliability of communications could be traded. Regarding the fact that energy consumption in synchronization, connection establishment, and signaling (in the cellular networks) is comparable with, or even higher than, the energy consumption in the actual data transmission (which is the only source of energy consumption in solutions leveraging grant-free access over unlicensed spectrum) \cite{nbt,lif_com}, IoT connectivity over cellular networks is also expected to move towards more relaxed radio access schemes, as investigated in \cite{azari2021energy}.
  The \gls{lpwa} IoT solutions over the unlicensed spectrum, along with cellular \gls{nbiot} and \gls{lte-m} solutions, are expected to share 60 percent of the IoT  market among themselves, a number that is expected to grow over time, and is expected to bring new revenue streams to network providers. Hence, {the competition  between IoT connectivity technologies (with low to medium reliability and datarate requirements\footnote{Later in Table \ref{tab1}, we will see that cellular solutions have no competitor in  medium to high reliability and datarate performance regimes. }) is becoming intense \cite{mag_all}}.
 In the sequel,  SigFox and \gls{lorawan} are introduced as two dominant \gls{lpwa} IoT solutions over the unlicensed spectrum. In these solutions, data is collected by local \glspl{ap}, which the point of access to the Internet for IoT devices. These \glspl{ap} are connected to the Internet infrastructure through wired or wireless backhauling. The IoT end-user accesses  the gathered data by devices through the IoT server, and potentially sends commands back to them, again through the gateways.  Both \gls{lorawan} and SigFox use the \gls{ism} spectrum band for communications (detailed description of frequency bands could be found in the following subsections as well as Table \ref{tab1}). Over the \gls{ism} bands,  any device can communicate as  long as it complies with the fair spectrum use regulations \cite{mag_all}. The regulations vary from one sub-band to another, from one region to another, and are mainly restricting the effective radiated power, number of transmissions per device per day, as well as the spectrum access method \cite{int2}. Based on \cite{int2}, the spectrum access can be either listen-before-talk,  or  the duty cycle. In the former, devices start transmissions in case the sensed channel is free, while in the latter,  devices start transmission once it has data to transmit by considering the constraint that it cannot transmit in more than $x$\% of the time, where $x$ is determined from the regulations.  The communications protocols used in the lower layers of SigFox and \gls{lorawan} are discussed in details in the sequel.
 
\subsubsection{{SigFox}}\label{SSG}
Founded in 2009 in France, SigFox is the first widely announced proprietary IoT technology. The SigFox technology  aims at providing low-cost low-power IoT connectivity over the \gls{ism} band. Thanks to its ultra-narrowband signals, as low as 100 Hz, SigFox is able to cover large outdoor areas, and even is able to provide high link budget for indoor coverage, as shown by experimental results in \cite{mads}. Once a SigFox device has a packet to be transmitted, one copy of the packet is transmitted immediately, and two other copies of the packet are sent consequently with a semi-random frequency hopping over a 600 KHz system bandwidth (6000 times more than the signal bandwidth) \cite{stars,int2}. In the first releases, SigFox was only supporting uplink communications; however, from 2015  it also supports uplink-initiated downlink communications. In other words,  once a  device transmits data and requests for a response from the application, the \gls{ap}  sends the acknowledgment (ACK) as well as the queued data waiting for the device from the server side \cite{lif_com}. As per other grant-free radio access solutions,  SigFox doesn't require pairing for starting communications. This on the other hand means that there is no need for  message exchanges (and energy consumption) in synchronization, authentication, and connection establishment \cite{int2,stars}. The duty cycle for a SigFox \gls{ap}  is 10\%, while the duty cycle for devices is 1\%. This means that a SigFox \gls{ap} can transmit data at most in 10\% of the time, as compared to a devices which can do it in 1\% of the time. The difference in duty cycling comes from the fact that \gls{ap} and devices are working on different sub-bands of the \gls{ism} band. Hence, they comply with different regulations. From this duty cycling puts  an upper-limit of  140 messages of 12-bytes per device  per day, which makes SigFox unfavorable for  applications with more data exchange requirements (also for applications with immediate Ack messages per packet, or with delay-bounded availability constraints). Monitoring of the status of the network and manage operations of the \glspl{ap} is managed at the core of SigFox networks.  SigFox modules are available at low cost, e.g. 2\$ per module \cite{cost},  to be used in various IoT devices. The cost comparison in \cite{cost} shows that the cost of a SigFox module is almost one-third of a \gls{lora} module (which is more complex than a SigFox module, as we will see in the sequel), and one-fifth of a cellular IoT module (which is expensive due to many complex functions embedded to comply with cellular standards).

\subsubsection{{\gls{lorawan}}}\label{SLOR}  \gls{lorawan} leverages \gls{lora} communications in its lower layers of protocol stack. \gls{lora} is utilizing a propitiatory \gls{css} as a modulation to mitigate  the potentially severe interference on the unlicensed bandwidth and has been claimed to be robust against multi-path fading and Doppler shifts to some
 extent \cite{LoRa2015Modulation}. In contradictory with the SigFox technology, the \gls{lora} gateway is open source, i.e. may build her own \gls{lorawan} gateway (e.g. on top of micro-controllers like  Arduino), connect it to the Internet, and use it for long-range data gathering from \gls{lora}-compatible IoT devices (or even built her own devices again on top of micro-controllers). 
 
 The \gls{lorawan} servers are managing the allocation of \glspl{sf} to connected devices in a centralized way , in order  to balance the cell-wide data-rate and reliability of  communications. \glspl{sf},  denote the number of chirps used to encode a bit, and are ranged from 7 to 12. In \gls{lora}, the higher the chirp rate is, the better reconstruction of the received signal is achieved, while on the other hand, an increase in the \gls{sf} stretches the  transmission time \cite{blenn2017lorawan}, and hence, increases the probability of collision with another data transmission (increased transmission time).  Regarding the ever-increasing rate of deployment of IoT solutions over the \gls{ism} band, and  the uncontrolled access to radio resources over this spectrum, a high level of  resilience to sporadic interference has been introduced in \gls{lora}.  In fact, the main feature of \gls{css}, which is used in the \gls{lora} modulation as a simplified version of a satellite communication protocol, is that signals with different \glspl{sf} can be almost distinguished and received simultaneously, even if they are transmitted at the same time and on the same frequency  channel \cite{bankov2017mathematical}.  \gls{lorawan} 3 classes in which, communications parameters and capabilities are different. Especially in class A, downlink transmission is possible after  data transmission of a device  during two non-overlapping and consecutive dedicated time intervals, which is called receive windows in the context of \gls{lorawan} \cite{bankov2016limits}.
 
 \subsubsection{Comparison of different technologies}\label{CALL}
Table \ref{tab1} represents a comprehensive comparison of enabling solutions for  wide-area IoT communications, i.e. SigFox and \gls{lorawan} (over the unlicensed spectrum), and \gls{lte-m} and \gls{nbiot} (over the licensed spectrum). A graphical illustration of the throughput/coverage comparison has been depicted in Fig.~\ref{comp_fig}. In this figure, one further observes  which areas of range and data-rate (as two main performance indicators) have been covered by different technologies.

\begin{table*}
\centering
	\caption{A comparison of different wide-area IoT enablers}
	\setlength{\tabcolsep}{3pt}
	\resizebox{1.7\columnwidth}{!}{%
			\begin{tabular}{p{2.9 cm}p{.1 cm}p{3.2 cm}p{.1 cm}p{3 cm}p{.1 cm}p{3.5 cm}p{.1 cm}p{4 cm} }\\
			\toprule[0.5mm]
     \textbf{Feature} &&  \textbf{SigFox} &&  \textbf{\gls{lorawan}}  &&  \textbf{\gls{lte-m}} &&  \textbf{\gls{nbiot}} \\[1ex] \hline\hline
     \Tstrut Frequency (MHz) &&  \gls{ism} (865-924)  && \gls{ism} (433/868/915)   && Licensed (410,$\cdots$,5900)  &&Licensed (410,$\cdots$,5900); also out of \gls{lte}-band   \\[3ex]\hline
   \Tstrut Signal \gls{bw}    && 100 Hz   &&125-500 KHz   && 1-20 MHz (downlink); 1.5,5 MHz (uplink)   &&  180 (\gls{lte} band); 200 (GSM band) KHz \\[3ex]\hline
 \Tstrut uplink/downlink support&&yes/limited   &&yes/yes   &&yes/yes   &&yes/yes   \\[3ex]\hline
 \Tstrut Packet size&&8 (downlink); 12 (uplink) bytes    &&19-250 bytes   && 2984-6968  bits   && 16-2536 bits   \\[3ex]\hline
 \Tstrut Modulation&&GFSK   &&\gls{lora} (proprietary)   &&  B/QPSK;16-64QAM &&QPSK (uplink/downlink); BPSK (uplink)    \\[3ex]\hline
 \Tstrut Radio access&& Random FDMA/TDMA    && Random FDMA/TDMA   && FDMA (downlink); \gls{sc-fdma} (uplink)   &&   OFDMA (downlink); \gls{sc-fdma} (uplink) \\[3ex]\hline
 \Tstrut \gls{bw} constraints&& Fair use of \gls{ism}: 14 packets/day  &&Fair use of \gls{ism}: 30 sec/day (uplink); 10 packets/day (downlink)   &&Pay as you go   && Pay as you go    \\[3ex]\hline
 \Tstrut Tx power (dBm)&&14, 22  (uplink); 27, 30  (downlink)   && 14-30  (region-dependent)  &&20     && 14, 20-23      \\[3ex]\hline
 \Tstrut Downlink sensitivity  && -130 dBm    &&-137  dBm   &&-132  dBm   &&-141    dBm \\[3ex]\hline
 \Tstrut Link budget (dB)&&163   &&154 (uplink); 154-157 (downlink, region-based)   &&155   &&164 (20-23 dBm Tx power); 155 (14 dBm Tx power)    \\[3ex]\hline
 \Tstrut Indoor outage  (20 dB loss) \cite{madsl}&& 10\% (6 km  \gls{isd};  2.5\% (4 km \gls{isd})   &&18\% (6 km  \gls{isd});  5.5\% (4 km \gls{isd})   && -   &&   8\% (6 km  \gls{isd});  1.5\% (4 km \gls{isd})\\[3ex]\hline
 \Tstrut Data rate&& 100 bps (uplink/downlink), 600 bps (downlink)    && 0.25-50 kbps (dependent on \gls{bw}, SF) \cite{lormag}  &&  3-7 mbps (uplink); 4 mbps (downlink) &&  127 kbps (uplink); 159 kbps (downlink) [Rel 14] \\[3ex]
	\bottomrule[0.5mm]
	\end{tabular}
}
	\label{tab1}
\end{table*}

%\subsection{Vision for Future  Technologies}
%\label{enablingaccesstech}

\subsection{Domain of Future Access Solutions}\label{sec:domainoffutureaccesssolutions}

\subsubsection{New Spectra}
Allocating new frequencies is a straightforward solution for increasing the number of connections in the network. A good example is the introduction  of 
\gls{mmwave} bands (30-300
GHz) in \gls{5g} networks. Similarly, \gls{thz} band (0.1–10 \gls{thz})  and optical spectrum are expected to be exploited in beyond \gls{5g} for ultra-high data rates \cite{tripathi2021millimeter,haas2020introduction,chowdhury2018comparative}. However, due to their wide bandwidth and high penetration loss, \gls{mmwave} frequency bands are %more appropriate for eMMB use cases but not \gls{mmtc}. The latter needs a high coverage which cannot be achieved in \gls{mmwave} bands. Therefore, usually, \gls{rf} bands, which can cover higher ranges, are used for massive IoT connections.  That is, new high-frequency bands are 
generally not considered  for massive connectivity, but they can be used for broadband IoT and critical IoT.

%For the sake of completeness we explain \gls{mmwave} in the following. 

%Besides the above-mentioned enabling solutions, IoT communications also rely on several other technologies such as data processing technologies (cloud computing and big data) as well as access-related technologies \gls{d2d} and mmWave \cite{arshad2018recent}. We discuss the latter in this subsection.  

\paragraph{mmWave IoT}
\label{mmwave}

High ranges of radio frequency, well above the single-digit gigahertz frequencies used for previous generations, are subject of technology development and deployment  in the time of \gls{5g} mobile networks and beyond.  Currently, the \gls{mmwave} bands includes several gigahertz  of licensed and unlicensed bandwidth in the 24-43 GHz range, as well as above 57 GHz, which has been made available by different regulatory bodies across the world \cite{federal2016use}.

Propagation characteristics for  \gls{mmwave} bands have been studied extensively for urban and indoor scenarios \cite{6834753,yoo2017measurements,chizhik2020path,du2018suburban}. The main challenge that is inherently coupled with \gls{mmwave} is  higher atmospheric absorption, causing large propagation loss. Moreover, \gls{mmwave} faces sever blocking due to limited amount of diffraction around objects. Thanks to small wavelength, tens of antenna elements can fit on a single small-sized array and enable high gain beamforming. Nevertheless, the high propagation loss  limits the communication range to several tens of meters up to a few hundred meters, which makes \gls{mmwave} well suited for indoor networks. Consequently,  ubiquitous coverage in \gls{mmwave} requires a sufficiently high density of radio \glspl{ap} and enabling  multi-link communications such that the  user can simultaneously communicate with multiple radio \glspl{ap}, also known as \glspl{trp} \cite{8570917}. Such densification of access points can improve the chances for a user to experience a non-blocked link to at least one \gls{trp} at a time, especially in environments with dens and moving clutter of blocking objects.

The wide  spectrum bandwidths available in \gls{mmwave} has made it mostly known as a solution for   significantly high data rates of multigigabits \cite{6834753}. Recently, other aspects of \gls{mmwave}, such as the large frequency diversity and beamforming gain potentials, have raised the attention of researchers towards enabling \gls{mmwave} \gls{urllc} in \gls{iiot}   \cite{vu2017ultra,semiari2019integrated}. Specifically, the large \gls{mmwave} bandwidths and contained delay spread offers the attractive possibility to shrink the \gls{tti} to a small fraction of what the lower band technologies can tolerate, which enables seemingly impossible over-the-air latency of tens of microseconds \cite{Gilberto:2018}.  Additionally, for industrial use cases, the large outdoor to indoor penetration losses, that eliminate legitimate interference and illegitimate  jamming  from outside the factory, makes the \gls{mmwave} bands attractive \cite{Ibra2106:Beam,mazgula2020ultra}. 

\paragraph{Waveform  and Multiple Access Design in New Spectra}

Evidently, the choice of waveform design and multiple-access technique for \gls{mmwave} deployments  is still  being investigated as an open problem. This owes to the differences in radio propagation as well as the hardware design challenges in high frequencies. \gls{ofdm} has long been adopted in cellular technologies, thanks to its low complexity of frequency-domain processing and high spectral efficiency. It further offers extensive flexibility in multi-user access. Impact of hardware impairments such as phase noise, non-linear power amplifiers,  and \gls{adc} could minimize the performance. In fact, using single-carrier waveform such as  single-carrier frequency domain equalization can be beneficial for \gls{mmwave} \cite{7499114}. 

Alternatively, \gls{uwb} transmission using carrierless  spread spectrum can offer simplicity of transceiver  and ideal performance for short-range \gls{mmwave} communications \cite{Gilberto:2018}. This becomes more evident when considering use cases such as industrial wireless control, where a number of actuators need to communicate periodically with a controller in close vicinity, at sub-milliseconds cycle time. In such cases, transmission using low-power \gls{uwb} can take advantage of wide \gls{mmwave} bands in favor of short latency while minimizing the interference footprint.

\subsubsection{Adaptive Network Topology}\label{d2dsec}

In the design of a large network of \gls{iot} sensors, e.g., the likes of \glspl{wsn} in agricultural, industrial, and urban sensory networks, low deployment cost has been a key driver \cite{5136720,3gpp22866}. On the other hand, \gls{iot} devices and sensors are, for the most part, expected to be inexpensive, low-power -e.g., capable of $>10$ years of continuous operation on a single battery \cite{3gpp22866}- and hence, low-range in communication reach.  Inevitability, this creates a dilemma for  densification of \glspl{ap} in network design: a low density of  network \glspl{ap} can severely reduce deployment costs, yet, it makes it virtually impossible for all the \gls{iot} nodes to have a dependable access to the network. One can easily imagine such a dilemma in the context of \gls{iot}-enabled smart agriculture, where  sensor  networks are deployed over thousands of acres of space to measure  temperature, humidity,  nutrition and PH levels of soil, etc., and help farmers with improved crop efficiency at \emph{lower costs}~\cite{8211551}. This has mainly driven wireless engineers to enable mesh-type of network topology for low-cost wireless \gls{iot} networks. Essentially, all the communication nodes in the network take part, not only in communicating  their own data, but also in supporting the communication of other nodes' data. Several existing technologies in the unlicensed band, such as the Zigbee and \gls{lora}  networks have built upon the concept of   \gls{wsn} to enable low power mesh networks among low-cost sensors. Nevertheless, research and development activities in \gls{wsn} are ongoing with a strong focus for utilizing \gls{d2d} relaying in sensor network topologies. Particularly, fulfilling \gls{qos} requirements in a \gls{wsn} has generally been studied for networks with rather homogeneous \gls{qos} across the sensors \cite{Mahmood_2015,Murugaanandam_2019,Zhang_2017,Lindsey_2002,9448986}. The future \glspl{wsn} are envisioned to include a variety of sensors with rather heterogeneous \gls{qos} requirements and battery-life capabilities, which motivates an adaptive network topology design.

\paragraph{D2D Relaying for Extended Coverage and Improved Efficiency}

Utilizing \gls{d2d} relaying has been long studied and specified for various wireless technologies. The goal has  been to extend coverage, improve the battery life of sensors, and enable out-of-coverage communication between devices \cite{dahlman20164g}. For example, in the context of the \gls{3gpp} standard developments, several relay-based specifications have been made available in the past years. This includes the proximity service specifications -service for devices within close proximity of each other, especially in public safety communication services- in Release 12 \cite{3gpp23303}, and the studies on \gls{ue} relays, self-backhauling and integrated backhauling in Release 14 and Release 15 of the specifications \cite{3gpp36746,3gpp38874}. Similarly, communication between vehicles was addressed in vehicle-to-vehicle  and vehicle-to-everything studies in Release 16. More recently, those attempts are extended towards enabling enhanced multi-hop relaying for extensive coverage and improved energy efficiency for Release 17 and beyond \cite{3gpp22866}.  The  studies in \cite{3gpp22866} offer interesting insights into the future of \gls{d2d} relaying for wireless networks, including use cases such as in-home traffic, smart city, low-power wearables, and smart factories, to name a few. For instance, in the  wireless deployments of the future, different facilities can be modeled as environments having variant topologies with multiple walls, long corridors, lots of metallic shelves and other static or moving objects -e.g., moving fork lifts, agricultural machines, slow-moving trains in the warehouses, and moving cranes in harbors. 
Meanwhile, to be able to thoroughly monitor the situation in such environments one can imagine massive wireless sensor networks with heterogeneous \gls{qos} requirement being deployed, including sensors that are installed in places with poor coverage. The argument is that mesh type of relaying will help in increasing coverage for these low-power sensors, which are expected to also have a long battery life.

\paragraph{Cooperative \gls{d2d} Relaying for Reliability}
In the context of industrial automation, a similar point on the benefits of network-device cooperation can be made, although from a different perspective. In such scenarios,  actuators   of the same or different production machines communicate together under extreme reliability and latency requirements, while co-existing with a large number of sensors that monitor the production process and periodically reporting sensory information to update the controller about the state of the machine \cite{mahmood2019six,adeogun2020towards,Xu_2017}. Overall, sensors and actuators in an industrial production line are under much more severe scrutiny -compared to, e.g., agricultural sensor networks- with regard to reliability and timeliness (a.k.a, latency) of their communication with the production line controller. On the other hand, the communication range is  not an issue in such deployments with nodes  typically being in close vicinity of the controller \gls{ap}. The challenge is however to guarantee reliability at five nine figure (i.e., $99.999\%$ of the time) and higher, up to nine nines  \cite{Gilberto:2018}. Therefore, it is vital to exploit  the available sources of diversity in the network in favor of link reliability. Aside from the limited amount of time and frequency diversity  -limited by physics of the environment in form of coherence time and frequency- the network should exploit spatial diversity and multi-path diversity through cooperation among the nodes \cite{swamy2015cow,zand2012wireless}. In fact, it is shown that relying solely on frequency diversity to achieve $10^{-9}$ error rate requires an impractically high \gls{snr} values in realistic channel conditions \cite{swamy2015cow}.

Spatial diversity in the form of multiple antennas at the transmitter and receiver sides is a promising way to collect  diversity gain. This has been widely studied in the literature around \gls{mimo}~ \cite{brahmi2015gc}. Such solutions are powerful in transforming the harsh distribution of typical fading channels (e.g., Rayleigh distribution) into more benign distributions where the channel is most of the time above the detection threshold. In fact, with the increasing deployment of  massive \gls{mimo}, in practice,  the impact of fast fading can be largely under control, hence, cellular communication technologies typically rely on the channel hardening effect of \gls{mimo} to mitigate the effect of small-scale fading \cite{Hochwald2004channelhardening}. However, the large-scale variations in the channel, particularly in the highly dynamic industrial environments, can cause severe challenges for reliable communication system design. For example, the large size (relative to wavelength) moving machines and cranes  in an industrial  environment can create a shadowing effect by blocking a line-of-sight link, or, completely change the propagation characteristics of the environment by creating strongly reflected paths that could cause both constructive and destructive impact on  link signal strength.

\paragraph{Adaptive Network-Device Cooperation}
For the above-mentioned reasons, it is highly favorable to adopt cooperation among devices in the network in form of \gls{d2d} relaying and multi-hop transmission, which unleashes an abundant source of spatial diversity in favor of reliability. In essence, for a network with a large number of devices to be served, the devices with a \emph{strong} momentary link to the controller \gls{ap} can communicate directly with the \gls{ap}, while the other devices with \emph{weak} link quality must be served by a cooperative transmission from the \gls{ap} and the strong devices. Such a technique was first introduced in  \cite{swamy2015cow}. Although the analysis in \cite{swamy2015cow} was focused on mitigating the impact of fast fading, it motivated the use of cooperative \gls{d2d} relaying for ultra-reliable applications by demonstrating $10^{-9}$ system probability of error---i.e., the probability of at least one device failing to receive a downlink packet or delivering an uplink packet in each cycle time---for a network with 30 devices, at practical \gls{snr} value of 5 dB.  Interestingly, such reliability level was achieved without \gls{csi} at the transmitter side, therefore, not relying on link-adaptation and pre-coding. On the other hand, such cooperative transmission  techniques utilize several relays to forward a message to a weak device, thus, inevitably, increasing  the interference footprint  of each cell \cite{Arvin:2019}. More importantly, without the use of \gls{csi}, radio resources are inefficiently exploited, resulting in loss of spectral efficiency \cite{9206086}. Other techniques in the literature are proposed to utilize the partial knowledge of \gls{csi} at the transmitter, e.g., in the form of \gls{cqi}, to identify devices with weak channel conditions and to deploy the cooperative relaying resources  in an \emph{on-demand} fashion, only for the devices with momentary poor coverage. Channel-awareness for allocating resources for each transmission and enabling on-demand cooperative relaying  is the advent of techniques with adaptive network-device cooperation \cite{9206086}.  This increases  the spectral efficiency while improving the system-level reliability by reducing the interference footprint of the relaying process.

\section{Channel Code Design}
\label{sec:channelcode}
\begin{figure*}[t]
    \centering
    \begin{subfigure}{.19\textwidth}
    \includegraphics[width=\columnwidth]{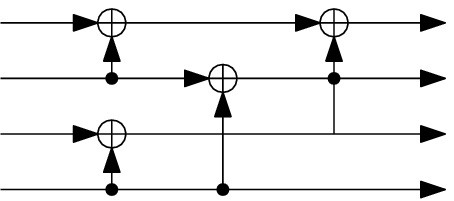}
    \caption{Polar encoder of length 4.}
    \label{fig:polarencoder} 
    \end{subfigure}
    \begin{subfigure}{.19\textwidth}
    \centering
    \includegraphics[width=\columnwidth]{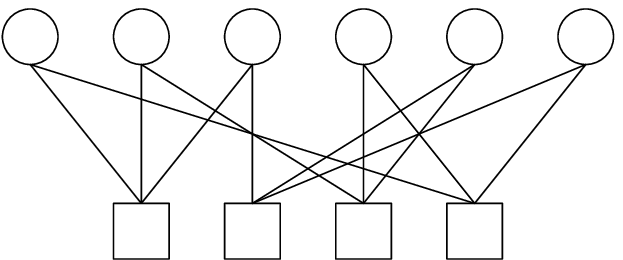}
    \caption{Tanner graph of \gls{ldpc} code.}
    \label{fig:ldpc} 
    \end{subfigure} 
    \begin{subfigure}{.24\textwidth}
    \centering
    \includegraphics[width=\columnwidth]{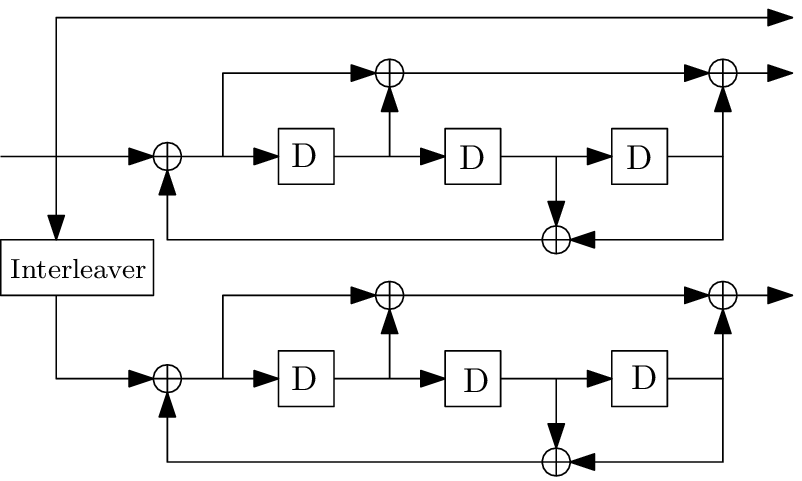}
    \caption{Turbo code encoder.}
    \label{fig:turbo}
    \end{subfigure}
    \begin{subfigure}{.34\textwidth}
    \centering
    \includegraphics[width=\columnwidth]{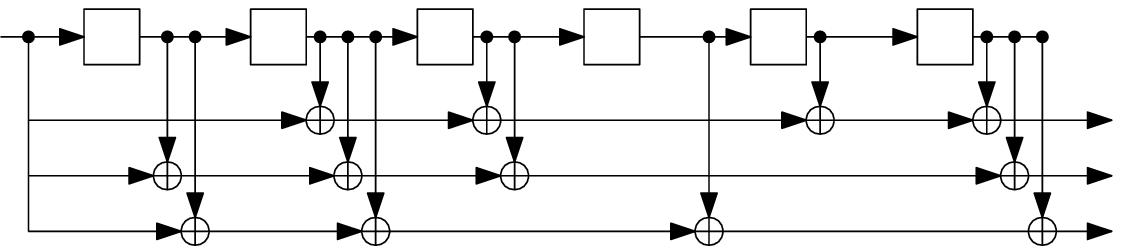}
    \caption{TBCC encoder.}
    \label{fig:tbcc}
    \end{subfigure}
    \caption{The encoder of channel code candidates for IoT.}
    \vspace{-2ex}
\end{figure*}
As discussed in Section~\ref{kpi:latencyreliability}, the next generation of IoT systems need to guarantee various service requirements in terms of latency and reliability. This has posed challenges in the design and implementation of channel coding techniques, which have been previously designed and optimized for long block lengths, mostly used for human-based communications. Interested readers are referred to \cite{kanj2020tutorial} for a detailed discussion on the \gls{phy}  design for wide-area IoT, which includes but is not limited to characteristics and the scheduling of downlink and uplink physical channels at the NB-IoT base station side and the user equipment side. In what follows, we summarize the channel coding challenges for IoT systems. 
\begin{itemize}
    \item \emph{Design of short block length codes:} In many IoT applications, devices send only a small packet of data in a sporadic manner. The payload data could be very short, ranging from only a few to up to a few hundred bits. On the other hand, to guarantee the delay requirement of delay-sensitive applications, it is necessary to use shorter packets and accordingly shorter time-to-transmit intervals. However, if the block length decreases, the coding gain will be reduced and the gap to Shannon’s limit will increase. This is mainly due to the reduction in channel observations that comes with finite block lengths and is not due to the channel code. Moreover,  most of the existing codes with efficient decoders, such as \gls{ldpc} or Turbo codes, show a gap to finite length bounds. This is often due to the suboptimal decoding algorithms.
    \item \emph{Design of efficient decoding algorithms:} Many IoT applications demand ultra-low latency. Despite the improvement of the round-trip delay from around 15ms in \gls{4g}   to around 50$\mu$s in \gls{5g}, the receiver components, including the decoder, still require significant improvement in terms of delay to meet the latency requirement. For example, the successive cancellation decoding technique of polar codes imposes a significant delay due to the successive nature of the decoder. Parallel decoding techniques,  such as the belief propagation algorithm, are favorable, but they work well for only sparse codes on the graph, such as \gls{ldpc} codes. 
    \item \emph{Achieving lower block error rates:} The design target for channel code in the \gls{lte} standard was to achieve the \gls{bler} of $10^{-2}$, however, many IoT applications require high reliability. That means that the \gls{bler} should be less than $10^{-5}$ and even lower than $10^{-9}$ for industrial IoT. Achieving such a low \gls{bler} with efficient decoders is very challenging, especially when we want to perform as close as possible to the finite-length bounds. Some codes, like \gls{ldpc} codes, show error-floor under practical decoders, which limit their performance for low-latency applications.
    \item 1-bit granularity in the codeword length and rate should be maintained for channel codes. This is essential for enabling retransmission techniques and achieving low code rates. Accordingly, efficient error detection algorithms should be developed paired with modern \gls{mac} protocols. 
\end{itemize}

\subsection{State-of-the-Art Channel Coding Techniques for IoT}
Several channel coding techniques have been considered for IoT. The main requirements are for a low-complexity encoder and decoder with shorter block size information and low modulation orders to satisfy low-power requirements. Candidate channel coding techniques for IoT are polar code, \gls{ldpc} code, Turbo code, and \gls{tbcc}. 

\subsubsection{Polar codes} The Polar code was introduced by Arikan in 2008 \cite{Arikan2009} and was recently standardized for the \gls{embb} control channel in \gls{5g} \gls{nr}. The code can achieve the capacity of the binary-input memoryless channel under low-complexity \gls{sc} decoding. The polar code can be realized as a recursive concatenation of a base short block code designed to transform the transmission channel into a set of virtual channels with different levels of reliability. The information bits are put only into the reliable channels and foreknown bits into the unreliable channels, also referred to as Frozen Set. The encoder is through polarization which is given by the kernel $T_2=\left(\begin{array}{ll}1&0\\1&1\end{array}\right)$. For longer inputs, the transformation is obtained through the Kronecker product of $T_2$ with itself. Fig. \ref{fig:polarencoder} shows the polar encoder of length 4. The decoder sets all frozen bits to zero, and starts the decoding of information bits from the highest reliable channel in a successive manner. %SC List (SCL) decoding is a modification of SC decoding which uses $L$ concurrent decoding paths. SCL has shown a significant improvement over SC in the finite-length regime. Further improvements can be made by using a \gls{crc} code, which in turn adds to the decoding complexity.
\begin{figure*}[t]
    \centering
    \begin{subfigure}{.49\textwidth}
    \includegraphics[width=0.9\columnwidth]{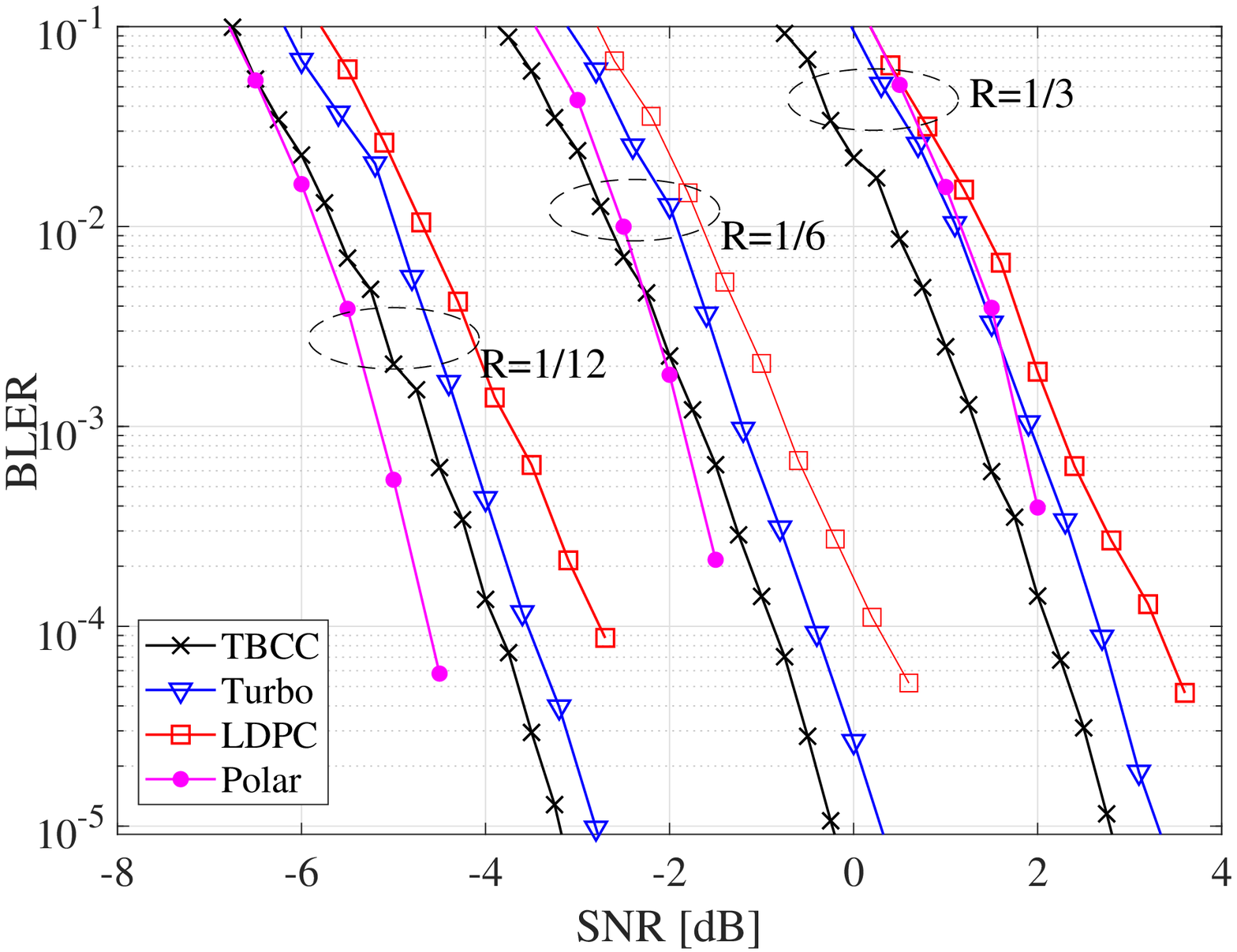}
    \caption{Very short information block length, $k=40$.}
    \label{fig:shortbler} 
    \end{subfigure}
    \begin{subfigure}{.49\textwidth}
    \centering
    \includegraphics[width=0.9\columnwidth]{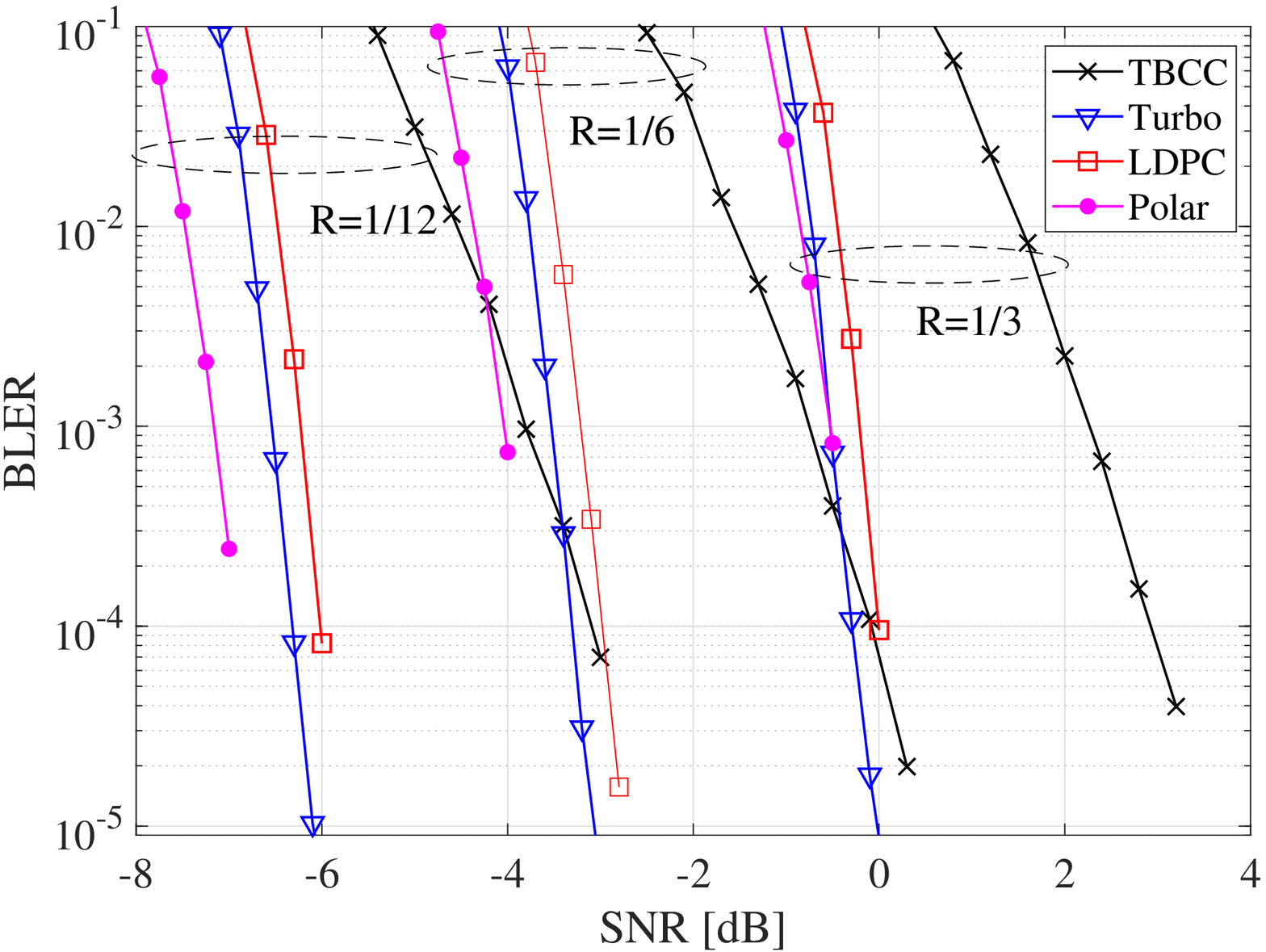}
    \caption{Moderate information block length, $k=600$.}
    \label{fig:modbler}
    \end{subfigure}
    \caption{\gls{bler} performance of candidate channel codes for IoT. Turbo and \gls{tbcc} codes are taken from the \gls{lte} standard. For polar code, \gls{crc} was added and the list size 32 was used for decoding. For \gls{ldpc}, the layered offset min-sum decoder was used with the maximum iteration number of 25. }
    \label{fig:BLER}
\end{figure*}
\subsubsection{\gls{ldpc} codes} The \gls{ldpc} code was first proposed by R. G. Gallagher \cite{gallager1962low} in 1962 and rediscovered later in 1997 \cite{mackay1997near}, when a practical iterative decoder based on the \gls{bp} was designed. \gls{ldpc} codes under the \gls{bp} decoding have shown to perform very close to Shannon's limit \cite{richardson2001capacity}. An \gls{ldpc} code is mainly characterized by its sparse parity check matrix (also shown as Tanner graph in Fig. \ref{fig:ldpc}), which facilitates low complexity decoding. The design of \gls{ldpc} codes is mainly to find the degree of nodes in the Tanner graph and then construct the parity check matrix in such a way to avoid short cycles. The \gls{bp} decoding involves exchanging beliefs (most commonly log-likelihood ratios) between nodes along the edges of the Tanner graph in an iterative manner until convergence. A variant of \gls{ldpc} codes, called the protograph-based Raptor-like \gls{ldpc} code, which is a concatenation of a high-rate \gls{ldpc} code and a low-density generator matrix code, has been recently standardized for the \gls{5g} \gls{embb} data channel due to their excellent performance at moderate-to-long block lengths and their low-complexity decoding. These codes however do not perform well in the short block length regime and show an error floor, which limits their application for short packet communications with latency and reliability constraints \cite{shirvanimoghaddam2018short}.

\subsubsection{Turbo codes} Turbo codes were proposed in 1990 and were adopted in the data channel in \gls{lte} mobile communications. It is also the channel coding technique for \gls{nbiot}. The turbo code is usually a parallel concatenation of two recursive convolutional encoders (see Fig. \ref{fig:turbo}), which are separated by an interleaver. The decoder of Turbo codes runs in an iterative manner by exchanging extrinsic information between two constituent decoders. The Turbo code with the iterative decoder is capable of performing within
a few tenths of dB from Shannon’s limit. However, it shows a gap to the finite length bound at short and moderate block lengths. Moreover, when the code rate is below $1/3$, the performance is further degraded. One advantage of Turbo codes is the capability to support 1-bit granularity for all coding rates and for a full range of block sizes. 

\subsubsection{TBCCs} \gls{tbcc} has been used in the control channel of the \gls{lte} and \gls{nbiot} data channel due to its low complexity encoding and decoding and outstanding performance at short block lengths. The encoder of a \gls{tbcc} is shown in Fig. \ref{fig:tbcc}. Unlike convolutional codes which require some termination bits to force all the states to the zero state, \gls{tbcc} does not terminate states to zero, therefore achieving better rates. The decoder relies on the Viterbi decoder which estimates the maximum likelihood sequence using the code trellis.

Figure \ref{fig:BLER} shows the \gls{bler} comparison between candidate channel codes for IoT. As can be seen, \gls{tbcc} outperforms other codes at very short block lengths (Fig. \ref{fig:shortbler}) however its performance degrades at lower code rates. Polar codes show excellent error performance at both short and moderate block lengths without any error floor. However, they suffer from relatively high-complexity list successive cancellation decoder. The decoder also introduces significant delay which might not be tolerable for delay-sensitive applications. Turbo codes which are being used in \gls{lte} and \gls{nbiot} are performing reasonably well at moderate block lengths, which suggests that they can be a favorable candidate due to their fast and very-low complexity decoders. 

\subsection{Low-capacity Communications}
The current strategies for serving a massive number of IoT are mainly based on narrowband technologies. For example, \gls{nbiot} will serve up to 50,000 \gls{mmtc} devices in one cell and \gls{mmtc} devices in very remote areas. The effective channel \gls{snr} can be as low as -13dB or with a capacity 0.03 \cite{abbasi2020polar}. In other words, the communication is over a very low-capacity channel. Another scenario is wideband communications, e.g., millimeter-wave in \gls{5g} and Terra-Hertz in \gls{6g}, which have higher frequencies and much larger bandwidths. Under a constant power, we will have a higher data rate but the individual bits will be going through channels with very small \gls{snr} \cite{abbasi2020polar}. Current techniques, such as repetition (in \gls{nbiot}) and \gls{harq} with fixed-rate channel codes are inefficient in these scenarios. 

Authors in \cite{abbasi2020polar} showed that concatenating the state-of-the art codes at moderate rates with repetition codes may be a practical solution for various applications, concerning ultra-low power devices. A polar-coded repetition strategy was also proposed with the aim to limit the encoding and decoding complexity. In \cite{Shirvani2016Low}, Raptor codes were designed for very low \gls{snr} channels, where the degree distribution was optimized to achieve the rates efficiencies as high as 96\%. The codes are however, mainly designed and optimized for asymptotically long block lengths and their performance dramatically degrade at short blocklengths. The decoder of Raptor codes at low \glspl{snr} also requires a very large number of iterations to converges, which might not be tolerable for applications with latency constraints. 

The decoding in low-capacity communications is also challenging due to the fact that the received symbols are not reliable. In particular, the \glspl{llr} are very small, that is the iterative message passing decoders, in case of LPDC or Raptor codes, require a large number of iterations to converge. Moreover, the successive interference cancellation even with large list sizes may not perform well when the \glspl{llr} are very low. Increasing the list size significantly increases the decoding complexity which is not feasible for latency-constrained scenarios. 

\subsection{Channel Coding for \gls{csi}-free Communications}
Most of existing channel adaptation techniques rely on the \gls{csi} usually sent from the receiver to the transmitter via a feedback message. In adaptive modulation and coding (AMC) the transmitter chooses a fixed-rate channel code with a proper rate and the modulation scheme  in order to maximize the spectral efficiency according to the \gls{csi}. There are three main problems with this approach which limits its efficiency in future IoT applications. First, obtaining \gls{csi} to all IoT devices is not practical especially in massive IoT scenarios with sporadic traffic. Second, even if the \gls{csi} could be obtained for IoT devices, it is not practical to frequently update it due to random traffic, mobility, and activity of devices. Third, obtaining \gls{csi} requires the transmitter to send pilot sequences to the receiver and then receive the \gls{csi} via a feedback message. This is time-consuming and adds significant overhead which will be further challenging in massive IoT scenarios with a very large number of devices. Moreover, this overhead and the delay associated with obtaining \gls{csi} in most cases violate the latency requirements of many delay-sensitive applications \cite{chen2018ultra}.

Retransmission techniques can be used to adapt to the channel condition without the \gls{csi} at the transmitter side. In particular,  \gls{harq} can achieve a fine-grain matching for channel conditions with multiple retransmissions and acknowledgments. However, these approaches are mainly effective when channel variations are slow. Severe path loss, high blockage and fluctuating characteristics of wireless channels at mm-Wave, and higher frequency bands which are expected to be used for the \gls{5g} and beyond of mobile standards \cite{chen2018ultra}, have questioned the suitability of current adaptive transmission techniques to meet the reliability and latency goals of future networks. 

\begin{figure}[t]
    \centering
    \includegraphics[width=\columnwidth]{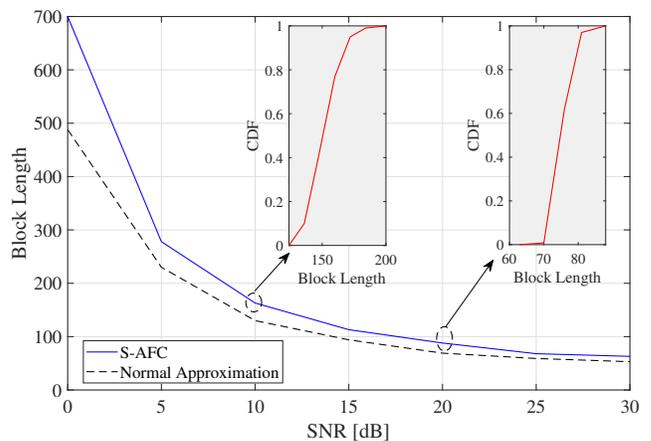}
    \caption{Average block lengths of \gls{afc} versus \gls{snr}. A (256,192) \gls{ldpc} code was used as the precode for \gls{afc} and the target \gls{bler} is $\epsilon=10^{-4}$.}
    \label{fig:afc}
\end{figure}

Rateless codes have gained interests in recent years due to their automatic channel adaptation features \cite{shirvanimoghaddam2018short}. Most of these codes have been designed for long message lengths and the encoding mostly involves random generations of parity symbols. Examples include Raptor codes \cite{shokrollahi2006raptor}, Strider \cite{gudipati2011strider}, rate compatible modulation (RCM) \cite{cui2011seamless}, \gls{afc} \cite{shirvani2013AFC}, and Spinal codes \cite{perry2012spinal}. The aim is to achieve a fine granularity of spectral efficiency and seamless link adaptation without the feedback of channel condition.  Although several attempts \cite{abbas2019safc,huang2019pegAFC,Zhang2020exitAFC,xu2018high,abbas2019performance} have been made to optimize these codes for short message lengths, the codes still suffer gaps to finite block length bounds, such as the normal approximation bound \cite{polyanskiy2010channel}, when practical decoders with bounded complexity are being used \cite{shirvanimoghaddam2018short}. Fig. \ref{fig:afc} shows the performance of short \gls{afc} \cite{abbas2019safc} in terms of the average block length to achieve a target \gls{bler} of $10^{-4}$. As can be seen, this code performs close to the normal approximation bound in a wide range of \glspl{snr}, however, further improvement at low \glspl{snr} is required. It is also important to note that at low \glspl{snr} the block length required to achieve the desired level of \gls{bler} varies significantly, which might limit the use of the code for delay-sensitive applications. 

\begin{table*}[!t] 
    \centering
        \caption{Comparison of channel code candidates for IoT applications.}
    \label{tab:FECcomparison}
    %\scriptsize
    \footnotesize
    \begin{tabular}{p{2.4cm}|p{2.65cm}|p{5cm}|p{5cm}}
         \textbf{Code}& \textbf{Decoding}& \textbf{Advantages} & \textbf{Disadvantages}  \\
         \hline
         \hline
         \Tstrut \Bstrut \gls{ldpc} codes& Belief propagation decoding, min-sum decoder, max-sum decoder&Excellent performance at long-block lengths, efficient decoder, parrallelization of the decoder to reduce latency & \Tstrut  Error floor at short and moderate block lengths, the \gls{bp} decoder is suboptimal for short codes, 1-bit granularity cannot be easily achieved without loosing performance \\
         \hline
          \Tstrut \Bstrut  Polar codes&Successive cancellation, List decoding, Stack decoding, Belief propagation&Can achieve very low \gls{bler}, No sign of error floor, 1-bit granularity can be achieved, & The successive decoder introduces delay, the list decoding is complex and not practical for low-power devices, Large list size should be used to achieve very low \gls{bler}\\
         \hline
          \Tstrut \Bstrut Turbo codes&Iterative soft decoding&Very fast encoder and decoder, Parallel architecture can be used to boost the decoding speed, 1-bit granularity can be easily achieved, close-to-Shannon performance at large block lengths& Loose performance at short and moderate block lengths, Large gap to the finite-length bounds as short block lengths \\
         \hline
          \Tstrut  \Bstrut  \gls{tbcc} & Viterbi decoding&Fast encoder and decoder, can achieve very low \gls{bler} at very short block lengths& The performance degrades at moderate block lengths or rates lower than 1/3, large memory order should be used to achieve lower \gls{bler}, The decoding complexity exponentially increases with the memory order\\
         \hline
          \Tstrut \Bstrut Algebraic codes (Hamming codes, Golay codes, Reed-Muller codes, \gls{bch} codes, Reed-Solomon codes) & \Gls{osd} \cite{Fossorier1995}, Chase decoding, Berlekamp–Massey decoding, majority-logic decoding, maximum-likelihood decoding&Fast encoders, large minimum Hamming distance, can achieve very low \gls{bler} without an error floor& The decoding is very complex, the decoding performance degrades when low-complexity decoders are used, \gls{osd} can achieve near ML performance but the complexity is huge especially at low rates and moderate block lengths\\
         \hline
         \Tstrut \Bstrut  Fountain codes (Raptor code \cite{shokrollahi2006raptor})& Message-passing decoding&Codes can automatically adapt to the channel without channel state information, the message passing decoder can be implemented efficiently& The code performs well when the block length is very large, the code has been optimized for erasure channels and over wireless channels the code design is channel dependent, Codes do not perform well at short block lengths, the decoding complexity is large at low code rates\\
         \hline
         \Tstrut \Bstrut Rate Adaptive codes (\gls{afc} \cite{Shirvanimoghaddam2013}, RCM \cite{cui2011seamless}, Spinal \cite{perry2012spinal})& \gls{bp} for \gls{afc} and RCM, and Viterbi-like decoding for Spinal codes&Can adapt to the channel condition automatically without channel state information, can be optimized for short block lengths, achieve the finite-length bound over a wide range of \glspl{snr}, iterative belief propagation decoding can be applied (for \gls{afc} and RCM)& The decoding complexity can be large, the output signal are usually from large constellations\\
         \hline
         \Tstrut \Bstrut Real/complex BCH-\gls{dft} codes \cite{Marshall,wolf1983redundancy,blahut2003algebraic} &  
             Coding-theoretic approach \cite{blahut2003algebraic,takos2008determination} and subspace-based approach \cite{rath2004subspace,vaezi2014generalized}&   Improved performance as code length reduces. Can be used for channel coding, joint source-channel coding, and distributed source coding \cite{vaezi2014thsis,rath2004subspace,gabay2007joint}, good performance for erasure detection \cite{rath2004subspace1,goyal2001quantized,vaezi2013frameJ}& 
To be used in digital communication systems, the output of this class of codes should be quantized and converted to bits. Both decoding algorithms are sensitive to quantization error and the performance drops if the number of quantization levels is not big enough.
         \\
                  \hline         
    \end{tabular}
\end{table*}

\gls{bch}-\gls{dft} codes \cite{Marshall,wolf1983redundancy,blahut2003algebraic} are another class of analog codes that have been applied in various applications in channel coding, joint source-channel coding, and distributed source coding \cite{vaezi2014thsis,rath2004subspace,gabay2007joint}. These codes can be used for 
erasure detection too \cite{rath2004subspace1,goyal2001quantized,vaezi2013frameJ}.
Classical decoding for this class of codes is called the coding-theoretic approach \cite{blahut2003algebraic}. To be able to use these codes in digital communications, the codewords undergo a quantization process. This process introduces quantization error to each element of the codeword. This, in turn, deteriorates the decoding performance. Subspace-based method \cite{rath2004subspace,vaezi2014generalized} is an alternative decoding approach that is more robust to the quantization error. 
 The decoding performance still may not be acceptable when the number of quantization levels is small or the code length is large. As such, these codes are more useful for short block length. The drawback of the short block length is that the code capacity move away from Shannon limits.

\subsection{Decoder Design}
Although algebraic codes, such as \gls{bch} and Reed-Solomon codes, have large minimum Hamming distance and accordingly excellent \gls{bler} performance with the maximum-likelihood (ML) decoding, their decoding complexity has made them impractical for IoT applications. On one hand, most of the near-ML decoding algorithms, such as the \gls{osd} \cite{Fossorier1994}, are very complex and even their best implementations introduce significant latency. On the other hand, their implementations are not power-efficient, therefore cannot be embedded into low-power devices. Researchers have recently proposed several approaches to reduce the complexity of the \gls{osd} algorithm, which are mainly focused on reducing the number of test error patterns when searching for the best candidate codeword. In particular, sufficient and necessary conditions have been introduced to terminate the decoding early or skip the test patterns without encoding them \cite{Chentao2021OSD,choi2020fast}. There are still however issues that need to be resolved. First, although these approaches can significantly reduce the average decoding complexity, the worst-case scenario is still very complex and the receiver might need to search through a large number of test patterns. Second, most of the \gls{osd}-based approaches run in iterations and it seems hard to devise efficient algorithms for parallelization of the process. However, approaches like segmentation-discarding \cite{Chentao2019SDOSD} and efficient tree-based search algorithms \cite{Chentao2020PBOSD} might help to further reduce the complexity. 

The \gls{bp} decoding algorithm of \gls{ldpc} codes involves a complex non-linear function in the check-node processing. This leads to large implementation complexity. Several simplified alternative algorithms, like min-sum \cite{fossorier1999reduced}, normalized min-sum \cite{chen2005reduced}, and off-set min-sum \cite{chen2005reduced}, have been proposed and being widely used in practice thanks to their inherent parallelism for hardware implementation. The simplification however leads to a significant performance loss, especially at low rates, short-to-moderate block lengths, and when the decoding iteration number must be kept low, which are the scenarios of interest for IoT \cite{wu2018decoding}. Moreover, the \gls{5g} \gls{ldpc} codes have numerous degree-1 variable nodes, which are very prone to be erroneous \cite{cui2020design}. Several approaches have been proposed in the literature to improve the decoding complexity of \gls{5g} \gls{ldpc} codes. These includes, linear-approximated \gls{bp} \cite{sun2018hybrid}, adapted min-sum \cite{le2019adaptation}, and improved adapted min-sum \cite{cui2020design}. Further reduction of the complexity is necessary for the decoder of the \gls{ldpc} to make them a realistic solution for IoT applications. 

Polar codes can be efficiently decoded using \gls{sc} decoding, however the performance significantly degrades at moderate-to-short block lengths. To solve this problem, the \gls{sc} List decoding \cite{chen2012list} can be used which improved the decoding performance but also scale the complexity by a factor $M$, which is the list size. For \gls{5g} Polar codes usually a large list size, larger than 32, should be used to achieve \gls{bler} of less than $10^{-4}$. The stack \gls{sc} decoding \cite{niu2012stack} has been also proposed to improve the decoding performance, which works similar to the Viterbi decoder. A hybrid list and stack \gls{sc} decoding was also proposed in \cite{li2012adaptive}, which slightly reduces the complexity. The performance however degrades when stack-size is short. Further improvement in the decoding performance can be obtained by using \gls{crc} bits \cite{niu2012crc}, which also increases the complexity.  The serial nature of the \gls{sc} decoding limits the speed of the decoder, which has implications for low-latency communications. 

The decoders for polar and \gls{ldpc} codes require the channel state information at the receiver side. When such information is not available, the decoding fails. The design of algorithms capable of handling limited or no \gls{csi} at the receiver is therefore essential for IoT applications \cite{mahmood2020white}. Furthermore, the current error detection schemes are mainly based on \gls{crc} codes, which adds significant overhead when short packets of data needs to be sent. For massive IoT applications, the design of strong error detection algorithms without \gls{crc} bits is required \cite{mahmood2020white}. 

Table \ref{tab:FECcomparison} shows a comparison between different channel coding techniques for IoT applications.

%--------------------------------------------------------
%-------------------------------------------------------
%-------------------------------------------------------

\section{Massive Connectivity}
\label{sec:NOMA}

 As explained \secref{sec:KPIdensity},  \gls{5g} and \gls{6g} networks are expected to support $10^6$ and $10^7$ devices/km$^2$, respectively. A big percentage of the new connections will be due to massive IoT applications, i.e., the \gls{mmtc} use case. We also indicated that traffic generated by such devices is sporadic and has a small payload which is different from those of \gls{embb} and \gls{urllc} applications. Further, \gls{mmtc} devices
 have limited power while they may need to reach
 a very far \gls{bs} (e.g., smart agriculture sensors which are typically deployed in rural areas). There is also a fast-growing number of IoT devices in \gls{embb} and \gls{urllc} applications. 
 
 Considering all use cases, a range of solutions are proposed to address the diverse range of devices connecting to \gls{5g} and beyond networks. 
One popular trend is moving from orthogonal to non-orthogonal design  in the waveform, random access, and multiple access designs \cite{vaezi2018book}. 
In particular, non-orthogonal multiple access (NOMA), non-orthogonal random access, and non-orthogonal waveform design have been extensively studied during the past several years. All in all, massive connectivity solutions can be categorized as

\begin{forest}
	for tree={
		align=left,
		edge = {draw, semithick, -stealth},
		anchor = west,
		font = \small\sffamily\linespread{.84}\selectfont,
		forked edge,          % for forked edge
		grow = east,
		s sep = 0mm,    % sibling distance
		l sep = 8mm,    % level distance
		fork sep = 4mm,    % distance from parent to branching point
		tier/.option=level
	}
	[Massive \\ connectivity \\ solutions
	[Random  \\ access \\based [ unsourced \\ random access][grant-free \\ random access] ]  [Multiple \\ access \\based [code domain \\ NOMA (uplink)][power domain \\ NOMA (downlink)] ] 	 [New \\spectra [optical]  [THz] [\gls{mmwave}]] 	]
\end{forest}
and we are going to elaborate focus on multiple access and random access in this section.

\subsection{Multiple Access Solutions}

Existing cellular networks, including \gls{4g}    \gls{lte}, have been designed based on non-overlapping radio resource allocation techniques. For example, in \gls{lte} networks,  a  \gls{rb} which is 180kHz, cannot be allocated to more than one user, i.e., it  cannot be shared with other users. This approach in resource allocation has two major drawbacks in massive IoT: 1) with exploding number of massive IoT is it not affordable to allocate one dedicated \gls{rb} to each device, 2) massive IoT users do  not deplete one \gls{rb} and, thus, such resource allocation may not be efficient. 
In addition, IoT communication is uplink-dominated and  uplink communication
of  massive low-rate IoT devices requires
a very different set of technologies than
those designed to serve humancentric communications  in \gls{4g}   and previous generations of cellular networks.

Due to the above issues and the fact that an unprecedented number of new IoT devices are projected to be connected to wireless networks\footnote{Cisco  visual networking index   projects  %more than 50\% annual data growth  and
	about 20\%   annual growth in the machine to machine connections  until 2021 \cite{Cisco}. Ericsson 2018 mobility report forecasts 30\% annual growth in IoT  devices between 2017 and  2023 \cite{ericsson}.},  
the trend in communication system design has recently  changed from orthogonal to non-orthogonal. This paradigm shift is happening  in various \gls{mac} and \gls{phy} technologies such as waveform design, multiple access, and random access \cite{vaezi2018book}. In particular, NOMA techniques, unlike their orthogonal counterparts, revolve around letting two or more users share the same \gls{rb}. Hence, NOMA is a promising approach to accommodate upcoming users, in general, and IoT users, in specific. 
%In addition, NOMA enhances the spectral efficiency, and improve the user-fairness in wireless networks and can be combined with various technologies successfully \cite{vaezi2019interplay}. 

NOMA  techniques
are inherently  different for uplink and downlink  communications. The difference becomes noticeable when it comes to IoT communication, where a massive number of
uncoordinated devices transmit small packets with low
data rates in the uplink. It should be noted that  the downlink is mostly focused on serving humancentric communications, which typically requires large  packets and
high data rates. In the following, we discuss the uplink and downlink NOMA techniques.

		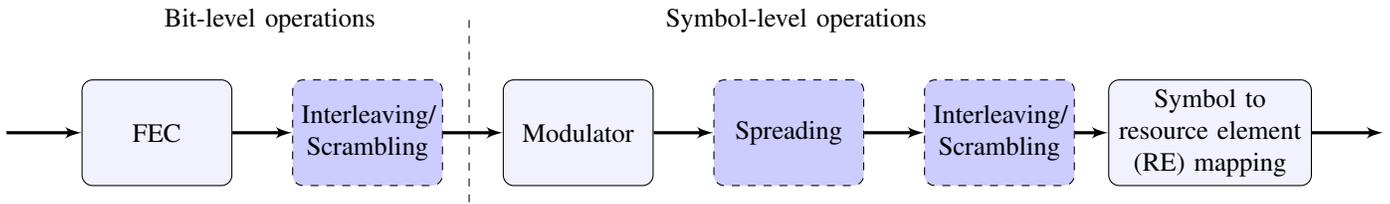
\begin{figure*}
	\begin{tikzpicture}[node distance = 2cm, auto]
	\node [block] (st1) {FEC};
	\node [block2, right of=st1, node distance=2.8cm] (st2) {Interleaving/ Scrambling};
	\node [block, right of=st2, node distance=2.8cm] (st3) {Modulator};
	\node [block2, right of=st3, node distance=2.8cm] (st4) {Spreading};
	\node [block2, right of=st4, node distance=2.8cm] (st5) {Interleaving/ Scrambling};
	\node [block, text width=7em, right of=st5,  node distance=2.8cm] (st6) {Symbol to resource element (RE) mapping};
	\path [line] (-2,0) -- (st1);
	\path [line] (st1) -- (st2);
	\path [line] (st2) -- (st3); 
	\path [line] (st3) -- (st4);
	\path [line] (st4) -- (st5);
	\path [line] (st5) -- (st6);
	\path [line] (st6) -- (16.3,0);
	%\draw[snakeline]  (-2,0) -- (2,0) ;
	%\draw (-3,-3) grid (3,3);
	\draw[dashed]  (4.15,1.5) -- (4.15,-1);
	%% draw 1 unit down from st3.south then horizontally left and up to reach st2.south.
	%\path[line,dashed] (st3.south) -- +(0,-1) -|  (st2.south) node[pos=0.25,below]{Blah2} ; 
	\node[] at (1.5,1.5) {Bit-level operations};
	\node[] at (8.5,1.5) {Symbol-level operations};
	\end{tikzpicture}
	\caption{General channel structure of a NOMA transmitter. Each NOMA scheme may have one or more of the dashed blocks.}
	\label{fig:NOMA}
\end{figure*}

\subsubsection{Uplink NOMA}
In the uplink, NOMA embraces a variety  of non-orthogonal transmission techniques, These schemes are devided into \textit{power-domain} NOMA and \textit{code-domain} NOMA \cite{vaezi2018book,NOMA3GPPNR}.  Power-domain NOMA is another name for Shannon capacity-achieving \gls{sic}-based coding  for the multiple access channel. Examples of power-domain uplink NOMA can be found in \cite{kara2018ber,shin2017relay,zeng2019energy,schiessl2020noma} 
Code-domain NOMA, however, embraces a number of new techniques which will be discussed in the following. 
% The list includes but not limited to sparse code multiple access (SCMA),  interleave division multiple access (IDMA), 
% \gls{pdma}, and low-density signature  coding(LDS) \cite{hoshyar2008novel,hoshyar2010lds}.

\subsubsection*{Transmitter Structure} 
Resource \textit{overloading}  is the common theme of these non-orthogonal access methods which allows  the  number of supportable users to be more than the number of available resources.
Then, a NOMA scheme needs a signature to differentiate users and cancel inter-user interference. 
The structure of a generic NOMA transmitter is depicted in Fig.~\ref{fig:NOMA} \cite{NOMA3GPPNR,YuanIndustry}. Different NOMA schemes apply their signatures in one or more of the dashed blocks in Fig.~\ref{fig:NOMA} \cite{chen2018toward,YuanIndustry,NOMA3GPPNR}. 

\subsubsection*{Categories/Schemes} 
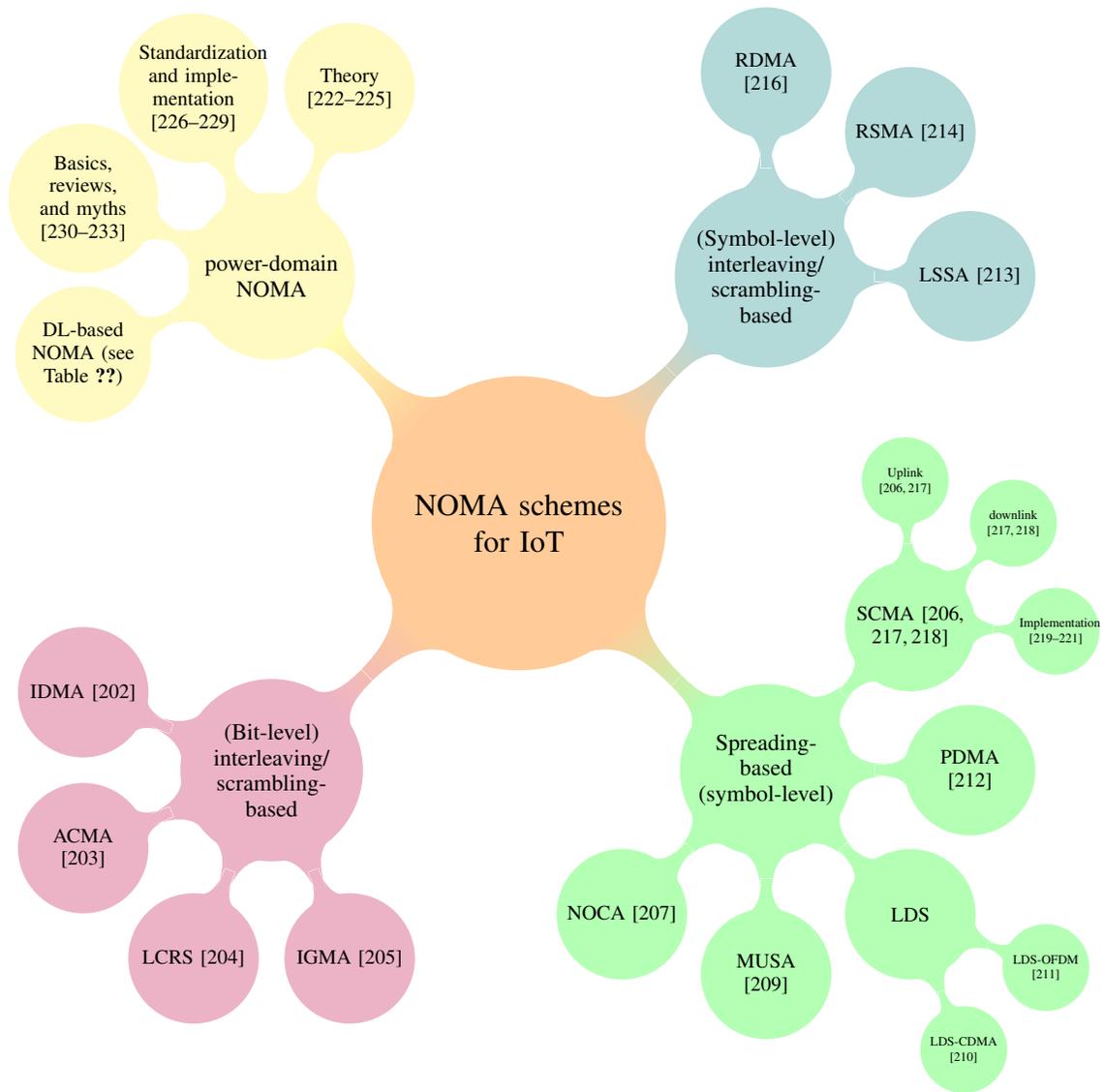
\begin {figure*}%[!hbtp]
\centering
\begin{tikzpicture}[mindmap, grow cyclic, every node/.style=concept, concept color=orange!40,
level 1/.append style={level distance=4.8cm,sibling angle=90},
level 2/.append style={level distance=2.8cm,sibling angle=45},
level 3/.append style={level distance=2cm,sibling angle=47}
]

\node{NOMA schemes for  IoT}
%child [concept color=blue!30] {node {Hybrid/other}
%	child { node {IGMA \cite{hu2018nonorthogonal}}}
%	child { node {SDMA }}
%	child { node {LPMA}}
%}
child [concept color=purple!30] { node {(Bit-level) interleaving/ scrambling-based}
	child { node {IDMA \cite{ping2006interleave}}}
	child { node {ACMA \cite{ACMA}}}
	child { node {LCRS \cite{LCRS}}}
	child { node {IGMA \cite{hu2018nonorthogonal}}}
}
child [concept color=green!30] { node {Spreading-based (symbol-level)}
	child { node {NOCA \cite{NOCA}}}
	child { node {MUSA \cite{yuan2016multi}}} 
	child { node {LDS}
		child { node {LDS-CDMA \cite{hoshyar2008novel}}}
		child { node {LDS-OFDM \cite{hoshyar2010lds}}}
	}
	child { node {PDMA \cite{PDMA}}}
	child { node {SCMA \cite{nikopour2014scma,tang2016low,Ma2019}}	
		child {node {Implementation \cite{sun2019ldpc,liu2016fixed,ghaffari2019toward}}}
		child { node {downlink \cite{tang2016low,Ma2019}}}
		child { node {Uplink \cite{nikopour2014scma,tang2016low}}}
	}
}
child [concept color=teal!30] { node { (Symbol-level) interleaving/ scrambling-based  }
	child { node {LSSA \cite{LSSA}}}
	child { node {RSMA \cite{RSMA}}}  		
	child { node {RDMA \cite{RDMA}}}
}
child [concept color=yellow!30] {node {power-domain NOMA}
	child { node {Theory \cite{Cover,ElGamal2011network,vaezi2019noma,shin2017non}}}
	child { node {Standardization and implementation \cite{NOMA3GPP,yuan2016non,Benjebbour,vanka2012superposition} 
	}} 
	child {node {Basics, reviews, and myths \cite{saito2013non,dai2015non,liu2017nonorthogonal,vaezi2019non} }
		% child {node {}}
	}
	%  child {node {MIMO-NOMA \cite{wu_noma_wpcn_cnsmp_4,lv2018millimeter,almasi2019lens,ding2016mimo}}} 
	child {node {DL-based NOMA (see Table~\ref{table:dlIoT})}}
%	child {node {Uplink NOMA \cite{kara2018ber}}}
};
\end{tikzpicture}
\caption{Various NOMA schemes proposed in \gls{3gpp} \gls{5g} RAN meetings. Code-domain NOMA (mostly uplink) is categorized into scrambling-based, spreading-based, interleaving-based,  hybrid, and others.  Power-domain NOMA (downlink) is in a separate category.  }
\label{fig:NOMAchild}
\end{figure*}

Code-domain NOMA schemes can be categorized in various ways. 
As can be seen in Fig.~\ref{fig:NOMA}, NOMA-related operations can be done in the \textit{bit-level} (before modulator) or \textit{system-level}.   
For this reason, as also shown in Fig.~\ref{fig:NOMAchild},  NOMA schemes may be categorized  as follows:

\begin{itemize}
	\item \textbf{Bit-level interleaving/scrambling:} 	An interleaver or a scrambler at bit-level is used to distinguish the users in this case \cite{ping2006interleave}. 
	% Release 15 NR already supports a bite-level scrambler for randomization which may be exploited for NOMA too. 
	A bit-level scrambler incurs less processing delay and memory requirement compared to bit-level interleaver \cite{NOMA3GPPNR}. Examples of this category are 
	\gls{idma} \cite{ping2006interleave},  \gls{acma} \cite{ACMA}, and  \gls{lcrs} \cite{LCRS}. 
	Some NOMA schemes apply more than one of the above methods. For example, \gls{igma} \cite{hu2018nonorthogonal} uses both bit-level interleaving and  sparse mapping.

	\item \textbf{Symbol-level spreading:}  The majority of NOMA schemes fall in this category where a spreading sequence is used as a signature to distinguish the users. Notable examples are:  \gls{scma} \cite{nikopour2014scma}, \gls{noca} \cite{NOCA}, non-orthogonal coded multiple access (NCMA) \cite{NCMA},  \gls{musa}  \cite{yuan2016multi},  \gls{lds} \cite{hoshyar2008novel,hoshyar2010lds}, and \gls{pdma} \cite{PDMA}.
	
	\item \textbf{Symbol-level interleaving/scrambling:} A symbol-lever interleaver/scrambler is used to distinguish the users. 	
	Low code rate and signature-based shared access  \cite{LSSA}, \gls{rsma} \cite{RSMA,Crsma}, and \gls{rdma} \cite{RDMA} are examples of this category.

	%			
	%			\item \textbf{Hybrid and others:} Some NOMA schemes apply more the one of the above methods. For example, interleave-grid multiple access (IGMA) \cite{hu2018nonorthogonal} uses bit level interleaving and  sparse mapping.  
	%			spatial domain multiple
	%	access (SDMA), Lattice partition multiple access (LPMA) \cite{fang2016lattice}.
\end{itemize}

\begin{table*} [h]
\centering
%	\begin{minipage}[b]{0.9 \textwidth}
\centering
\begin{threeparttable}[b]		
	\caption{Complexity and overloading for selected code-domain NOMA schemes} 
	\label{tab:size_shape}
	\begin{tabular} { p{0.18 \textwidth} p{0.08 \textwidth} p{0.08 \textwidth}  p{0.08 \textwidth} p{0.08 \textwidth} p{0.08 \textwidth} p{0.08 \textwidth} p{0.08 \textwidth}  } 
		\toprule
		\rowcolor{Lightgray}& SCAM    &  \gls{idma}  &  \gls{musa} & \gls{igma} & \gls{pdma}    &  \gls{noca}  &  RSMA  \\        \cmidrule{1-8}  \addlinespace
		\rowcolor{Lightgray} Proposing company    & Huawei &  InterDigital  & ZTE & Samsung & CATT & Nokia & Qualcomm  \\ \addlinespace
		\rowcolor{Lightgray}Receiver complexity   & High &  Moderate  & Low & High & High &  Low  & Low \\ \addlinespace
		\rowcolor{Lightgray}Overloading factor    & Moderate  &  High & High & High &Moderate  &  Moderate & High  \\ \addlinespace       
		%			\rowcolor{Lightgray}Elongated  & 5\%  &  12\% & 4\% & 21\% \\ \addlinespace
		%			\midrule
		%			\rowcolor{Lightgray}Size fraction  & 19\% & 66\% & 15\% & 100\% \\
		\bottomrule
	\end{tabular}
\end{threeparttable}
%	\end{minipage}
\end{table*}

Symbol-level spreading-based NOMA schemes are mostly inspired by \gls{cdma}.  
\gls{cdma} is a multiple access technique in which data symbols are spread over a group of user-specific  mutually-orthogonal codes.
\gls{lds} is a version of \gls{cdma} in which spreading codes have a low density, i.e., a small fraction of  code's elements  are non-zero \cite{hoshyar2008novel,hoshyar2010lds}. 
This allows using near-optimal \gls{mpa} receivers with
practically feasible complexity. 
In spite of its  moderate detection complexity, \gls{lds} suffers from performance degradation  for  constellation sizes larger than QPSK.

\Gls{scma} is a multi-dimensional codebook NOMA scheme in which the encoder directly maps incoming data bits to \gls{scma} codewords selected from
a layer-specific codebook where each codeword represents a spread transmission layer \cite{taherzadeh2014scma}.  While in  
\gls{cdma} (and \gls{lds}) spreading and bit to symbol mapping are
done separately, \gls{scma} merges these two steps and directly maps
incoming bits into a spread 
codeword within the \gls{scma} codebook sets.

Due to the sparsity of codewords, like \gls{lds}, \gls{scma} enjoys the low complexity reception
techniques \cite{sun2019ldpc}. Additionally, it takes advantage of additional
degrees of freedom in the design of multi-dimensional constellations
and  outperforms \gls{lds} \cite{taherzadeh2014scma}. 
More importantly, sparser \gls{scma} codewords can tolerate more
overloading to enable massive IoT connectivity in the uplink.  A sparser code, however,
will result in lower coding gain. \Gls{scma} can also be used for the downlink \cite{nikopour2014scma,tang2016low,Ma2019}, but the complexity decoding is still very high
for low-cost IoT devices and this limits the capability of \gls{scma} to support massive IoT connectivity in the downlink \cite{Ma2019}. 

Table~\ref{tab:size_shape} compares the complexity and overloading factor for a selected group of the above \gls{noma} schemes listed in Fig.~\ref{fig:NOMAchild}.  The schemes with higher overloading and lower complexity are preferred to others.

\subsubsection{Downlink NOMA} 
In the downlink, \gls{noma} reaps the benefits of the \textit{broadcast
channel} (BC) in which all users can transmit their messages 
at the same \gls{rb}   \cite{Cover,ElGamal2011network,vaezi2019noma}. In  a single-cell, \gls{siso} system, such transmission is theoretically optimal and \textit{superposition coding} is used at the \gls{bs} and \gls{sic} is used to cancel the inter-user interference introduced by  all users with weaker channel gains while treating the interference of stronger users as noise. As such, all users except the one with the weakest channel gain will need to apply \gls{sic} for decoding. 
Although \gls{sic}  has been implemented in smartphone size devices \cite{benjebbour2018outdoor}, it  is too complex for simple IoT devices.  As a result, although some papers  propose downlink \gls{noma} for IoT users \cite{wu_noma_wpcn_cnsmp_4,ding2016mimo,lv2018millimeter,almasi2019lens,jia2019design},  due to the \gls{sic} complexity, practically it is hard to justify more than one IoT device in each group of \gls{noma}.    
% if achieving capacity of the BC is a goal,
This may limit the large-scale implementation of \gls{noma} for IoT devices.

In Release 13 and 14 \gls{lte}, multi-user superposition transmission (MUST) was studied and specified as a
downlink \gls{noma} scheme \cite{NOMA3GPP}. 
However, 
MUST was not borrowed for \gls{5g} \gls{nr} downlink since the gain from \gls{noma} was marginal in massive \gls{mimo} setting, and the study was terminated in October 2016 \cite{YuanIndustry}. As a result, Release~15 \gls{nr} networks are largely OMA-based. 
%  Despite its well-established theory,  there have been several myths and  misunderstandings about \gls{noma} \cite{vaezi2019non}. 
Also, the fact that power-domain \gls{noma} requires a relatively complex receiver, the application of \gls{noma} to IoT networks is not trivial and needs careful studies. With the recent trend of applying \gls{dl} to communication systems, in general,  and to \gls{noma} \cite{mohammadi2018deep,liu2018deep,doan2019power,gui2018deep}, in specific, the chance of coming up with viable \gls{noma} solutions  for IoT networks increases. As an example, in \cite{zhang2019deep}, it is shown that DL-based precoding is advantageous in terms of delay and complexity and is more  suitable for applications.

 It is worth highlighting that \gls{sc-sic} is not an optimal solution when  nodes have multiple antennas or the network has more than one cell. Particularly, \gls{sc-sic} is not optimal in  \gls{miso}  \cite{clerckx2021noma} and \gls{mimo}, and secure \gls{mimo} systems \cite{vaezi2019noma}. 
 Despite the fact that the capacity of \gls{mimo}-BC is known \cite{weingarten2006capacity}, many \gls{noma} researchers have applied \gls{sc-sic} to this channel \cite{wu_noma_wpcn_cnsmp_4,lv2018millimeter}, calling it \gls{mimo}-\gls{noma}, which is sub-optimal.
 Recent works have tried to address this deficiency by focusing on optimal capacity-achieving schemes for this channel. The optimal schemes are still non-orthogonal but not based on \gls{sc-sic}. Rate splitting and \gls{dpc} are among these schemes and are necessary for the spectral efficiency of \gls{miso} and \gls{mimo}-BC channels \cite{vaezi2019noma,qi2020secure,clerckx2021noma}.

\subsection{Random Access Solutions}\label{RAS}
Packet flow through an \gls{lte} network needs coordination between the \gls{bs} and the device. A device can be in an \textit{idle} or \textit{active} state \cite{vaezi2017radio}. 
When the device is idle, it is only listening to control channel broadcasts.
Access protocols come into play when the device wakes up, i.e., its state  changes from idle to active \cite{vaezi2017radio}. To wake up from an idle state, the device must synchronize with a nearby  \gls{bs} by sending a request for radio resources.
For this purpose, an access protocol is used
to coordinate the data exchange between multiple active devices and the \gls{bs} \cite{laya2013random}.
Since the activity of the devices is random,
 \textit{random access} protocols are used in
cellular IoT \cite{chen2020massive}.   
  Random access protocols can be  divided into three types:
  %into grant-based, grant-free, and unsourced random access.

	 \begin{forest}
 	for tree={
 		align=left,
 		edge = {draw, semithick, -stealth},
 		anchor = west,
 		font = \small\sffamily\linespread{.84}\selectfont,
 		forked edge,          % for forked edge
 		grow = east,
 		s sep = 0mm,    % sibling distance
 		l sep = 8mm,    % level distance
 		fork sep = 4mm,    % distance from parent to branching point
 		tier/.option=level
 	}
 	[Random  \\ access \\ protocols [ unsourced  ][grant-free  ] [grant-based  ] ]   	 	 
 \end{forest}

\subsubsection{Grant-based  random access} 
\begin{figure}
	\begin{center}
		\begin{tikzpicture}[>=latex]
		\coordinate (A) at (2,7);
		\coordinate (B) at (2,0);
		\coordinate (C) at (9,7);
		\coordinate (D) at (9,0);
		\draw[thick] (A)--(B) (C)--(D);
		\draw (A) node[above]{ \textbf{Device}};
		\draw (C) node[above]{ \textbf{\gls{bs}}};
		
		\coordinate (E) at ($(A)!.15!(B)$);
		\draw (E) node[left]{};
		
		\coordinate (F) at ($(C)!.3!(D)$);
		\coordinate (F1) at ($(C)!.35!(D)$);
		\draw (F) node[right]{};
		\draw[->] (E) -- (F) node[midway,sloped,above]{\verb$(1) Preamble Transmission$};
		
		\coordinate (G) at ($(A)!.45!(B)$);
		\coordinate (G1) at ($(A)!.5!(B)$);
		\draw (G) node[left]{};
		\draw[->] (F1) -- (G) node[midway,sloped,above]{\verb$(2) Random Access Response$};
		
		\coordinate (H) at ($(C)!.65!(D)$);
		\coordinate (H1) at ($(C)!.7!(D)$);
		\draw (H) node[right]{};
		%\draw (H) node[right]{\verb$LAST_ACK$};
		
		\coordinate (I) at ($(A)!.85!(B)$);
		\draw (I) node[left]{};
		\draw[->] (G1) -- (H) node[midway,sloped,above]{\verb$(3) Connection Request$};

		\draw[->] (H1) -- (I) node[midway,sloped,above]{\verb$(4) Contention Resolution$};
		\end{tikzpicture}
	\end{center}
	\caption{A grant-based random access protocol.} \label{fig:GBRA}
\end{figure}
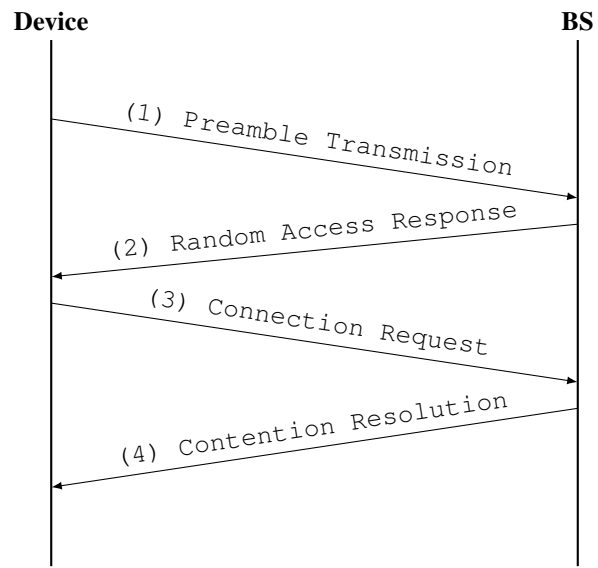

In a \textit{grant-based} random access protocol, to access the network  an active device needs to obtain permission from the \gls{bs} before sending its payload data. This procedure has four steps, as shown in Fig.~\ref{fig:GBRA}: 
\begin{itemize}
	\item [(1)] Each active device \textit{randomly} chooses a preamble out of a collection of orthogonal preambles and sends it to the \gls{bs} to  notify that the user has become active. 
	\item [(2)] The \gls{bs} sends a response as a grant for transmitting in the next step.
	\item [(3)] Each device receiving a response  sends a connection request  to ask for resource
	allocation for data transmission in the next step.
	\item [(4)] If the preamble was chosen by `only' one device, a connection resolution message is sent to that device  informing which resource is reserved for it. Otherwise, the connection requests collide, and the connection resolution message is not sent to any of those colliding devices. In this case, those devices need to restart the procedure.   
\end{itemize} 

This procedure may take 10 ms or more \cite{vaezi2018book}, which is high above the \gls{5g} and beyond networks' latency requirement. Further, as the number of active devices goes up, the probability of collision increases and this time becomes higher. Today's \gls{5g} \gls{nbiot} is a grant-based random access  \cite{liu2018sparse,chen2020massive}

\subsubsection{Grant-free  random access}  
A grant-free transmission, on the other hand, 
 eliminates the scheduling request for uplink data transmission. Unlike the grant-based case, the procedure starts with the \gls{bs} detecting the active devices and sending them a \textit{unique} preamble for all time slots. Next, the \gls{bs} receives metadata and uses it to estimate the channel \cite{liu2018sparse}.  This can reduce
 the access latency compared to the grant-based
access protocols. Device activity detection is the main challenge here as orthogonal preambles may not be available to all active devices  due to their massive number and the limited channel coherence time.

In \cite{liu2018sparse}, the device-activity detection is cast into a compressed sensing problem, and an approximate message passing algorithm \cite{donoho2009message}
is proposed to detect the active devices. 
This algorithm is used in several papers \cite{donoho2009message,liu2018massive,ke2020compressive} that apply the massive \gls{mimo} technology to boost the device-activity detection for massive IoT connectivity. Recent works \cite{li2019joint,cui2020jointly} have used DL-aided methods to design preamble sequences
and recover sparse signals effectively.
In \cite{ding2019comparison}, Zadoff-Chu sequences are proposed for preamble design for their good auto- and cross-correlation properties.
Several other approaches, such as Bayesian detection, 
joint device detection and channel estimation, 
 joint device and data detection,
 are proposed in the literature   \cite{chen2020massive}. 
With the exploding number of \gls{mmtc} IoT, massive device detection in grant-free random access is an important problem in  \gls{b5g}  wireless
networks. Several of these challenges are discussed in \secref{sec:future}.

 Also, grant-free \gls{noma} is extensively covered in \cite{shahab2020grant}.

\subsubsection{Unsourced random access}
Unsourced multiple access channel is referred to the 
multiple access problem introduced by Polyanskiy in
\cite{polyanskiy2017perspective}.
This work sets an information-theoretic basis for a typical \gls{mmtc} scenario
in which each of a massive number of uncoordinated users infrequently transmits a small payload (a b-bit message) to the \gls{bs}. 
 The \gls{bs} is interested only in
recovering the list of messages without being interested in the identity
of the user who transmitted a particular message.   
Due to the small size of the message, the block length is also small.
In \cite{polyanskiy2017perspective}, the number of
users is comparable to the channel block length.  Due to the unsourced, uncoordinated nature of the problem and
the small block lengths, the problem is very different from 
the traditional multiple access channel and, consequently, has fundamental limits and  coding schemes. 
The transmission is grant-free, but it is different from grant-based and grant-free random access protocols discussed earlier in that in unsourced massive random access each device does not get a unique preamble sequence, but all devices use  the same channel codebook (sequences).

Since the introduction of this new information-theoretic setting several 
advances and extensions have been made.  The first low-complexity coding  for the unsourced multiple access channel was described in \cite{ordentlich2017low}.
The extension to massive \gls{mimo} cases was introduced in \cite{fengler2019massive}. The authors in \cite{calderbank2020chirrup} proposed chirp detection with higher computational efficiency and lower energy per bit. 
Despite these advances,   many challenges remain to be addressed to bring unsourced massive random access into practice. More efficient codebook design and activity detection algorithms are two of them.

%-------------------------------------------------------
%-------------------------------------------------------

\section{IoT Security} 
\label{sec:security}

Billions of devices are being connected to the IoT through many different communication protocols. Regardless of the underlying connectivity standard, be it the licensed cellular spectrum or the unlicensed spectrum, we can view these devices as networked systems that are vulnerable a) through physical means or b) remotely, through the network and through software exploits.

Most types of communication standards (\gls{nbiot}/\gls{lte-m}/\gls{4g}/\gls{5g}, \gls{lora}, Sigfox) were not designed with end-to-end security in mind. Even though each of them provides some authentication and/or encryption solutions, they are not always applied correctly \cite{electronics9081195}. According to a recent IoT threat report, 98\% of IoT traffic is unencrypted when 100\% of the traffic should be encrypted \cite{Paloaltonetworks}. The most obvious attack vector is a man-in-the-middle attack where attackers can get in between devices or in between the device and the larger enterprise network. Even though IoT devices might not be the actual target of an attacker, they provide the first point of entry in order to launch more serious attacks on other systems on the network. 

Many IoT devices are intrinsically less secure than traditional personal computers or network servers. Limited by their size, memory and power constraints, these devices either do not have an operating system at all, or they can only run stripped-down versions of operating systems that do not have sophisticated security features. Intrusion detection algorithms and malware detectors are resource-intensive security tools that are normally not supported by IoT devices. 57\% of IoT devices are vulnerable to medium-or high-severity attacks, making IoT the low-hanging fruit for attackers \cite{Paloaltonetworks}.

\begin{figure*}[htbp]
	\centering
	\includegraphics[width=0.67\textwidth]{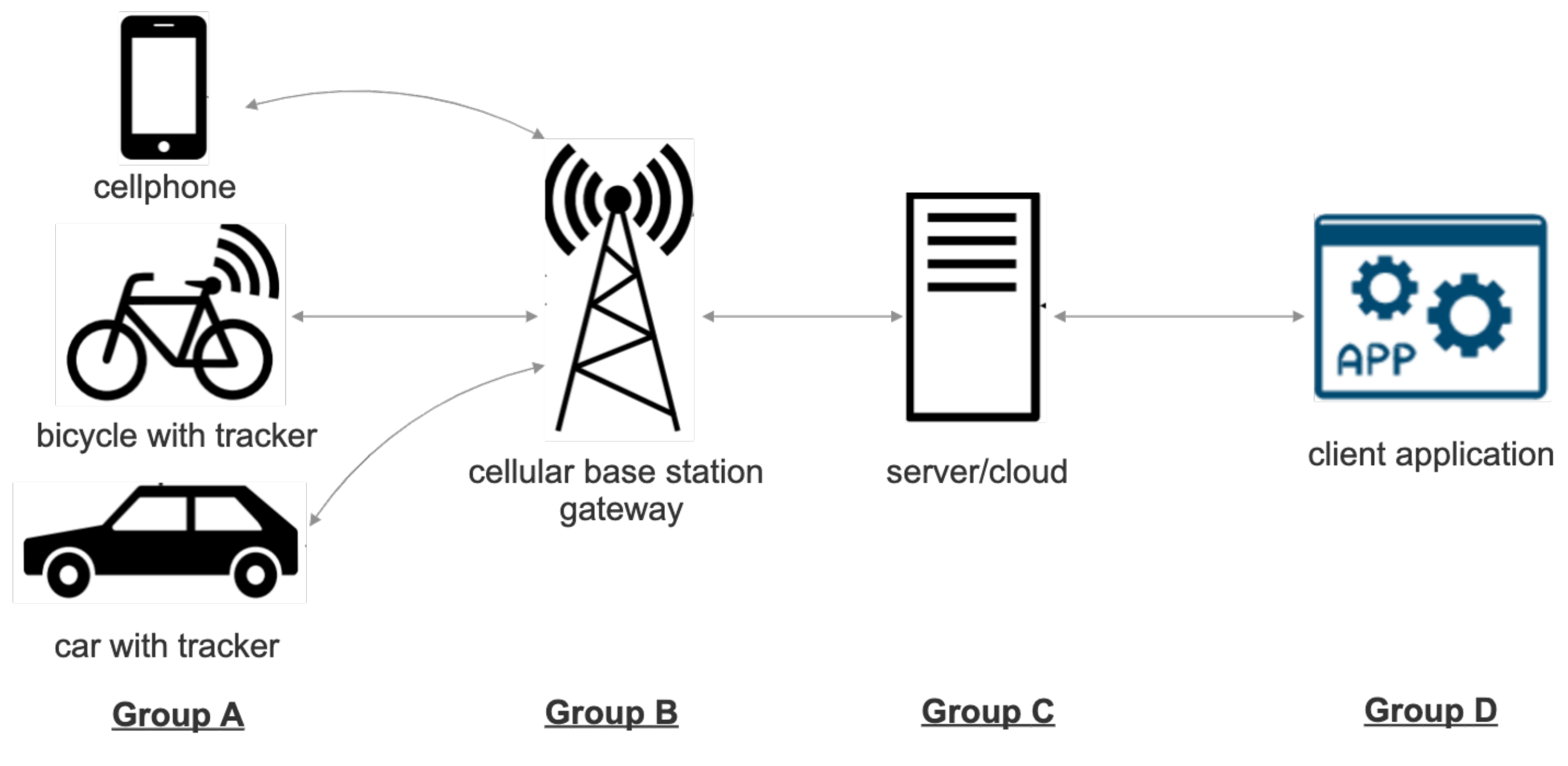}
	\caption{Overview of devices and communication links in a typical IoT deployment.}
	\label{fig:overview}
\end{figure*}

As shown in Fig.~\ref{fig:overview}, data from IoT devices (A) are typically transmitted wirelessly to a gateway/cellular base station/wireless access point (B). Gateways  use a backhaul network to connect  to a server/cloud (C) which processes the data transmitted by the IoT devices. After processing, information is transmitted using a \gls{tls} protected connection (HTTPS protocol) to the client's application (D). In a 100\% secure IoT environment data have to be protected both at rest and during transmission through all intermediate hops. Even if the communication link from the IoT device (A) to the gateway (B) is protected, after that, modern IoT data still has a long way to go to its destination.

\subsection{IoT Security Attacks Based on Their Characteristics}

Numerous cyberattacks in the form of eavesdropping, impersonation, key extraction, jamming have impeded the successful deployment of cellular and wide-are IoT. Password-related attacks continue to be prevalent on IoT devices due to weak manufacturer-set passwords and poor user practices. Recent reports on the IoT attack landscape indicate that attackers typically use botnets to conduct \gls{ddos} attacks or utilize malware that spread across the network \cite{7971869}. Major topics such as key management and authentication continue to play a central role in current research \cite{8168328}. 

\begin{figure*}[htbp]
	\centering
	\includegraphics[width=0.67\textwidth]{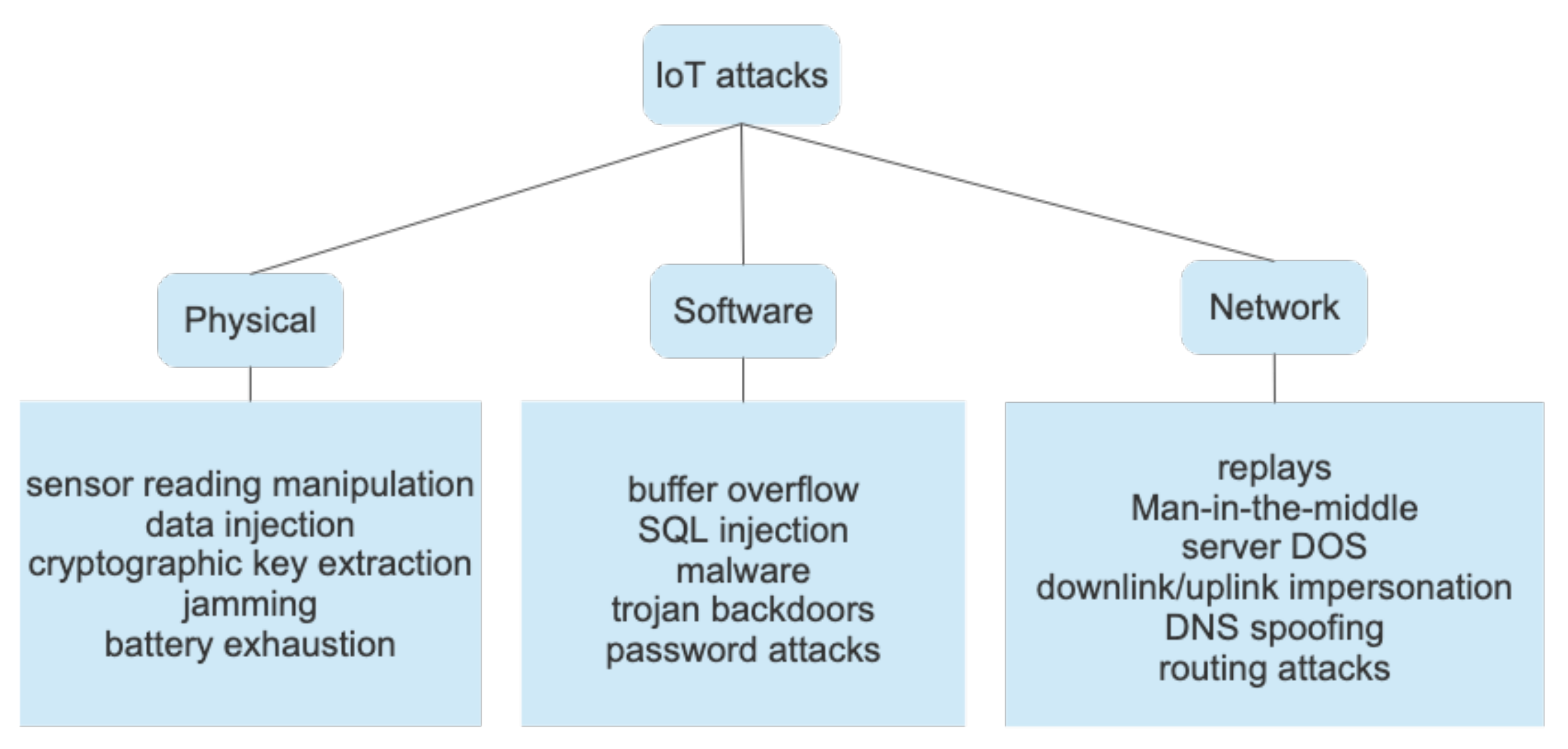}
	\caption{IoT attack classification}
	\label{fig:attacks}
\end{figure*}

In this section, we categorize the most prevalent attacks against cellular/wide-area IoT based on the characteristics of the medium used for each attack. Malicious entities a) either use physical means to compromise devices or intermediate gateways and serves, b) they attack the upper layer applications on the IoT device, or c) attack the network itself as shown in Fig. \ref{fig:attacks}. The last type of attack is a major concern moving forward as the network of devices scales up; attacks can now propagate from one device to another and eventually have a more devastating effect on the IoT infrastructure.

\subsubsection{Physical}

Physical security is a necessary condition for the security of an IoT device. Any sophisticated cryptographic algorithm will no affect if the device can be simply disconnected or tampered with. Physical types of attacks either target the environment to produce malicious sensor readings \cite{ms_lora} and cause false alarms, they target the hardware/software of the IoT device assuming that the attacker has physical access to the device, or they target the gateway that connects the IoT device to the rest of the network \cite{7985777,smart_campus,8766430,Chacko_2018}. Such attacks include stealing cryptographic keys from the non-volatile memory of the device \cite{8730781} and injecting false sensor signals that confuse the control logic of the system. The attacker could also target the gateway physically, in order to block the communication signals between the IoT device and the backhaul network. Jamming attacks are one of the most serious problems for IoT security since attackers can transmit radio signals near the devices with enough power that interrupts the radio communications. A low-cost micro-controller board like an Arduino board with a radio module can be used to perform jamming attacks. The main objective of the attacker is to transmit a wide-band signal with a higher \gls{snr} than the user \cite{DBLP:journals/corr/abs-1712-02141}. Battery exhaustion attacks and disconnection attacks in \gls{lorawan} are other types of attacks with physical manifestation \cite{8366983,9191495}.

\subsubsection{Software}

Software attacks on any device, not only on cellular/wide-area IoT, start from the operating system. If the attacker manages to bypass the operating system they can access parts of the filesystem that are private or they can execute untrusted code. Buffer overflows and \gls{sql} injections are some common exploits that lead to malicious code execution and breaches in privacy \cite{buffer,cite-key}. It is worth noting that some IoT devices can not even support an operating system because of their size and memory constraints.

Under the same category, we have malware that can lead to accessing, copying, or altering of information, and backdoor trojans that can be inserted to constantly leak data \cite{AbdulGhani2018ACI}. Password based attacks account for approximately 15\% of IoT threats, poor user practices that disable strong passwords and the use of weak manufacturer-set passwords contribute greatly to the insecurity of IoT devices \cite{Paloaltonetworks}.

As we move to more decentralized network deployments (e.g. \gls{5g}) software is not only at the core of IoT devices, but at the core of the network infrastructure as well. At the same time, the security checks that earlier technologies were employed at a centralized level, are no longer possible. To add flexibility, networks have become programmable and virtualized in software. This increases the potential threats at the network level itself, since software testing is an undecidable problem \cite{Davis1965-DAVTUB} and there will always be loopholes that a determined attacker can exploit. Even though resources are considered isolated in sliced networks, there are concerns that isolated network partitions can still be accessed at the hardware level \cite{7926920}.

\subsubsection{Network}

Technologies like \gls{lorawan}, Sigfox and \gls{nbiot} have several authentication and data encryption schemes to protect data between the IoT device and the gateway \cite{electronics9081195}. However, recent publications reveal that wide-area IoT deployments are still susceptible to impersonation and availability attacks. One example is a replay attack, where a malicious entity captures and stores a duplicate of a genuine network packet that is used to authenticate the legitimate user, and then at a later point in time, the attacker ``replays" this network packet to the system in order to get services that are only available to authenticated users \cite{Chacko_2018}. To perform this attack in wireless networks, the malicious entity should know the communication frequencies and channels to sniff data from transmission between devices \cite{7985777,smart_campus}. Routing attacks against low power wide-area networks are a well-studied topic as well \cite{doi:10.1155/2013/794326}.

An attacker can also target the link between the gateway and the server \cite{8366983,8766430}. Using a man-in-the-middle attack the malicious entity can take over the communication channel, and either delay/disrupt the message exchange, or modify the transmitted messages.  Another target could be the wide-area network server itself. Multiple service requests can lead to a denial of service attack that overwhelms the server's resources and causes the server to malfunction.

\gls{4g}   and \gls{5g} networks have made significant improvements to their security compared to former cellular technologies, like mutual \gls{aka} of the device and the cell tower. Control plane communication is kept confidential by applying encryption, but they are also protected against malicious modifications by using integrity protection schemes. Actual content information (data plane) is however only encrypted, not integrity protected. \gls{5g} networks enable integrity protection for content data – but this is only optional and not mandatory. At the same time, we have to consider the fact that non-standalone \gls{5g} deployments inherit all the security vulnerabilities of their predecessors since they are using older cellular technologies as their network core (\gls{lte}, \gls{4g}).

An example of an attack that can affect \gls{5g} IoT was first published in \cite{8835335}. The attacker takes advantage of the fact that there is no integrity protection on this channel, changes crucial parts of the encrypted transmitted information so that when those parts are decrypted the original message is no longer the same. Without knowing any encryption keys the attacker can perform an attack called \gls{dns} spoofing; i.e. changing the IP address of the \gls{dns} query issued by the target user equipment to the IP of a malicious \gls{dns} server. Of course, the malicious server is operated by the attacker, so the \gls{dns} request will be routed to the attacker. Such an attack could be very effective, overcoming the basic security capabilities of LTE and \gls{5g}, using the fact that no integrity protection was included.

Cellular denial-of-service (DoS) attacks have been one of the most popular attacks on cellular networks; attackers impersonate cell towers utilizing ``authentication and key agreement fake towers" and convince devices to connect by providing the best signal. Once the device connects to the fake tower it is no longer available since it does not receive service.A  recently published authentication attack was also reported in \cite{rupprecht-20-imp4gt} and is based on a reflection attack where the response from a device includes a copy of a specific part of the request. The objective of the attacker is to modify the IP addresses of the victim IoT device on the uplink or the network server on the downlink. Now the attacker can get access to any service, assuming the victim’s identity (uplink impersonation) or assuming the identity of any legitimate service provider (downlink impersonation).

\subsection{SDN and Blockchains in IoT}

One of the first studies that highlighted the importance of \gls{sdn} and blockchains in IoT is \cite{KOUICEM2018199}. With billions of devices being connected to the same network, we need flexibility in resource allocation, network management, and a variety of new services that enhance the user's experience.  At the same time, these emerging technologies can address the issue of key management, confidentiality, integrity, and availability in IoT systems \cite{7324414,7096257,8664092}. Blockchain and \gls{sdn} can bring additional benefits to IoT security in terms of flexibility and scalability \cite{8960432,iotsdn,8993716,8784168}.

\subsubsection{SDN}
\Gls{sdn} has emerged as a novel networking paradigm that revolutionized network deployments. It allowed for efficient network management, as well as the ability to launch new services with minimal effort. Programmability in software, offers great flexibility and scalability in the network infrastructure. In the pre-\gls{sdn} world, the control and data plane were coupled in networking devices. With \gls{sdn}, devices like switches, routers or IoT devices themselves do not have routing tables and access control lists; they merely forward packets, while intelligence has moved to the \gls{sdn} controller. The controller can now run security services in a centralized manner and take all the routing decisions using protocols like Openflow. \cite{ethane,openflow} Devices in the \gls{sdn} architecture handle packets based on flow tables setup at the \gls{sdn} controller. 

An \gls{sdn} deployment can help resource-constrained IoT devices by optimizing the resource allocation problem. At the same time, many security issues can be overcome by moving security intelligence to IoT gateways. In \cite{7575858}, authors presented an \gls{sdn} architecture for IoT devices based on Openflow, where IoT gateways, by analyzing network traffic, can identify abnormal system behavior and pinpoint which devices are misbehaving or which devices could potentially be compromised. Then, the gateway is entrusted with the task of taking appropriate action like a firewall would do, block the packet, forward the packet, and even classify it in terms of quality of service requirements.  

Alternatively, the work in \cite{7841889} moves security intelligence to some of the less resource-constrained IoT devices that are connected to the same network. The authors make the assumption that within the IoT infrastructure devices come with different capabilities, some have low power constraints but some have a robust operating system. The devices that are less resource-constrained in terms of computation and power consumption can assist by authenticating other IoT nodes and enforcing security rules using the Openflow protocol. In this way, secure, end-to-end connections can be established between IoT nodes, and man-in-the-middle attacks can be avoided.

%\subsubsection{Blockchains}
\subsubsection{{IoT security, privacy and trust}}
Blockchain is a novel type of database that stores data in a unique way. It stores data as blocks that are chained together, such that any modification in intermediate blocks will result in an invalid chain. This concept can be applied in many modern application domains including cellular, wide-area, and non-terrestrial IoT deployments. The advantage of this new technology is that any transaction in the cyber-worlds can be validated in a distributed way using a consensus algorithm run by a set of users in the network called miners. Because of this property Blockchains can improve the security of IoT devices. If IoT devices store their data in a Blockchain misbehaving IoT devices will be singled out by the whole group. 

As we can see in recent literature Blockchain-based IoT applications can take advantage of integrity and privacy by design, which is very important for the successful deployment of cellular and wide-area network \glspl{iot} at scale \cite{industrial_iot,KHAN2018395,8960432,9024627}. Every piece of data that is part of a transaction is hashed as a block to protect its integrity. At the same time, public key cryptography is used, where each node in the \gls{iot} network gets its own set of public/private keys. They can use their private key to authenticate a transaction, since the only node who could have possibly made the transaction is the owner of the private key. They can also use another node's public key to encrypt a short transaction message and make sure that the only node that can decrypt it is the owner of the corresponding private key. As \gls{iot} continues to grow, security by design is the only way forward. \gls{iot} devices and particularly sensor-based \gls{iot} devices communicate location/temperature/vital signs; many data points per second are collected for data analytics and monitoring purposes. It is important to have an underlying infrastructure, like a Blockchain, that creates permanent records of those data sets. This, not only ensures the reliability of data but the overall security of the exchanged messages in terms all confidentiality, integrity, and authentication. Of course, public key cryptography is as good as its key exchange. Traditional ways of storing keys and exchanging certificates do not scale with the volume of \gls{iot}, so this is an area that future research needs to address.

%--------------------------------------------------------
%-------------------------------------------------------
%-------------------------------------------------------

\section{Deep Learning for IoT} \label{sec:AI}

\Gls{ml} algorithms try to learn the mapping from an input to output
from data rather than through explicit programming.  \gls{ml} uses  algorithms that iteratively
learn from data to describe data or predict outcomes, i.e., to produce precise models based on that data.
As the training data becomes larger, it is then possible to produce
more precise models. 
There are many learning algorithms and applications.
Careful hand-crafted  \textit{features} are a critical component of the model building process in traditional \gls{ml} algorithms. That is, a domain expert should 
extract and select the key features of the input before feeding them to the model.  \textit{Feature engineering} not only needs considerable expert knowledge but is also a very time-consuming process.

Three types of \gls{ml} tasks can be considered: supervised learning, unsupervised learning, and reinforcement learning. 

\begin{itemize}
	\item \textit{Supervised learning} is the task of inferring a classification or
	regression from labeled training data.\footnote{Classification algorithms are used to predict/classify the \textit{discrete} values such as an email is spam or not whereas regression algorithms are used to predict the \textit{continuous} values such as price, salary, etc.  }
		\item \textit{Unsupervised learning} is the task of drawing inferences from unlabeled training data, i.e., 
	 input data without labeled responses.
		\item  \textit{Reinforcement learning}  is the task of learning how to interact with an environment by taking a sequence of actions to
	maximize cumulative rewards.
\end{itemize}

\subsection{Neural Networks}

\begin{figure}[!t]
	\centering
	\includegraphics[width=2in]{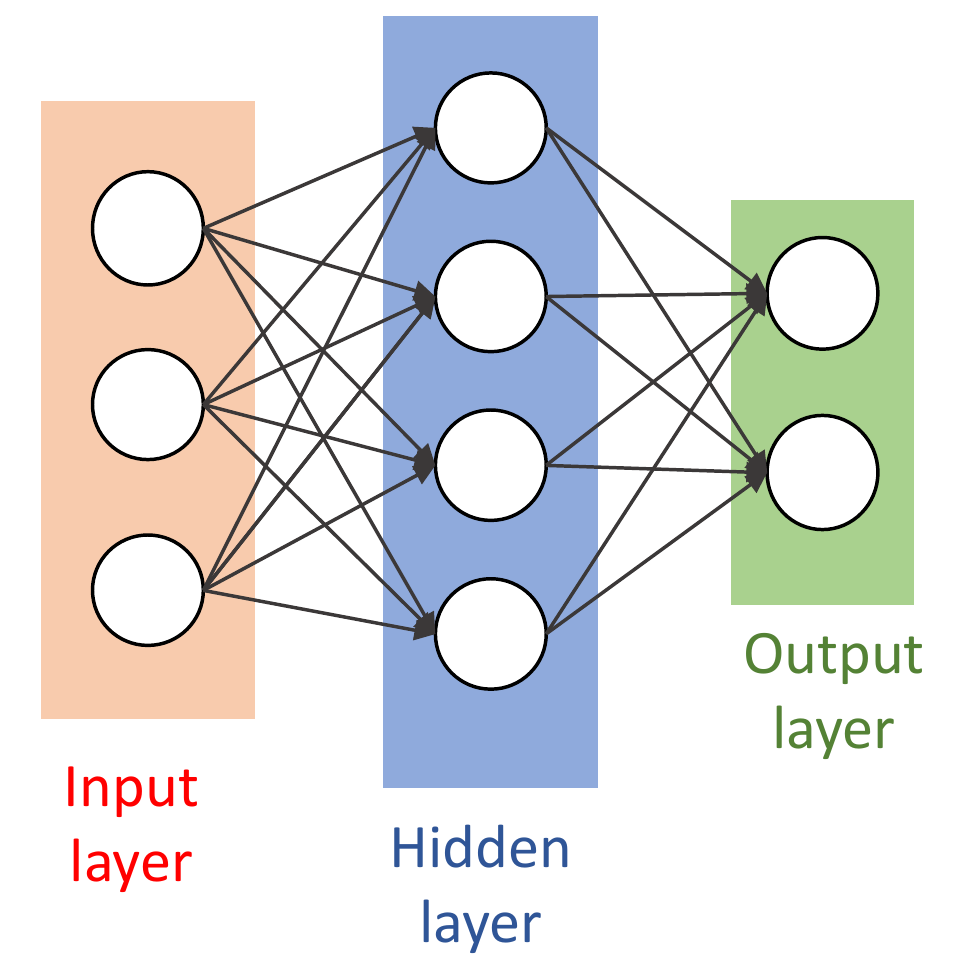}
	\caption{An artificial neural network. } \label{fig:ANN}
\end{figure}

\Gls{ann}, or simply neural networks, are learning algorithms (computing systems) that mimic the operations of a human brain to find the input-output mapping from large amounts of data. 
As shown in Fig.~\ref{fig:ANN}, the \gls{ann} consists of three layers: input layer, hidden layer, and output layer. Each layer consists of multiple \textit{neurons} and tries to learn certain weights. The input layer accepts the inputs, the hidden layer processes the inputs, and the output layer produces the result.

 A single neuron can be imagined as a  \textit{regression classifier}, which is a supervised \gls{ml} algorithm. 
The neuron receives the inputs, multiplies them by their weights, adds them up, and then it applies the \textit{activation function} to the sum.
The activation function is used to normalize the output.
It returns the output in the range $[0,\,1]$ which can be thought of as a probability. There are various activation functions. \textit{Linear} activation function,  \textit{sigmoid} function, \textit{softmax},  and  \gls{relu} \cite{xu2015empirical} are examples of activation functions.

In supervised learning, a neuron learns the weights based on the inputs and the desired outputs through an optimization process in various phases. 
 First  \textit{forward propagation} occurs where the input training data crosses the entire neural network and the output is predicted.
 Next,  a \textit{loss function} is used  to estimate the loss (or error) between the correct output and the predicted one. It measures how good/bad our prediction result is. The next phase is  \textit{backward propagation} in which the calculated loss is propagated backward to all neurons in the hidden layer, such that all the neurons in the network receive a loss signal that describes their relative contribution to the total loss. 
This information is then used to update the weights of the  network through the \textit{gradient descent} algorithm such that the next loss becomes smaller, and thus a better model is obtained. This process is iterated until we get a `good' model, i.e., the loss approaches  to zeros.

 Due to their capacity to  learn weights that map any input to the output, \glspl{ann} can learn any nonlinear function.  This is the main advantage of \glspl{ann} which also makes them be  known as universal function approximators. 
 It should be highlighted that \gls{ann} ideas have been around for a long time. The main limitation of \gls{ann} is that it may over-fit easily, is slow to train, and is not great for images, time series, and large numbers of
 inputs-outputs. Today,
 \gls{ann} is a broad term that encompasses any form of DL model.

A basic three-layer \gls{ann} is comprised of 
 \begin{itemize}
 	\item \textit{Input layer:}
 	data is a vector, and each input neuron collects one feature/dimension of the input data
 	and passes it on to the hidden layer.
 \item 	\textit{Hidden layer:}
 	 each hidden neuron computes a weighted sum of all the neurons from the previous layer (input layer) and passes it through a \textit{nonlinear activation function}.
 \item	\textit{Output layer:}
 	Each output neuron computes a weighted sum of all the hidden neurons and passes it through a (possibly nonlinear) \textit{threshold function}.
 	
 \end{itemize}

\begin{figure}[!t]
	\centering
	\includegraphics[width=2.5in]{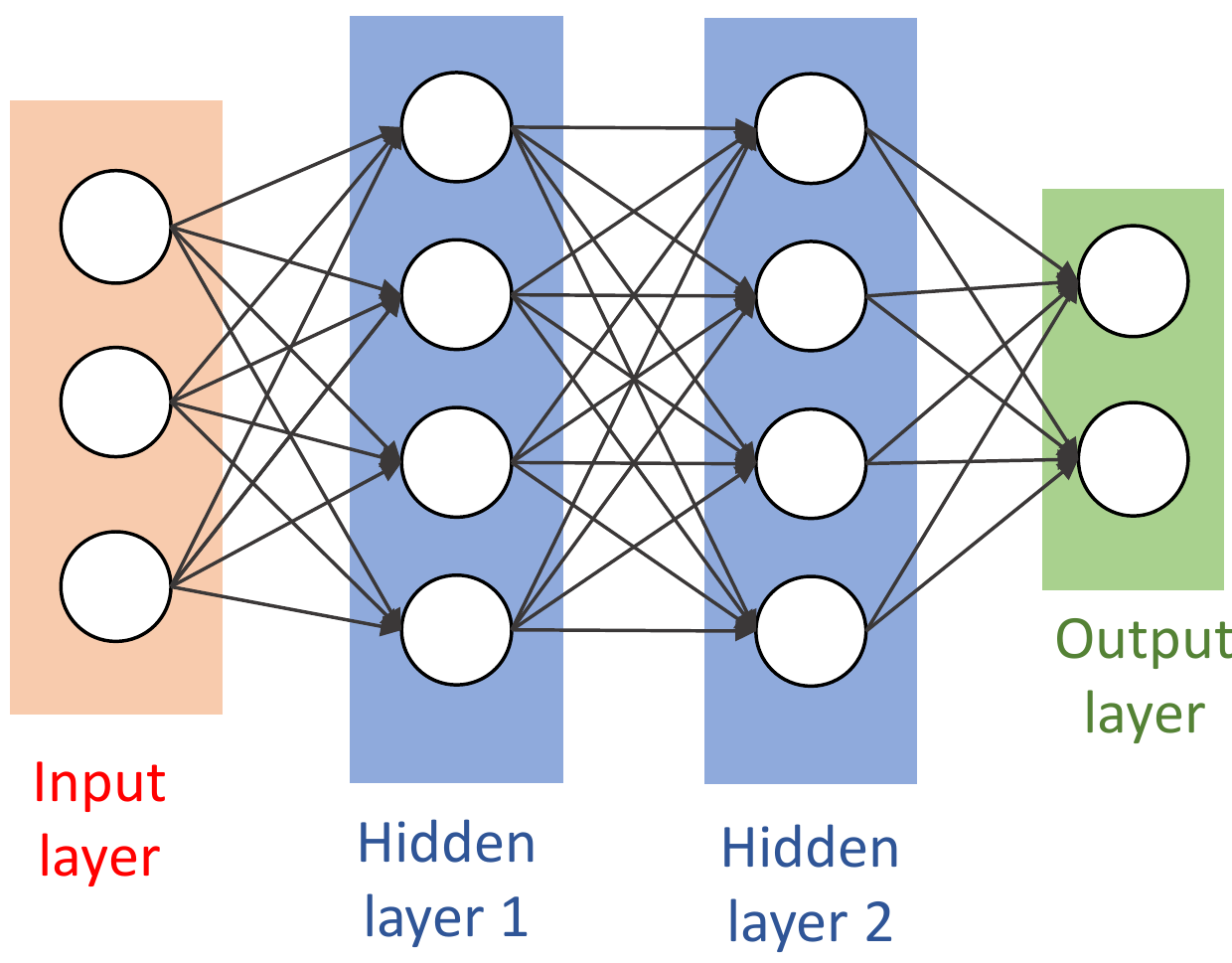}
	\caption{A deep neural network.} \label{fig1d}
\end{figure}

\subsection{Deep Learning Models}

  Deep learning (DL) algorithms are a subset of \gls{ml} algorithms.
 \glspl{dnn} are \glspl{ann} with some level of complexity, usually, networks with two or more  hidden layers. %(layers between the input and output). 
 \glspl{dnn} perform well in tasks like \textit{regression} and \textit{classification}.\footnote{The former deals with predicting a continuous
 value whereas the latter predicts the output from a set
 of finite classes.} They can learn hidden features from raw data \cite{lecun2015deep}.
Each hidden layer trains on a set of features received from the previous layer and can recognize more complex features. So, a deeper network can result in more complex features and more accurate learning.  
 
 \glspl{dnn} are changing the way we interact with the world.  
 Before introducing the most common \gls{dnn} models, we discuss what makes DL different from \gls{ml} algorithms.  
There are two main reasons why researchers and experts tend to prefer DL over \gls{ml}: 
\begin{enumerate}
	\item DL algorithms are capable of learning much complex  \textit{decision boundaries}. Hence, in general, DL models improve accuracy/performance.
	\item DL automates the process of \textit{feature engineering} or at least reduces the need for that.
	Recall that feature engineering is a key step in the model building process. It  also needs expert knowledge and typically is very time-consuming.
\end{enumerate} 

Therefore, compared to the conventional \gls{ml} techniques, DL models improve the accuracy and reduce 
the need for hand-crafted features \cite{lecun2015deep}. These make DL attractive to many people in different fields, not just machine leaning experts.  

In addition to the above advantages which apply to almost any applications,  embedding  DL models to \gls{5g} and beyond mobile networks  has several other justifications and advantages such as \cite{zhang2019deepTut}
\begin{itemize}
	\item  There exist various DL models to learn useful patterns from unlabeled data, which is the most common type of data in today's mobile networks \cite{alsheikh2016mobile},  in an unsupervised manner. \Gls{rbm} and \gls{gan} are two examples of such DL methods.  
	\item  A single DL model can be trained to satisfy multiple objectives, without a need for complete
	model retraining for each objective \cite{georgiev2017low,zhang2020multi}. This reduces computational and memory requirements of
	the systems which are particularly important in \gls{iot} networks. 
	\item Mobile networks generate large volumes of heterogeneous data. \glspl{dnn} not only can handle such data but also benefit from it since training with big data prevents model 	over-fitting. In contrast, 
	traditional \gls{ml} algorithms such as \gls{svm} may require storing all the data in memory which makes scalability difficult \cite{zhang2019deepTut}. 	 
\end{itemize}

\begin{table*} [h]
	\centering
	%	\begin{minipage}[b]{0.99 \textwidth}
	\centering
\scalebox{0.8}{ \begin{threeparttable}[b]		
		\caption{Different types of deep neural networks and their  characteristics} 
		\label{tab:DL}
		\begin{tabular} { p{0.06 \textwidth} p{0.33 \textwidth} p{0.36 \textwidth}  p{0.3 \textwidth}} 
			\toprule
			\rowcolor{Lightgray} \textbf{Model} & \textbf{Description}    &  \textbf{Characteristics}  &  \textbf{Applications}   \\        \cmidrule{1-4}  \addlinespace
			%			
			%			\rowcolor{Lightgray} \gls{ann}    &  Artificial neural network, is a group of multiple neurons at each layer. \gls{ann} is also known as a feed-forward neural network. \gls{ann} ideas have been around a long time. today, \gls{ann} is a broad name that 
			%			includes many DL models including \gls{dnn} and \gls{cnn}.&  \begin{itemize}
			%				\item \gls{ann} consists of 3 layers: input, hidden and output layers.
			%				\item In an \gls{ann}, inputs are processed only in the forward direction. 
			%				\item Limitations: overfits easily, slow to train, not great for images
			%			\end{itemize}  &  \\ \addlinespace
			%			
			\rowcolor{Lightgray} \gls{dnn}    & Deep neural network is an \gls{ann} with some level of complexity.  Usually, a network with more than two hidden layers. &
			\begin{itemize}
			%	\item  A \gls{dnn} has at least two hidden layers
				\item  Dense connectivity in each hidden layer 
				\item Matrix multiplication in each layer
			\end{itemize}  & 
		   \\ \addlinespace
			
			\rowcolor{Lightgray} CNN &  A \gls{cnn}  applies multiple 2D filters to the input    to extract  features (convolution operations). A \gls{cnn}  is a \gls{dnn} that uses convolution instead of general matrix multiplication at least in one of its layers. Convolution has sparse connectivity.
			
			&  \begin{itemize}
				\item Convolution layers for feature extraction
				\item Less connection compared
				to \glspl{dnn}
				\item Simple (sparse connectivity) and efficient for image 
				%\item Processes input on fixed size vectors of data
				\item Takes a long time to train
			\end{itemize}   & 	\begin{itemize}
			\item Image recognition 			
			\item Video analysis 
			\item Document analysis
		\end{itemize}  \\ \addlinespace
			\rowcolor{Lightgray} RNN  & \Glspl{rnn} are the time series version of \glspl{ann} and are meant to process sequences of data.   LSTM (long short term memory) is the most common type of recurrent layer in which past information can flow through the model.  &  
			\begin{itemize}
				\item  Links outputs
				back to inputs over time
				\item  Good for time series learning
				\item Models and uses ``memory"
				\item Useful for extracting information from  time-dependent and time-correlated  data 
				\item Understands the temporal nature of data
			\end{itemize}  &
			\begin{itemize}
			\item Robot control
			\item Speech recognition/synthesis			
			\item Machine translation 
			\item Mobility detection
			\item Predict next value in a sequence 
		\end{itemize}
			\\		\addlinespace
			%			\rowcolor{Lightgray}RBM  &			restricted Boltzmann machine (RBM) &   \\
			%				\addlinespace
			%	\rowcolor{Lightgray}\gls{drl}  &  Deep reinforcement learning is an RL algorithm empowered with a \gls{dnn}. RL tries to follow  the human learning process by infraction with the environment.
			%Beside supervised and unsupervised learning algorithms, RL is the third main type of \gls{ml}.   &  	\begin{itemize}
			%	\item Learning is interactive
			%	\item Input and output given only when action 
			%	provided
			%	\item No given model or labels ahead of time
			%	
			%\end{itemize}  & \\ %
			\addlinespace
			\rowcolor{Lightgray}\gls{gan} &
			A generative adversarial network
			is composed of two networks: \textit{generator} and \textit{discriminator} networks. The former generates new data instances, whereas the latter   evaluates them for authenticity. %; i.e., decides whether data  belongs to the actual training dataset or not.
			& 	\begin{itemize}
				\item Train two networks at once
				\item Similar to \gls{drl} but does not require a value/reward
				function
				\item Can be used together  with  \gls{cnn}, \gls{rnn}, etc.
				
			\end{itemize}  &  	\begin{itemize}
			\item Generating photographs
			\item Text-to-image translation 	
			\item Image editing/inpainting
			\item Improving cybersecurity
			\item Tumor detection
		\end{itemize}
	\\	\addlinespace
			\rowcolor{Lightgray} AE  &
			An \gls{ae} is an \gls{ann} that learns to copy its input to its output. It tries to reconstruct  the original input   by minimizing the reconstruction error (the differences between  the input and reconstructed version of that)
			& 	\begin{itemize}
				\item \gls{ae} is an unsupervised learning technique
				\item \glspl{ae} are data-specific 	
				\item Same input and output size 
				%i.e., they are only  able to compress data similar to what they have been trained
			\end{itemize}   & 
			\begin{itemize}
			\item Anomaly detection
			\item Information retrieval 	
			\item Data denoising			\item image Inpainting
			\item \Gls{e2e} communication \cite{o2017deep} 
		\end{itemize}
   \\	\addlinespace
			
			\bottomrule
		\end{tabular}
	\end{threeparttable}
}
%		\end{minipage}
\end{table*}

%\begin{center}
%	\begin{tabular}{|c| c |c | c|} 
%		\hline
%		Model & Learning & Typical Input Type & IoT Applications \\ [0.5ex] 
%		\hline\hline
%		\gls{cnn} & Supervised & 2D image & xxx \\ 
%		\hline
%		2 & 7 & 78 & 5415 \\
%		\hline
%		3 & 545 & 778 & 7507 \\
%		\hline
%		4 & 545 & 18744 & 7560 \\
%		\hline
%		5 & 88 & 788 & 6344 \\ [1ex] 
%		\hline
%	\end{tabular}
%\end{center}

Besides their various advantages for mobile and \gls{iot} networks, there are also several limitations for the application of DL in these domains. In particular, DL models are vulnerable to adversarial examples, they heavily rely on data but data collection may be costly or infeasible in some cases due to the privacy constraints, they can be computationally expensive for many \gls{iot} device, and they have many hyperparameters and finding their optimal configuration is not trivial, and  DL algorithms  
have low interpretability  \cite{zhang2019deepTut}.

 Table~\ref{tab:DL} describes several types of NN that are used to solve DL problems and summarizes their salient characteristics. In the following we  describe several of these DL models as well as their application related to \gls{iot} networks. Nonetheless, 
% \subsection{Questions}
 several important questions need to be answered if \gls{iot} devices and mobile apps can effectively implement \gls{dnn} technology \cite{zhang2020implementation,hadidi2020toward,deng2020model}:

 \begin{itemize}
 	\item What \gls{dnn} structures can effectively process and fuse sensory input data for diverse \gls{iot} applications? 
 	\item How can  DL models  be efficiently deployed on resource-limited \gls{iot} devices? In other words, how can we reduce the resource consumption of \glspl{dnn} to make them suitable for \gls{iot} devices? 
 	\item  Deploying \glspl{dnn} on resource-constrained 
 	\gls{iot} devices is a big research challenge. If such \glspl{dnn} are  implemented on the cloud where powerful resources are available what new challenges will arise?  
 	 	\item How can we minimize the need for labeled data  in \gls{iot} learning?
 	 	\item What distributed learning methods are relevant to the \gls{iot} realm?
 	 	\item What is the state-of-the-art and the outlook  for \gls{dnn} accelerators that can operate in diverse \gls{iot} applications?
 \end{itemize}
  As an example, \glspl{dnn} can be used successfully in mobile apps provided that they can collect data from various \gls{iot} devices and analyze them energy-efficiently and accurately using minimal data labels. 
  While research in this field is very vibrant, there are already promising solutions out.
  We will present some of those solutions in this section.  
% Studies show that the semi-supervised strategy, called SenseGAN, greatly reduces the need for labeled data.

% 
%
%\subsection{Convolutional neural network (\gls{cnn})} are prevalent in image and video processing. Their other applications include speech recognition and understanding natural language processing. 
%By applying relevant filters, a \gls{cnn} is able to capture the \textit{spatial} and \textit{temporal} dependencies in an image.
%
%
%The \gls{cnn} is composed of one or multiple \textit{convolution} layers (the Kernel)  and \textit{pooling} layers followed by fully connected layers and activation functions. The convolution and pooling layers are for feature extraction whereas fully connected layers are for classification. While in \gls{ml} methods filters are hand-engineered, 
%with enough training, CNNs are able to learn these filters.
%The inputs are convoluted with the filter to  
%extract the high-level features (such as edges from the input image) required in the CNN. This processing is much less than other classification algorithms.  
%
%The input of a CNN is assumed to be a 2-D signal, e.g., an image. The image matrix is flattened to a vector 
%
%\begin{center}
%	\begin{tabular}{|c| c |c | c|} 
%		\hline
%		Literature & Model & Application & Contribution \\ [0.5ex] 
%		\hline\hline
%		1 & 6 & 87837 & 787 \\ 
%		\hline
%		2 & 7 & 78 & 5415 \\
%		\hline
%		3 & 545 & 778 & 7507 \\
%		\hline
%		4 & 545 & 18744 & 7560 \\
%		\hline
%		5 & 88 & 788 & 6344 \\ [1ex] 
%		\hline
%	\end{tabular}
%\end{center}
%
%
%

\subsection{Autoencoder}
An \gls{ae} is a neural network that is trained to learn efficient representations of the input data, that is, it learns to copy its input to its output. \glspl{ae} are mainly used for solving unsupervised problems. These networks first compress (encodes) the input and then reconstruct (decodes) the output from this representation by minimizing the differences between  the input and reconstructed version of the input. 
The point behind data compression is to get a smaller  representation of the input and passed it around. This compressed  representation should keep data quality such that people can recreate it when needed.\footnote{The idea is similar to \textit{distributed source coding} techniques \cite{xiong2004distributed,pradhan2003distributed,vaezi2014distributed} where a compressed version of a signal is transmitted to a receiver that is expected to reconstruct the original data from that. Hence, such techniques may be useful in this context.} 
Since \glspl{ae} are expected to create a copy of the input, they are great for fault detention, intrusion detection, disease diagnosis, and in general for anomaly detection as described below.  are  
\begin{itemize}
	\item \textit{Anomaly detection:}
	Anomaly detection (or outlier detection) refers to the process of identifying rare items or events in data sets that  do not
	conform to expected behavior i.e., differ significantly from the norm \cite{chandola2009anomaly}.  Anomaly detection is based on two assumptions: (i)
	anomalies occur very rarely in the data, (ii) they have significantly different features compared to the normal instances.

	Anomaly detection is important because in many applications  anomalies in data correspond to critical and  actionable information. 
	For example, an anomalous traffic pattern in a computer or an \gls{iot} device could mean that the device is compromised or hacked. 
	Applications of anomaly detection are wide and increasing. Examples include but are not limited to  fraud detection for credit cards, intrusion detection 
	for cyber-security, fault detection in industry, and medical diagnosis in healthcare. 
	DL techniques, including \gls{ae}-based ones, have been proposed to address security  threats such as  anomaly and intrusion
	detection  in  \gls{iot} networks \cite{chalapathy2019deep}. Some recent \gls{ae}-related works can be found in Table~\ref{table:deepAE}. 
	\item \textit{Intrusion detection:}
	An \gls{ids} is a device software  that monitors network traffic for suspicious activities and reports  such a harmful activity or policy breaching  to an administrator. Suspicious activities may also be collected centrally using a security information and event management system  to differentiate malicious activity from false alarms \cite{zarpelao2017survey}.
	\gls{ids}
	prevents \gls{iot} botnet propagation as well as outbound attack traffic. 
	There are various \gls{ids} systems such as network \gls{ids}, host \gls{ids}, protocol-based \gls{ids}, and hybrid. Yet, there are two main detection methods: \textit{signature-base}d detection and \textit{anomaly-based} detection. The former 
	detects attacks based on specific patterns (such as the number of bytes in the network traffic) or the already known malicious instruction sequence that is used by the malware. The latter, however, detect unknown malware attacks. It  classifies system activity as either normal or anomalous and this classification is based on heuristics or rules, rather than patterns or signatures. 
	
	As new malwares are developed rapidly, anomaly-based detection has become more and more important. Further, the fact that \gls{ml} and DL can be used for the classification of  activities \cite{aceto2019mobile}, has given momentum to this type of detection. \gls{ml}/DL can be used to create an activity model to declared an activity is suspicious or not. Such models can be trained based on the applications and hardware and thus can perform better than the signature-based \gls{ids} in practice. Various deep \gls{ae}-base \glspl{ids} have been introduced in recent years. We list several of them in Table~\ref{table:deepAE}.
%\item End-to-end communications
	
\end{itemize}

\begin{table*} [!t]
\small 	\caption {Applications of deep \gls{ae} in \gls{iot}} 
	\begin{center}{ 
		\begin{tabular}{|m{0.3in}|m{.6in}|m{1.1in}|m{3.5 in}|@{}m{0cm}@{}}
				\hline 
			\Tstrut	\textbf{Ref.} & \textbf{Model} & \textbf{Application} & \textbf{Contribution} \\ [0.5ex] 
				\hline\hline
				\cite{chen2020unsupervised},	\cite{azzalini2021minimally},	\cite{nazir2020autoencoder} & \gls{ae} & Industrial \gls{iot} & Anomaly detection for industrial robots \\ 
				%		\hline
				%	\cite{} & \gls{cnn} + AE& Industrial IoT  & Anomaly detection for industrial robots \\
				\hline
				\cite{yin2020anomaly} & \gls{rnn} + \gls{ae}& \Tstrut Sensor networks  & Anomaly detection for \gls{iot} time series \\
				\hline
				\cite{savic2021deep}  & \gls{ae}  & \Tstrut Smart logistics & Anomaly detection as a service
				in  cellular \gls{iot}   for both  \gls{iot} devices and core network \\
				\hline
				\cite{muna2018identification} & \gls{drl} + \gls{ae} & Smart city & \Tstrut Indoor localization based on Bluetooth low energy signals \\
				\hline
				\cite{muna2018identification} & \gls{ae} & Healthcare & \Tstrut Heart diseases diagnosis    \\
				\hline
				\cite{meidan2018n} & \gls{ae} & Generic & \Tstrut Intrusion detection  using \glspl{ae}  to learn normal behaviors of the \gls{iot} devices  \\ 
				\hline
				\cite{lee2020impact} & \gls{ae} &  WiFi-connected  \gls{iot}  & \Tstrut  Attack detection
				using deep \gls{ae} and feature abstraction based on WiFi networks database \cite{kolias2015intrusion} \\
				\hline
				\cite{hwang2019detecting} & \gls{ae} + \gls{cnn} & \gls{iot} gateways  & \Tstrut Intrusion detection   by
				examining as few packets as possible  \\
				\hline
				\cite{al2018deep} & \gls{ae} + \gls{svm} & Generic & \Tstrut Intrusion detection   \\ 
				[1ex] 	
				\hline

			\end{tabular}\label{table:deepAE}}
	\end{center} 
\end{table*}

\subsection{Deep Reinforcement Learning}

\Gls{rl} is a type of \gls{ml} where the system learns how to behave in an environment by performing actions and seeing the results.
More precisely, an \textit{agent}  learns from an uncertain \textit{environment} by interacting with it and receiving \textit{rewards} for performing \textit{actions},  as we do in our natural experiences. In other words,  trial and error is employed to train \gls{ml} models that make a sequence of decisions to attain a complex objective (goal) or how to maximize along a particular dimension.\footnote{
	\gls{rl} is typically modeled as a Markov decision process (MDP).
} For instance, the goal could be maximizing the points won in a game over many moves. 
We want the \gls{rl} agent to earn lots of reward.
The agent must \textit{exploit} actions already knows to obtain a reward. It must also explore/select untested actions to discover new reward-producing actions. Thus, the action selection is a trade-off between exploration and exploitation.

\gls{drl} combines \gls{ann} with a reinforcement learning architecture that enables  agents to learn the best actions   to attain their goals. It has successfully trained
computer programs to play games like AlphaGo and beat the
world's best human players. \gls{drl} has found applications in many fields like robotics, computer vision, natural language processing, energy, and transportation   \cite{li2017deep}. 
Compared to \gls{rl}, the \gls{drl} is much better for practical uses in many applications. For example, to achieve acceptable training times for practical uses in robotics, \gls{rl} typically needs hand-crafted policy representations which compromises the
autonomy of the learning process.  \gls{drl}, in contrast, relieves this limitation by training general-purpose neural
network policies \cite{gu2017deep}.

\begin{table*} [!t]
\small
	\caption {Some representative works on \gls{drl} in IoT} 
	\begin{center}{ 
			\begin{tabular}{|m{0.3in}|m{.6in}|m{1.1in}|m{3.5 in}|@{}m{0cm}@{}}
				\hline 
				\Tstrut \textbf{Ref.} & \textbf{Model} & \textbf{Application} & \textbf{Contribution} \\ [0.5ex] 
				\hline\hline
			 \cite{chinchali2018cellular}	 & \gls{drl} &  Broadband \gls{iot} &  \Tstrut A scheduler that can dynamically
				adapt to traffic variation to schedule IoT traffic \\ 
				\hline
				\cite{li2019deep} & \gls{drl} &   Mobile crowdsensing  & \Tstrut A framework for mobile crowdsensing
				in fog environments with a hierarchical scheduling \\
				\hline
				\cite{liu2019performance}  & \gls{drl}   & \Tstrut \gls{iiot} & \Tstrut Blockchain-enabled \gls{iiot} systems to  improve the scalability of the blockchain without sacrificing latency, and security  \\
				\hline
				\cite{muna2018identification} & \gls{drl} + \gls{ae} & Smart city &  \Tstrut Indoor localization based on Bluetooth low energy signals \\
				\hline
	\cite{gu2017deep} & \gls{drl}  & Robotics &  \Tstrut Asynchronous \gls{drl} for robotic manipulation trained on real physical robots  \\
[1ex] 	\hline
	\cite{doan2019power} & \gls{drl}  & IoT massive access &  \Tstrut \gls{drl} implemented in a real \gls{bs} for user pairing and  resource allocation in cache-aided non-orthogonal multiple access for IoT    \\
[1ex] 	\hline
	\cite{mao2016resource} & \gls{drl}  & Resource management &  \Tstrut The resource and task management states are transferred into images as the input of the \gls{dnn}.    \\
[1ex] 	\hline

	\cite{li2020distributed} & \gls{drl}  & Edge
	computing &  \Tstrut \gls{drl} algorithm is proposed for the computation offloading in the heterogeneous  edge
	computing server.
   \\
[1ex] 	\hline
	\cite{bu2019smart} & \gls{drl}  & Smart agriculture &  \Tstrut \gls{drl} is applied in the cloud layer of a smart farming system to make smart  decisions, like irrigation,  for improving crop growth.    \\
[1ex] 	\hline
			\end{tabular}\label{table:DRL}}
	\end{center} 
\end{table*}

Applications of \gls{drl} in \gls{iot} are diverse and numerous.
We list several representative works in Table~\ref{table:DRL}, and elaborate on some of them in the following. 
In general, \gls{drl} can be used in (i) perception layer (physical autonomous system), (ii) physical and network layer of \gls{iot} communication networks, as well as (iii) application layer (\gls{iot} edge/cloud computing) \cite{lei2020deep}. Robotic manipulation, autonomous driving, and energy storage management in smart grids are examples of the first one. 
Gu \textit{et. al} \cite{gu2017deep} have used \gls{drl} to learn complex robotic
manipulation skills on real physical robots.

In \cite{chinchali2018cellular}, 
a \gls{drl}-based  scheduler is proposed to  dynamically
adapt to traffic variation, and to reward functions set by network operators, to schedule \gls{iot} traffic to avoid congestion.  This work is motivated by \gls{iot} applications that need a high volume of data but have flexible
time and thus their traffic is not needed to be scheduled immediately when it originates. In \cite{azari2020machine}, a deep Q-learning solution for handover and radio resource management in cellular networks hosting aerial devices, drones, along with terrestrial users has been proposed. In \cite{liu2019performance} a framework of performance optimization for blockchain-enabled \gls{iiot} systems is proposed that aims to  improve the scalability of the underlying blockchain without affecting the system's decentralization, latency, and security. Doan \textit{et al.} \cite{doan2019power} applied \gls{drl} for resource allocation in cache-aided \gls{noma} for \gls{iot} and showed that \gls{drl} increases the probability of successfully
decoding at the users. 
In	\cite{muna2018identification} \gls{drl} is used for indoor localization based on BLE signals.	
	These are just a few examples of \gls{drl} applied to improve \gls{iot} communications in the physical or network layers.

Finally, \gls{drl} has also been used in the application layer
 for data
processing and storage.  An application layer agent may locate in the \gls{iot}  edge/fog/cloud 
 computing systems. Task offloading, resource allocation, and caching are examples of tasks in this category. 
  Applying \gls{drl} for resource management was first proposed in \cite{mao2016resource}, where 
 the authors transferred the resource and task management states into images as the input of the
 \gls{dnn}. 
   The work in \cite{doan2019power} applied \gls{drl} for resource allocation in cache-aided \gls{noma} is an example. 
Li \textit{et al.} \cite{li2019deep} have proposed \gls{drl} for mobile crowdsensing in fog environments.
\gls{drl} provides a scalable, elastic, and self-adapting method to adapt to varying fog computing structures and mobile
environments. 
In \cite {li2020distributed} a \gls{drl} algorithm is proposed for the computation offloading in the heterogeneous  edge
 computing server. The \gls{iot} device
 performs computation offloading for each task request
 according to the decision given by a \gls{drl} model. 
Lastly, in \cite{bu2019smart} a
 smart agriculture \gls{iot} system based on \gls{drl} is proposed. \gls{drl} is applied in the cloud layer to make smart decisions like controlling the amount of water
needed to be irrigated for improving crop growth environment.

\subsection{Distributed and Federated Learning for IoT}

  \gls{ml} algorithms build mathematical models from training data. Such models can be built in different ways: 
 
 \begin{tikzpicture}
 \node {ML algorithms}
 [edge from parent fork right,grow=right, level distance=30mm, sibling distance=7mm]
 child {node {Federated}}
 child {node {Distributed}}
 child {node {Centralized}};
 \end{tikzpicture} 
 
 Standard \gls{ml} algorithms collect data and train and validate the models in a \textit{centralized} cloud server. In centralized \gls{ml} all of the training and validation
 data resides in one location where the  model is validated, verified, and frozen before its deployment. Centralized \gls{ml}
 is not appropriate
 when training data is not in one location and there are sharing constraints. In such cases  distributed  \gls{ml}
 or Federated learning is preferred. 
In \textit{distributed} \gls{ml} the learning/inferencing processes can be distributed depending on the needs of the system to overcome
 computational/storage constraints. That is, the training
 data is moved for processing by multiple computers. 
  Distributed \gls{ml} often uses parallel processing techniques.
Federated \gls{ml} is decentralized learning in which, unlike distributed learning,  training data is not exchanged among the servers. This method is appropriate when there are concerns or constraints for sharing training
data, e.g.,  due to policy, security, insufficient network capacity (wireless networks). Distributed and federated learning approached have great potential for \gls{iot} networks as we discuss in the following.

 \subsubsection{Distributed learning for IoT} 
 Reducing signaling between \gls{iot} devices and the access network, e.g., over the unlicensed spectrum, results in decreasing energy consumption per data transfer for \gls{iot} devices at the cost of decreasing control over radio resource usage for the access network. For example, in the case of \gls{iot} solutions over the \gls{ism} band, there is almost no centralized control over the devices' transmissions. This results in no performance
 guarantee for \gls{iot} communications in this band \cite{mads,madse}. In order to benefit from the relaxed transmissions, and suffer as low as possible from the uncoordinated access, there is a need for distributed solutions able to adapt communications parameters of devices to the environment. 
 In recent years, there is an ever-increasing interest in
 leveraging machine learning tools in operation control of
 independent entities which have limited observability on their
 environments \cite{disAI}.
 In \cite{disAI}, distributed reinforcement learning at the devices  has been proposed for overload control in  serving massive \gls{iot} traffic over \gls{lte} networks. Self-organized clustering and clustered access for  \gls{iot}  have been proposed in \cite{azari2016energy}. In \cite{azari2018self,lorm}, the multi-arm bandit has been proposed for distributed management of \gls{iot} parameters, in which, devices learn how to avoid sub-channels suffering from a high level of  interference.  The presented approach in \cite{azari2018self} leverages two notions of external and internal regrets, for targeting energy
 consumption and collision probability, respectively, and is able to converge to the  centralized optimized approach, e.g. \cite{sandoval2019optimizing}, if network state (interference statistics) is constant for a longer period of time. 
 
 While regarding the limited computation/storage of IoT devices, deploying deep and complex \gls{ai} algorithms seems impossible on most IoT devices, this area of research is expected to attract more attention in future works by adapting \gls{ai} algorithms to the device constraints or splitting learning between the device and the network.

 \subsubsection{Federated learning for IoT}
  \gls{fl} is a new \gls{ml} framework  that  is \textit{distributed} over mobile devices. The core concept of \gls{fl} is to send the algorithm to the data rather than sending data from user devices to the server. With this,  there is no direct access to labeling of raw user data, and thus,
the model development, training, and evaluation in \gls{fl} are different from centralized \gls{ml}.
\gls{fl} was coined by Google \cite{mcmahan2017communication}
and has remained at the forefront of the field. 
In \gls{fl}, both the central server and edge devices contribute to the model training process collaboratively via the following steps:
 \begin{enumerate}
 	\item The central server chooses a model to be trained. 	
 	\item It sends the initial model to thousands of edge devices.
 	\item The devices train the model locally with their own data.  
 	\item The central  server gathers all the model results and generates a global model without accessing any data.
 	\item This process is repeated over and over again.  	
 \end{enumerate} 
Once the training is completed, test versions of the algorithm are sent to many devices to evaluate the performance on real data. This step  can also be iterated until an updated model is sent to all devices to replace the old ones. 

Privacy has been the main motivation for devolving the \gls{fl}. Due to its distributed model training,  \gls{fl} provides highly personalized models and does not compromise user privacy.
\gls{fl} has already been deployed in Google keyboard (Gboard) to improve word suggestions to the user without compromising their privacy. 
\gls{fl} however offers several other advantages. 
Security  and speed are other reasons that \gls{fl}  can work better in certain use-cases. 
Several   advantages and  limitations
of \gls{fl} are listed in Table~\ref{tab:FL}. 

When it comes to \gls{iot} applications, one main limitation of \gls{fl} is that it requires high local device power and memory to train the model. For this reason, \gls{fl} cannot be directly applied to sensors and resource-limited devices. However, with the rapid  increase in the computational and storage power of smart devices, there are several potential scenarios for adopting  \gls{fl} in  \gls{iot} applications 
in the future: 
\begin{itemize}
	\item \textit{Self-driving cars:}
	 it is expected to have are a large number of self-driving cars on our roads in the future.  Since \gls{fl} can reduce latency, a likely future scenario is to use \gls{fl} in those vehicles  to be able to rapidly respond to each other during safety incidents. 
	 \item \textit{UAVs:} 
	 similar to driver-less cars, \glspl{uav} are increasingly used in various applications such as for delivery services, emergency response, disaster relief, agriculture, waste management, telecommunication, and Internet access. 
	  Another likely scenario for using \gls{fl} is in \glspl{uav} to enable them to rapidly respond to each other during safety incidents since \gls{fl} can reduce latency.
	  \item \textit{Edge \gls{ai}:} Owing to the growth of smartphones and other smart devices,   
	   modern  networks generate a huge amount
	  of distributed data every day. 
	 Personalized features that are invaluable for many apps can be achieved on device. 
		 \item Mobile edge computing (MEC):  While cloud computing and cloud RAN \cite{vaezicloud,checko2014cloud} offer unquestionable economies of scale, MEC enjoys compute and storage resources close to the end users, typically within or at the boundary of operator networks. Benefits edge solutions include low latency, high bandwidth, device processing and  trusted computing and storage.
	Deep integration of computing in the network  is expected to blur  the line between the device, the network edge, and the cloud in the future, and as a result,  distributed algorithms and applications will flourish.  
	
	\item \textit{Robotics and smart industry:} Robotic is a key component of Industry 4.0. Robots will operate in increasingly complex environments in a cooperative manner. 
	Google is providing infrastructure essential to building and running robotics solutions for business automation.
	The literature on \gls{fl} for robotics is sparse. 
	Liu \textit{et al.} \cite{liu2020federated} have investigated an \gls{fl} scheme for cloud robotics. The proposed solution enables the cloud to fuse heterogeneous knowledge from local
	robots to develop guide models for robots. 
	It makes navigation-learning robots use prior knowledge	and adapt to a new environment.
	In \cite{yin2020fdc} a secure \gls{fl}-based data collaboration framework is proposed and validated the effectiveness of the proposed framework in a real \gls{iot} scenario.

\end{itemize}

Collaborative \gls{fl} is a new approach for resource-constrained \gls{iot}. The concept is similar to \gls{d2d} communication. \Gls{iot} devices that are not able to communicate with a \gls{bs} due to their limited resources send their local learning models to nearby devices rather than the \gls{bs}.  Some recent overview papers on federated learning for \gls{iot} networks can be found in  \cite{khan2020federated,du2020federated,nguyen2021federated}.

%	
% \textit{Smart Grid:} The smart grid is critical to building a secure, clean, and more efficient electricity networks. It is used to
%which can be used to remotely monitor and manage everything from lighting, traffic signs, traffic congestion, parking spaces, road warnings, and early
%
%

\begin{table*} [h]
	\centering
	%	\begin{minipage}[b]{0.9 \textwidth}
	\centering
	\begin{threeparttable}[b]		
		\caption{Federated learning compared to centralized learning} 
		\label{tab:FL}
		\begin{tabular} {  p{0.42 \textwidth} p{0.41 \textwidth} } 
			\toprule
			\rowcolor{Lightgray}   \textbf{Advantages}    &  \textbf{Limitations}    \\        \cmidrule{1-2}  \addlinespace			
			\rowcolor{Lightgray}  \begin{itemize}
				\item Enhances user privacy by keeping personal data on a person’s device
				\item Lower latency as the updated model can be used to make predictions on the user's device.
				\item Less power consumption at the central server as models are trained on user devices
				\item  Less data transfer and thus faster learning pace
				\item Deliver both global and personalized models
				\item Accelerating clinical research as  diseases dataset  can be unlocked with less privacy concerns 
			\end{itemize}  &
			\begin{itemize}
				\item Requires significantly more local device power and memory to train the model
				\item Requires higher bandwidth over  WiFi or cellular networks during training
				\item Insufficient bandwidth could increase latency and makes the learning process slower 
				\item Devices may drop out during the training process  by their owners
			\end{itemize}  \\ \addlinespace

		\end{tabular}
	\end{threeparttable}
	%	\end{minipage}
\end{table*}

\subsection{IoT-Friendly DL}
State-of-the-art deep learning models usually involve millions to billions of parameters and need  sophisticated computing and large
storage resources to evaluate and save such a huge number
of parameters. As an example, the number
of parameters in commercial \glspl{cnn} in computer vision applications is as high as $\sim 10^8$ \cite{denton2014exploiting}.
Such huge models are  computationally expensive for resource-constrained IoT and  edge devices (e.g., smart sensors,
wearables, smartphones, robots, drones, etc.), which have limited processing power, battery energy, and memory.  For this reason, many DL methods can hardly be used in such devices for training purposes. Even worse, resources may not be sufficient for the test phase where a pre-trained model is run on the device \cite{lane2015early}. 
Even if the devices have enough memory and computation resources, it is of great value to reduce the need for such resources to save energy, cost, and time. 
Therefore, there are growing interests in developing efficient \glspl{dnn}.

Several other research teams have recently shown that the complexity of \gls{dnn} models can be largely reduced without a significant drop in the accuracy. In the following, we discuss techniques used to achieve such results and illustrate some important results for each technique.  
The effort in this field can be categorized into network compression  and acceleration techniques. We elaborate on these in the following.

%Denil \textit{et al.} have shown that there is significant redundancy in the parameterization of many DL models \cite{denil2013predicting}. Specifically, they have shown that many of the weights of a network need not to be learned and  can be predicted. 

\subsubsection{Network compression} 
With negligible degradation of performance,  \glspl{dnn} can be compressed  in various ways including 
low-rank tensor approximations,
weight pruning,
quantization, building a compact model, and a combination of two or more of these techniques. We describe these methods  in the following. 	
\begin{itemize}
	\item [i.]	\textit{Low-rank tensor approximations:} Let us consider \gls{cnn} as an example.  It is well-understood that 
	the bulk of  computations is from convolution operations wheres
	most weights of \glspl{cnn} are in the fully connected layers \cite{denton2014exploiting}. The fully-connected layer can be view as a 2D matrix (or 3D tensor).  Approximating the matrices corresponding to those layers is a one compression method. 
	The approximation can be performed in different ways including matrix decomposition. In this method,
	\gls{svd} of those matrices is found and only singular values with large magnitude are kept. This low-rank approximation 
	will be useful if the singular values of the matrix decay rapidly, and it could decrease the computation complexity up to an order of magnitude \cite{denton2014exploiting}. Other matrix decomposition methods like QR decomposition and CUR decomposition are proposed in the literature but are less accurate compared to the \gls{svd} method \cite{deng2020model}. 	 
	It is worth mentioning that low-rank decomposition  can be also used for reducing the convolution layer.  The convolution wights can be viewed as a 4D tensor and low-rank up the evaluation of \glspl{cnn}  \cite{jaderberg2014speeding}. 
	There are also other side benefits for \gls{svd}. It can reduce memory footprint and decrease the number of learnable parameters 
	which may improve generalization ability.

	\item [ii.]	\textit{Pruning:} Consider an initially trained NN with a reasonable solution.
	One can remove all connections whose weight is less than a threshold and re-train the network until a reasonable solution is obtained \cite{han2015learning}. Such a method permanently removes less important links and converts a dense, fully-connected layer to a sparse layer.
	Pruning process can be iterated to increase the compression efficiency.  One may also use pruning with rehabilitation to improve learning efficiency \cite{guo2016dynamic}. In this method, there is no chance for the pruned connections to come back  and 
	incorrect pruning may cause severe accuracy loss.  
	The above pruning methods are magnitude-based. There are other methods like the Hessian-based method. The idea is to delete parameters with small  saliency, i.e., small effect on the training error \cite{lecun1989optimal}. 
	
	\item [iii.]	\textit{Quantization:} This method compresses the original network by
	reducing the number of bits required to represent each weight. Unlike pruning which permanently drops less important connections and thus reduces the number of weights, this approach compresses the weights. In addition to 
	weights,  other data objects such as activation, error, and gradient can be quantized. 		
	Quantization is proved to be an effective way for model compression and acceleration \cite{vanhoucke2011improving,gong2014compressing,wu2016quantized}.
	In the extreme case of the 1-bit representation of each weight  \cite{courbariaux2015binaryconnect} binary  neural networks are realized.  The accuracy of binary nets is not impressive and needs further improvement  \cite{tang2017train}.  Pruning, quantization, and other compression methods, e.g., source coding techniques, can be combined to increase the degree to which a  network can be compressed.  Han \textit{et al.}  \cite{han2015deep} proposed a deep compression  technique that can reduce the size and reduce
	the storage requirement of \glspl{dnn} more than an order of magnitude without losing accuracy.  
	This approach rains the initial network until a reasonable solution is obtained.
	It then prunes the network with a magnitude-based method until a reasonable solution is obtained.
	Next, it quantizes the network with k-means based method until a reasonable solution is obtained.
	Finally, it compresses the weight using Huffman coding.
	\item [iv.]	\textit{Compact model:}			
	Unlike the previous models that start with a dense model and then try to compress it,  the idea behind a \textit{compact model}  is to design small \gls{cnn} architecture from scratch while preserving acceptable accuracy 		
	\cite{sandler2018mobilenetv2,iandola2016squeezenet}. Various techniques are introduced for this purpose. For instance, SqueezeNet replaces 3x3 filters of AlexNet with 1x1 filters, which creates a compact model reducing filter parameters 9 times. It also reduces the number of input channels to 3x3 filters resulting in about 50 times  overall parameters reduction.  
	Similarly, MobileNetV2 \cite{sandler2018mobilenetv2} is a family of general-purpose computer vision \gls{dnn} designed with mobile devices in mind to support classification, detection, and more.
\end{itemize}

\subsubsection{Hardware acceleration} 
Designing specialized hardware for efficient processing of \glspl{dnn} is another active line of research that aims to minimize
energy  and storage required for DL
models in general, and in IoT devices in specific.
Such hardware is
named neural network \textit{accelerators}. Conventional \gls{dnn} accelerators were only based on the hardware architecture to
improve computation and reduce memory
requirements. 
Parallel computing and processing in memory are two main techniques to this end \cite{deng2020model}.   
However, with the exhaustion of parallelism
and data reuse, sole hardware optimization reaches its performance upper bound. 
To overcome this, recent works have investigated  
algorithm and hardware co-design. Algorithm-hardware codesign has great potential in reducing the hardware cost. For this reason,
modern accelerators have been exploited several of the network compression techniques discussed
earlier.

Combining \gls{ai} and edge devices with rich sensors, have great real-world applications in smart homes, autonomous driving, drones, and so on, and can dramatically expand the scope of \gls{ai} applications.
\gls{dnn} on IoT devices based on microcontroller units is, however, extremely challenging. The memory (\gls{sram}) of microcontrollers is about 3 orders of magnitude smaller than that of mobile phones and  5-6
orders of magnitude smaller than cloud GPUs \cite{lin2020mcunet}, making it impossible to run off-the-shelf \gls{dnn}  models, even the compact models like MobileNetV2 \cite{sandler2018mobilenetv2}. 

Efficient DL computing and model compression is now an active line of research. 
A comprehensive survey of model compression and hardware acceleration for neural networks can be found in \cite{deng2020model}.

\section{NTN Integration into IoT Networks} \label{sec:NTN}

% \hline

% Advantages of the UAV-assisted IoT networking: 1- reducing the IoT nodes energy consumption, 2- extending the coverage, 3- reliability enhancement, 4- One-demand wireless charging, 5- less shadowing, 

% Challenges: 1- limited on-board energy, 2- Resource allocation and communication scheduling, 3- trajectory design, 4- connectivity and control, 5- high mobility, For satellite: 6- unreliable satellite links, 7- delay, 8- limited resources

% \hline

{ A truly global and \gls{3d} connectivity of IoT cannot be realized by current cellular and wide-area networks. For example,} in rural areas such as mountains, forests, and oceans, terrestrial infrastructure may not exist to enable such connectivity. For these cases, satellites networks play an important role in providing global connectivity. In addition, satellites are of crucial assistance when terrestrial networks are overloaded or out of service for instance in disaster scenarios. The terrestrial cellular coverage can be extended and complemented by the inclusion of satellite networks. On the other hand, the more favorable \gls{los} propagation condition with satellites allows the IoT devices to lower down their transmit power and hence consume less energy. Therefore, satellite integration into IoT networks may provide significant advantages. Motivated by these facts, recent \gls{3gpp} study items include  providing \gls{nbiot}  via  satellite  systems  \cite{3GPP22822,liberg2020narrowband}.

Furthermore, increased use of \glspl{uav} in novel commercial applications is currently being enabled by recent advances in the design of cost-effective drone technology \cite{geraci2021will,vinogradov2019tutorial}. A swift and infrastructure-independent deployment of \glspl{uav} particularly can be beneficial for disaster and poorly covered remote areas. As compared to satellite systems, \glspl{uav} deployments are more flexible, less expensive, and closer to the ground IoT nodes. The communication latency with \glspl{uav} is lower and the throughput can be higher at the expense of a shorter footprint and lifetime \cite{azari2017ultra}. That is, \glspl{uav} complement satellite systems in supporting IoT networks.

{ In short, \gls{uav}-based  and satellite-based networks provide global, \gls{3d}, and continuous  coverage and are generalizations of cellular and wide-area networks. In this article, the term \gls{ntn} refers to both \gls{uav} and satellite systems. In the following, we classify the main roles of \glspl{ntn} integrated into IoT networks and elaborate on their particular advantages and challenges ahead.} 
%\footnote{The terms \gls{ntn} and \gls{3d} network are used interchangeably in this paper.}

\subsection{UAVs Integration into IoT Networks}

Three fundamental roles of \gls{uav}-aided IoT networks are as follows:

    \subsubsection{IoT data collection with UAVs} % \cite{Mozaffari8038869,Liu8620506,Yang8314824,zhan2019energy,fu2021trajectory,li2019joint,fu2020joint,saraereh2020performance,moheddine2019uav,saraereh2020performance,nomikos2020uav,motlagh2017uav}
    
    \Glspl{uav} can be deployed to collect data produced by the low-power IoT devices or wireless sensors distributed over a certain region such as in precision agriculture applications. Such data might be delay-tolerant information and hence does not need an instant connection to the central unit using permanent expensive ground infrastructure. Nevertheless, a real-time communication can also be facilitated by using \gls{uav} relays between IoT nodes and \glspl{bs}. In effect, \glspl{uav} can fly above the ground nodes and establish \gls{los} connection with both the ground nodes and the target \glspl{bs}. Therefore, \glspl{uav} are capable of addressing several challenges in IoT-centric scenarios. Specifically, the communication links turn to be more reliable, the coverage range is potentially extended, and the energy consumption of the IoT devices can be reduced. 
    In \cite{zhan2019energy}, a \gls{uav}-enabled IoT system is considered where a \gls{uav} collects data from IoT devices. The maximum energy consumption among all the devices' is minimized by jointly optimizing the devices transmit power, communication scheduling, and the \gls{uav} trajectory. In \cite{fu2021trajectory}, the authors design the trajectory of \gls{uav} for data collection using the \gls{rl} approach, in order to prolong the \gls{uav} flight time. To minimize the \gls{uav} data collection time, several other network parameters, including \gls{uav} height, velocity, and link scheduling, in addition to the \gls{uav} trajectory, are optimized in \cite{li2019jointT}. To improve the reliability between ground IoT nodes and \glspl{bs}, authors in \cite{fu2020joint,saraereh2020performance} study the performance of \glspl{uav} relaying collected data to \glspl{bs} and propose efficient transmission algorithms.
    
    In \cite{bayerlein2020uav}, an efficient path for maximizing the collected data under flying time and obstacle avoidance constraints is designed using a double deep Q-network approach. To maximize the energy efficiency of the IoT network, a deep \gls{rl} approach is adopted in \cite{cao2019deep} through joint optimization of the power and channel allocation to the IoT devices in the uplink. 
    
    \subsubsection{Localization of IoT devices with UAVs} % \cite{alshrafi2014compact,han2016survey,gezici2005localization,zanella2016best,al2014modeling,azari2018key,Sallouha2018aerial,sallouha2018energy,vinogradov2019tutorial}
    
    In various applications of wireless networks location-aware services are required. In effect, the location information enables a more meaningful process of the gathered data and allows to optimize the network performance accordingly. As for the outdoor use cases, one may consider the \gls{gps} where most of the time it provides satisfactory performance. However, such technology particularly for IoT devices is expensive and power-consuming. Moreover, \gls{gps} is vulnerable to jamming which might be a crucial drawback for a reliable and secure provision of services \cite{alshrafi2014compact}. Therefore, over the past decade, extensive efforts have been made in order to provide alternative solutions for localization purposes. 
    
    Among several localization methods \cite{han2016survey}, multilateration is of great importance for precisely determining the users' positions. For this method, at least three estimations of the node's distance from three different anchor positions are needed. The distance estimation is typically done using time-based or (received signal strength)-\gls{rss}-based techniques \cite{gezici2005localization}. In time-based techniques, the distance is estimated by multiplying the flight time by the light's speed. However, determining the flight's time is a tough challenge. This is because an adequately precise time synchronization is required between the transmitter and the receivers which is hard to acquire. However, time synchronization is not needed for an \gls{rss}-based solution. Moreover, \gls{rss} estimation functionality is available in the chipsets \cite{zanella2016best}. In the \gls{rss}-based method, the distance is estimated by using the \gls{rss}-distance function well represented by the path loss models. 
    
    \Gls{rss}-based localization, however, faces an important challenge when the anchors are on the ground due to significant variation around the mean received signal power, as known as the shadowing effect. In effect, a ground-to-ground communication link suffers from a high level of shadowing which results in a large estimation error of the distance between a node, to be localized, and the ground anchor. This fact imposes a low localization accuracy and limits the use of the \gls{rss}-based method \cite{zanella2016best}. 
    
    The characteristic of shadowing corresponding to different receiver's altitude has been studied in \cite{al2014modeling}. This shows that shadowing is less at higher altitude for ground-to-air communication, although the mean path loss behaves differently. Accordingly, the use of \gls{uav} anchors for \gls{rss}-based ground localization sounds like an interesting solution to mitigate the weakness of the \gls{rss}-based technique. In this manner, \glspl{uav} can act as aerial anchors to localize ground IoT devices. \gls{uav} anchors can combine the benefits of satellites with a good link probability of \gls{los} and the advantages of ground anchors with a short link length and hence might result in a higher RSS resolution. 
    
    Several of the recent studies have addressed the concept of using \glspl{uav} for ground localization \cite{azari2018key,Sallouha2018aerial,sallouha2018energy,vinogradov2019tutorial}. In \cite{Sallouha2018aerial} the authors considered the localization of IoT ground devices by using multiple static \gls{uav} anchors located at arbitrary locations. This study shows that the aerial anchors are capable of localizing the ground nodes with notably higher accuracy if the altitudes of the \glspl{uav} are adopted properly. Indeed, it is shown that the impact of \glspl{uav} altitudes is significant in an urban environment due to considerable variation of shadowing and mean power with the altitude. Statistically speaking, as the altitude of a \gls{uav} increases the communication link with the ground becomes more \gls{los} and hence the mean path loss decreases. However, at a certain altitude at which the link is almost always \gls{los} the more increase of the \gls{uav} altitude increases the path loss due to longer link length. We note that, in contrast to the mean path loss behavior, the shadowing, i.e. the fluctuation around the mean path loss, is becoming less at a higher altitude. In \cite{Sallouha2018aerial}, the authors for the first time showed that in an urban area there exits an optimum altitude at which the localization accuracy is the highest.
    
    The same report has investigated other important design factors effects such as \glspl{uav}' inter-distances and the number of \glspl{uav}. The results reveal that there is optimal \glspl{uav} inter-distances for the maximum localization accuracy in order to localize ground IoT devices uniformly distributed over a certain region. Also, as the number of \glspl{uav} increases the localization error drops sharply.
    
    The \glspl{uav} networking is constrained by their energy budget. As known well, the mechanical energy consumption of present \glspl{uav} are dominant over their electrical energy consumption and hence play a prominent role in their trajectory design and the lifetime of the network. Accordingly, it is of great importance to study such impact on the target quality of service. For this, first, a detailed description of \gls{uav}'s energy consumption model is presented in \cite{sallouha2018energy}, where it is shown that the velocity of \gls{uav} can be optimized for the minimum energy consumption. Interestingly, such an optimum velocity is not zero for a rotary-wing \gls{uav} and hence hovering the \gls{uav} may consume more energy than a mobile \gls{uav}. Then by taking such an issue into account, an optimized trajectory design of \glspl{uav} for minimum localization error is investigated. In addition, the data acquisition time and its role in the localization error have been studied.

    \subsubsection{IoT wireless energy supplier and information dissemination with UAVs} \label{wptmma}
    
    % Wireless Energy Supplier \cite{park2019uav,xu2018uav}
    % Information Dissemination \cite{wu2018joint}
    % SWIPT \cite{huang2019multiple,feng2020uav,jeong2020simultaneous,kang2020joint}
    
    Limited energy budget and connectivity problem of IoT devices in certain environments and disaster cases can be well addressed using \glspl{uav}. Compared to fixed infrastructure \glspl{uav} could be more suitable for \gls{wpt} due to their flexible deployment and mobility which bring them to the proximity of IoT nodes. Furthermore, due to interference issues or limited battery capacity, the ground IoT transmitters may lower their transmit power resulting in a low range of communication for some cases. In such a situation, rapid deployment of \glspl{uav} can assist to broadcast common and private files. 
    
    A \gls{uav} trajectory for \gls{wpt} is optimized in \cite{park2019uav} where ground terminals use the received energy to upload their data to the \gls{uav}. In \cite{wu2018joint} a joint optimization of \gls{uav} trajectory and power control along with the communication scheduling of \glspl{uav} serving ground terminals maximizes the minimum rate of downlink communication.  In \cite{feng2020uav,jeong2020simultaneous,huang2019multiple,kang2020joint,feng2020uav}, the authors study the deployment of \glspl{uav} for simultaneous data and energy transfer to IoT devices. For these, the trajectory and resource scheduling of \glspl{uav} are optimized in \cite{feng2020uav} to prolong the IoT network lifetime. The authors in \cite{jeong2020simultaneous} study simultaneous data and energy transfer to the IoT devices using \glspl{uav}. In this work, the \gls{uav} transmit power for each stream of IoT nodes and its trajectory are optimized to increase the minimum rate of communication with the IoT devices.

To enable such \gls{uav}-assisted IoT networking and to improve the network functionality, one needs to provide reliable and secure communication between drones and ground stations, particularly when drones require a beyond visual \gls{los} tetherless connectivity for control and/or communication. This should be performed by a wireless technology suitable for the target \gls{uav} application.

% The candidate technologies  such as WiFi and cellular networks
% (i.e., \gls{4g}-\gls{lte}, \gls{5g}). The choice of wireless technology
% may depend on diverse factors such as
% required security, reliability, and system responsiveness.

% In the following we specify the communication requirements of UAVs.

To support \glspl{uav} in large scales, the existing  point-to-point \gls{uav} communication with a ground station over an unlicensed band fails to satisfy the important criterion for \gls{uav} communication and networking \cite{azari2019cellular}. Specifically, related to \gls{uav} command and control and a beyond visual \gls{los} connectivity, a \textit{high coverage and continuous connectivity} is needed to guarantee reliable control and tracking of human-driven or autonomous \glspl{uav}. In addition, \textit{low latency} communication is essential to enable a robust remote control. Moreover, associated with payload communication, a \textit{high throughput} communication link is to be established in order to exchange data and stream video signals. \gls{uav}'s communication technology should also be \textit{secure} to protect the data and should be capable of controlling interference. Additional features include \textit{system scalability} for supporting the prompt growth of \glspl{uav} and \textit{regulation compliance}. Cellular technology seems to be an adequate choice for satisfying the aforementioned requirements. The feasibility of using the existing cellular networks and the corresponding \gls{lte} and \gls{5g} technologies for \glspl{uav} connectivity have been investigated in  \cite{azari2019cellular,3gpp2017,azari2017coexistence,azari2020uav}. 

% \mahdi{last paragraphs may be shortened.}

\subsection{Satellite Integration into IoT Networks}
%References: \cite{liberg2020narrowband,marchese2019iot,soret20205g,shi2020joint,dai2020uav,cheng2019space}

Satellite systems enable sustainable  global coverage and hence significantly increase the IoT applications. A review of the \gls{nbiot} and \gls{ntn} integration in \cite{liberg2020narrowband} investigates how the \gls{nbiot} can be modified to support satellite networks for IoT services. In a densely deployed IoT networks, the use of satellite for IoT traffic offloading through \gls{leo} satellites is an alternative solution which is investigated in \cite{soret20205g}. In \cite{shi2020joint} \glspl{uav} acting as gateways transfer the collected IoT data to the satellite. In this paper, the problem of limited resources in integrated ground-air-space IoT networks is considered. To deal with this problem, spectral efficiency is maximized through the optimization of the \gls{uav} bandwidth allocation and gateway selection. In \cite{dai2020uav} a mobile IoT node and its improved connectivity through the integration of \gls{uav} relay and satellite system are presented. Improved performance of the system is shown through the  optimization of \gls{uav} trajectory and power allocation, under the \gls{uav} energy constraint. 

Using the \gls{drl} approach an energy-efficient channel  allocation  algorithm  in  \gls{leo}  satellite  IoT  is  presented  in \cite{zhao2020deep}. An  optimization  problem  is  formulated  in \cite{cui2020latency} in order to  minimize the energy cost and latency affected by user  association  and  resource  allocation  for  computing  and communication. To  handle  the  complexity  of  the  problem,  the \gls{drl} technique is partially used to solve joint user association and offloading decision sub-problem.

\section{Remaining Challenges and Future 
Directions}\label{sec:future}

The IoT has already have a big impact on our society and new applications are being introduced almost on a daily basis. With a vision of global, 3D connectivity and  with the help of leading technologies like \gls{ai}, \gls{5g} and beyond cellular networks, and cloud and edge computing, the IoT is expected to produce increasingly promising results
in every aspect of the world around us from health to agriculture and industry.  To this end, several important research
issues remain to be addressed in the future. We conclude our
survey by discussing these challenges and pinpointing key  problems that should be addressed in order to fully utilize the IoT to  improve quality of life and generate new
revenue. 

\subsection{Limited Battery Lifetime}
\begin{figure*}
 	\begin{forest}
 		for tree={
 			align=left,
 			edge = {draw, semithick, -stealth},
 			anchor = west,
 			font = \small\sffamily\linespread{.84}\selectfont,
 			forked edge,          % for forked edge
 			grow = east,
 			s sep = 0mm,    % sibling distance
 			l sep = 8mm,    % level distance
 			fork sep = 4mm,    % distance from parent to branching point
 			tier/.option=level
 		}
 		[Solutions for  \\ prolonging \\ the battery \\lifetime						%%%%%%%%%%%%%%%%%%%%%%%%%%%%%%%%%
  			[  Wireless \\ power transfer
 	            [By base station
 	                        [[Beyond 5G\\ {\color{blue}Section \ref{SWPT}}]]]
 	            [By auxiliary devices\\ (e.g. UAVs)
 	                        [[Device \\dependent\\ \color{blue}Section \ref{wptmma}]]]]
 		%%%%%%%%%%%%%%%%%%%%%%%%%%%%%%%%%
 	 		[ Energy \\ harvesting 
 		 		[Battery-less 
 		 		                [[Device\\ dependent\\ {\color{blue}Section \ref{SWPT}}]]]
 		 		[With battery\\for saving energy
 		 		                [[{Device\\ dependent\\\color{blue}Section \ref{SWPT}}]]]]
 		%%%%%%%%%%%%%%%%%%%%%%%%%%%%%%%%%
 		 [Energy \\saving 
 		 		 [In data transmission\\ from device
     		    [Increased sensitivity \\of the BS's receiver [NB-IoT \\LTE-M\\ {\color{blue}Section \ref{SNBT}}]]
     		    [Increased link budget\\ by signal repetition [NB-IoT\\ {\color{blue}Section \ref{SNBT}}]]
     		    [Increased link budget\\ by narrow bandwidth  [NB-IoT\\LTE-M\\SigFox\\LoRaWAN\\ {\color{blue}Section \ref{CALL}}\\ {\color{blue}Section \ref{SUL}}]]
     		    [Increased resilience \\to interference by\\ chirp spread spectrum [LoRaWAN\\ {\color{blue}Section \ref{SLOR}}]]
     		    [Increased resilience \\to interference by\\ time/freq hopping [SigFox\\ {\color{blue}Section \ref{SSG}}]]
     		]
     		[In data reception\\ at device
     	    	[Increased link budget \\by signal repetition [NB-IoT\\ {\color{blue}Section \ref{SNBT}}]]
     	    	[Time-window based \\data reception   [PSM (NB-IoT/\\LTE-M)\\SigFox\\LoRaWAN\\ {\color{blue}Section \ref{SNBT}}\\
     	    	 {\color{blue}Section \ref{SUL}}]]
     	    ]
            [In network access
                [Grant-free access\\ [SigFox\\LoRaWAN\\Beyond 5G\\ {\color{blue}Section \ref{SGFA}}]]
                [Contention resolution\\ over RACH;\\Compressed sensing [4G/5G\\ {\color{blue}Section \ref{SENH}} \\ {\color{blue}Section \ref{RAS}}   ]]
                [Narrowband RACH and \\new preamble setup [NB-IoT\\ {\color{blue}Section \ref{SNBT}}]]
                [Capillary networking;\\UAV-assisted IoT;\\D2D communications[4G/5G\\{\color{blue}Section \ref{d2dsec}}] ]
            ]
     		[In idle mode
     		[Wake-up receiver \\ assisted by network [Beyond 5G\\ {\color{blue}Section \ref{SWUR}}]]
     		[DRX; eDRX; PSM [4G/5G\\ {\color{blue}Section \ref{SENH}}\\{\color{blue}Section \ref{SNBT}}]]
     		]
 		]
 		]
 	\end{forest}
 \caption{Classification of various solutions targeting  a long battery lifetime for IoT devices.} \label{fig:class}
 \end{figure*}
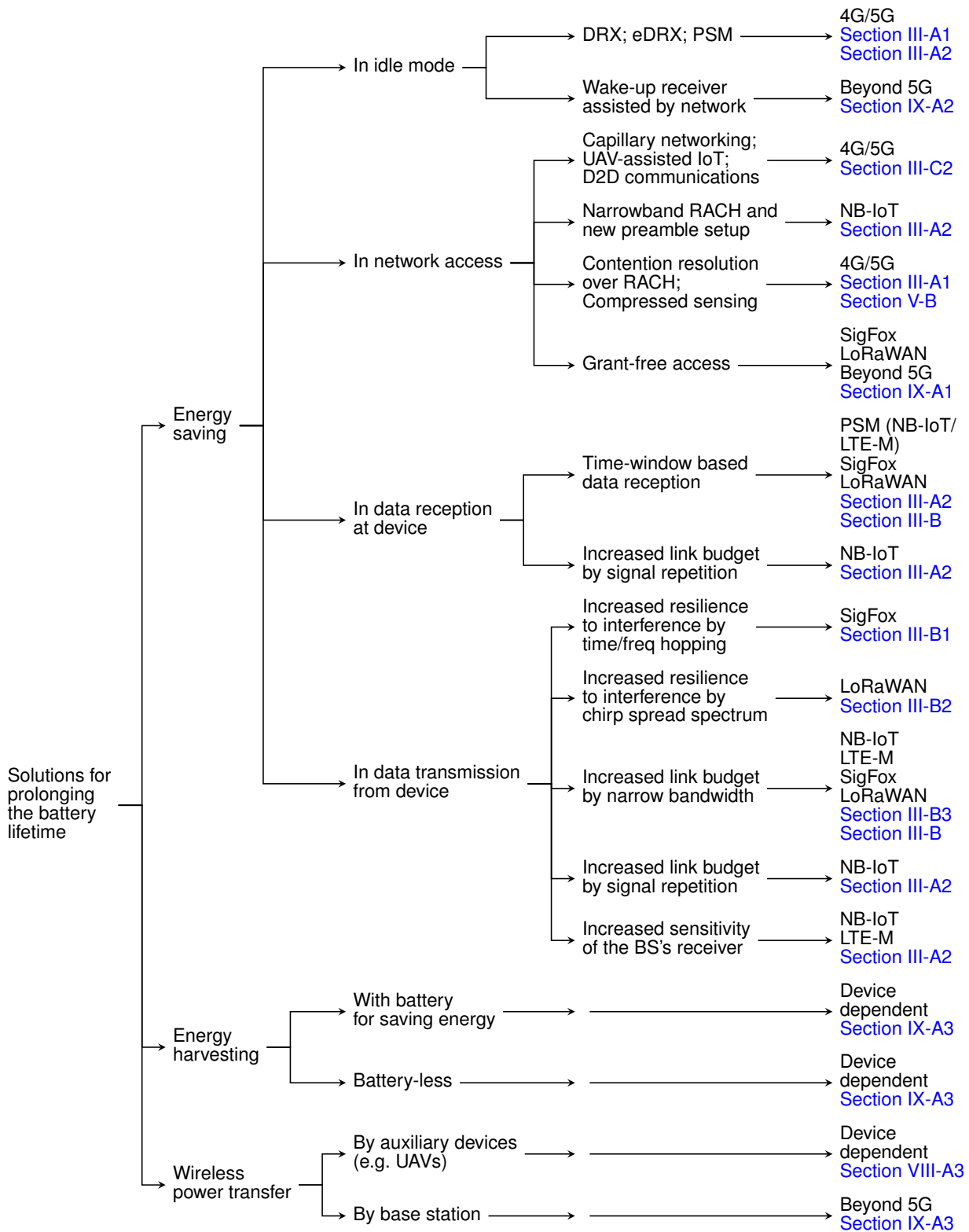

With the introduction of \gls{nbiot} technology, the expected battery lifetime of IoT devices has been  significantly improved  in comparison with the \gls{lte} networks. However, the preliminary performance analyses show that there is still a gap between the expected battery lifetime in \gls{nbiot} networks and the \gls{5g} expectations \cite{nbt,mnbt}. This is mainly because the  energy consumption in synchronization, connection establishment, resource scheduling, and waiting for the control signals from the \gls{bs} are still there in the \gls{nbiot} standardizations \cite{nbt,mnbt}. Potential directions for resolving this problem include (i) relaxing the  resource reservation procedure for short-packet IoT communications, i.e. grant-free radio access \cite{gf}; wake-up receiver for IoT devices; and energy transfer to IoT devices and energy harvesting from environment, which are investigated in the sequel. Furthermore, Fig. \ref{fig:class} presents an overview of schemes reviewed in this work which aim at increasing battery lifetime of IoT devices. 

\subsubsection{{Grant-free radio access}}\label{SGFA}
\gls{gfa} to radio resources has been already implemented in IoT technologies that operate over the unlicensed spectrum, such as SigFox and  \gls{lorawan}~\cite{mag_all}. In \gls{gfa}, once a packet arrives at the device, it is transmitted without any handshaking, resource reservation, or authentication process.
This type of access is opposed to the legacy grant-based access schemes in existing cellular networks. As part of the efforts shaping the \gls{5g} \gls{nr}, the use of non-orthogonal multiple access schemes, including \gls{gfa}, has attracted profound attention in recent years ~\cite{lv2018millimeter,noma,jsacS}.  {This is due to the fact that a recurrent element in various beyond \gls{5g} visions is to move} towards zero-cost zero-energy IoT communications \cite{zce}. Then, as \gls{gfa}  can potentially offer reduced complexity and  energy consumption for the end-devices, it could be a  complementary access solution for IoT communications in future wireless access networks, along with the legacy grant-based access (GBA) schemes.
Performance of communication with \gls{gfa} has been investigated in a number of recent works \cite{zgf1,hsgk,anf,dlgf}, and also in the context of the \gls{3gpp}  \cite{gf3}.
In~\cite{zgf1}, \gls{gfa} in massive \gls{mimo}  systems has been investigated and the analytic expressions of success probability for conjugate and zero-forcing beamforming techniques have been derived. 
In~\cite{disgf}, a novel distributed \gls{gfa} scheme has been presented, in which, a cell is divided into different layers based on the pre-determined inter-layer received power difference from devices. Then, the transmission of each device is adapted to the layer that device belongs (close to the repetition class in the context of \gls{nbiot}).
The performance of a massive deployment of devices leveraging \gls{gfa} has been investigated in \cite{anf}, and approximate expressions for outage probability and throughput have been derived. Regarding the extra complexity needed in \gls{gfa} receivers, deep learning powered receiver for increasing the reliability of \gls{gfa} systems has been investigated in~\cite{dlgf}.
Furthermore, \gls{ml}-powered radio resource provisioning for hybrid grant-free/grant-based  resource management has been investigated in~\cite{ragf}.
In the context of the \gls{3gpp} , the set of radio resources that could be allocated to \gls{gfa} communications, and the choice of modulation and coding scheme (MCS) have been investigated in ~\cite{gf3}. Furthermore, regarding the fact that \gls{gfa} is a complementary solution to the legacy GBA solution, the impact of \gls{gfa} on GBA communications has been  investigated in \cite{gf3}. 

\subsubsection*{Challenges of \gls{gfa}}
In the following, we describe some unexplored research problems related to \gls{gfa} which are investigated in this work. 
 \paragraph
{Reliability of \gls{gfa}}
 Once access to radio resources is shared among a set of devices which are sporadically active, the first concern to be addressed is the reliability of communications.  Especially, when we consider wide-area IoT networks, supporting heterogeneous IoT devices, it is important to investigate the probability of success in grant-free communications for different transmission protocols of different traffic types.  In \cite{2d}, outage probability for \gls{gfa} has been studied by assuming a constant received power from all contending devices.  The success probability in grant-free radio access  has been  analyzed in \cite{sic,mey} by assuming a Poisson point process  distribution for IoT devices over the service area. Performance of  SJD and \gls{sic} in detection/decoding of collided packets in a grant-free IoT network has been investigated in \cite{mahyar}. Regarding the fact that reliability is amongst the main KPIs for communications, and \gls{gfa} communications are vulnerable to collisions, more complex and advanced receivers, are expected to be explored in future research works \cite{dlgf,gf,azari2021energy}. Especially, an advanced receiver leveraging carrier frequency offset of IoT devices for collision decoding has been presented in \cite{gf,azari2021energy}.

\paragraph{Coexistence analysis in \gls{gfa}}
For \gls{gfa} IoT solutions, especially over the unlicensed spectrum, coexistence analysis is of crucial importance. For example, in the deployment phase of a new IoT service, when the choice of  IoT connectivity solution, including \gls{lorawan} and SigFox, must be done,  such analysis enables feasibility check for each connectivity option in presence of already existing IoT communications. The state-of-the-art research on performance analysis of \gls{gfa} IoT networks has been mainly focused on  performance analysis in homogeneous scenarios \cite{2d,sic}, e.g. single technology case. Hence, there is a lack of research on performance analysis with multi-type IoT devices and heterogeneous  communications characteristics, which is investigated here.
 
 \paragraph{Scalability of \gls{gfa}}
Another important challenge with grant-free radio access IoT connectivity is the scalability analysis. The scalability analysis is indeed of crucial importance for IoT solutions over the  licensed radio resources due to the scarcity of radio resources, and the crucial impact of the amount of allocated radio resources on the \gls{qos} of communications. For example, the scalability analysis can shed light on the extent up to which, \gls{qos} of communications could be preserved by increasing the number of devices that have access to the shared resource and is currently an active study item in \gls{3gpp}  standardization \cite{gf3}. Apart from the radio resource, the required density of the \glspl{ap} as well as the required device's consumed energy per packet\footnote{for transmit power or transmission of multiple replica}, is tightly coupled with the required \gls{qos}, and the rate of increase in each of them by increasing the \gls{qos} demand is of crucial importance. Therefore, there is a need to develop a model to investigate such a relation between the provisioned resources and \gls{qos} demand, which is investigated here.

\paragraph{Device operation control in \gls{gfa}}
Finally, in \gls{gfa},  due to the reduced signaling between devices and the network, the fine-grained control over the operation of each device is missing. Then, tuning the communications parameters of devices\footnote{including transmit power, data rate, spreading factor, and etc.}, for achieving a level of \gls{qos} is challenging. For example, the analysis in \cite{gf,azari2021energy} presents that increasing the number of replicas per packet  increases both reliability and energy efficiency of communications, beyond that level and up to the next level it tunes a tradeoff between them, and beyond the next level it deteriorates both of them. In contrast to such crucial impact, the operation control  for \gls{gfa} IoT networks is missing in the literature, and the state-of-the-art IoT networks usually use predetermined communication parameters, e.g. 3 replicas per packet  in SigFox \cite{mag_all}. \\

\subsubsection{Wake-up receiver for IoT}\label{SWUR}
Since most IoT devices are battery-driven, enhancing energy efficiency has been always a crucial KPI for network designers.  A novel scheme to realize long-lasting IoT devices is to equip them with  wake-up receivers (WuRs). A WuR is a module, integrated into the main circuit or along with that circuit, which can awake the device on-demand.  WuRs usually consist of an \gls{rf} demodulator, sequence decoder, digital address decoder, and are provided with a unique authentication address for paging \cite{wurj1}.  The WUR constantly intercepts and decodes from a specific radio channel, transmitting the entire wake-up calls until it recognizes its unique address to awaken the device. 
The presence of an ultra-low-power WUR capable of harvesting its adequate energy from ambient sources such as wireless, thermal, light, and vibration energy is a promising method to maintain the preserved battery power and increase the device's lifespan. Towards this end, the energy consumption of the WuR circuit must be as low as possible. On the other hand, the reliability of WuRs, including (i) the risk of false alarm (making device awake when another device has been paged); and (ii) and risk of missed alarm (missing the paging from the network when data is waiting for the device or input from the device is needed), are to be investigated.
Finally, although the literature on making efficient demodulators is mature, there is a lack of research on fast yet low-power and reliable address decoders. As this module is continuously monitoring the received ambient engines of the environment for potential paging of the device, and  the IoT market demands longer device addresses for the sake of security and expansion of IoT networks, the contribution of energy consumption of address decoding to total WuR's energy consumption is of crucial importance \cite{wurj}. 

\subsubsection{Energy harvesting and wireless power transfer for IoT}\label{SWPT}
While replacing batteries of devices in general is time and cost consuming due to the need for human intervention, in some remote IoT applications, access to devices is limited. As an example, one can consider the multitude of IoT devices embed in today's modern airplanes, where most of devices are under the inner decoration of the aircraft. In order to avoid the difficulty in replacing the device or battery in such applications, harvesting energy is considered as a promising solution \cite{sanislav2021energy}. Broadly speaking, energy could be harvested from different natural sources, including thermal and solar,  and radio signals like TV and radio broadcasting \cite{haridas2018opportunities}. When a device relies on harvested energy as the main power source, its performances  ties closely to the statistics of  energy harvesting and utilization (battery-less operation). This is because for natural sources like solar and wind, it is difficult to predict the time and the amount of available energy. Hence, employing energy buffer, i.e., battery for energy storage, might be necessary  to compensate the unpredictability of the energy. In contrast to the energy harvesting approach, wireless power transfer relies on intended radio signals, e.g. from the base station, for charging batteries of devices. In Section \ref{wptmma}, prior art on \gls{wpt}, especially through UAVs, has been presented. The major challenge with the above schemes, both from natural and radio transmissions, comes from the fact that the rate of saving energy by existing technologies are low in comparison to the rate of energy consumption in activity and data transmission of devices. Hence, devices need to store the harvested energy for a longer time, to a sufficient level, and during this time interval, devices won't be accessible by the network, and hence, it incurs delay in communications.  Apart from these challenges, and if the potential delay in accessibility of the device is acceptable,  the received energy from nature or base station can enable a device to last for a long time without human intervention. In \cite{liu2020rach}, performance of energy harvesting IoT devices in random access over NB-IoT networks has been investigated. In \cite{sanislav2021energy}, recent advances in energy harvesting  for IoT have been presented, and  future research challenges that must be addressed to realize the large-scale deployment of energy harvesting solutions for IoT have been investigated. An energy harvesting circuit has been presented in \cite{eshaghi2020energy}, which is  compatible with the mmWave frequencies of 5G technologies. The  spatio-temporal variations in the amount of harvested energy by IoT devices have been investigated in \cite{haridas2018opportunities}, and the  challenges to be dealt in enabling wide usage of energy harvesting in NB-IoT networks have been presented.

%%%%%%%%%%%%%%%%%%%%%%%%%%%%
\subsection{mmWave Access} 

For future \gls{mmwave} access, the choice of waveform design and multiple-access technique are among the major open problem. The challenges root  in differences in hardware design  in high frequencies. It is crucial for the realization of low-cost, efficient and reliable  \gls{mmwave} access to explore and innovate solutions based upon and beyond the existing ones, including \gls{ofdm}-based and  carrier-less \gls{uwb} transmission. 

In presence of analog beamforming with beam-based access, high-gain beams are known to be beneficial to the link budget and achievable data rate, however, they bring about some challenges too. For instance, current \gls{mmwave} access points equipped with phased arrays can illuminate one 'beam' at a time, imposing a large delay in serving a multi-user network system, where the users are spatially spread and cannot be served by the same beam without loss of efficiency. Moreover, in dynamic IoT systems, for examples in a highly flexible factory floor deployment, tracking the best beam to serve each user becomes a tedious and time-consuming task with a large imposed overhead. Thus, there is a clear trade-off between the gain of narrow beams and the ease and rapidity of beam selection of wide beams. It is beneficial to jointly optimize the beam design and beam selection according to the dynamics of the network as well as the dynamics of the traffic.

Additionally, the development and use of intelligent reflective surfaces in shaping the propagation environment in high frequency bands, despite a large number of publications on the topic, is still a big open area of research. Especially in indoor deployments, such  reflective surfaces have shown a great deal of potential for controlling the propagation environment \cite{kaina2014shaping}. Nevertheless, proper hardware   development technology and algorithm design to deliver on the promise of low-cost low-energy intelligent reflective surfaces require extensive research and development effort.

\subsection{Network-Device Cooperation} 

The future research and development for adaptive network-device cooperative schemes, particularly in the context of \gls{urllc} use cases, is envisioned to pay more attention to the following areas.

\subsubsection{Low-latency multi-hop transmission}
Traditionally, relays have  been considered as a means for extending coverage, less for low-latency applications. There is an inherent and unavoidable added latency from two-hop (and multi-hop) relaying. The main bottleneck for low-latency multi-hop relaying is in the processing (e.g., in case of decode-and-forward relaying) that is expected at the relay nodes, which can take several milliseconds down to a fraction of a millisecond, depending on the hardware capabilities.  With  rapid  advances in \gls{ue} capability categories and coding algorithms, the processing times for decoding and re-encoding a packet are expected to become shorter. For example, one can easily confirm that a user plane latency of 1 ms  over two-hop transmission  is achievable  with \gls{3gpp} Release 15 considerations for \gls{ue} capabilities and radio access numerology. Deploying relaying techniques with less processing overhead, e.g., denoise-and-forward relaying, can further reduce the expected delay per relaying hop \cite{zhao2013denoise}. Similarly, relaying delay can be reduced by full-duplex relaying operation \cite{liu2015band}.

\subsubsection{Distributed cooperative transmission}
To enable cooperative relaying from multiple devices in the network, distributed multi-antenna transmission schemes, such as distributed \gls{cdd} for \gls{ofdm} modulation, are required \cite{bauch2006cyclic}. In \gls{cdd}, by applying conscious cyclic shifts to the \gls{ofdm} symbol transmitted from each relay node, the spatial diversity from distributed relays is transformed into frequency diversity, which can be picked up by error-correcting codes such as the \gls{ldpc}, which is utilized by \gls{5g} \gls{nr} for data channels. Despite the extensive literature on such distributed cooperative transmission techniques, we expect to see emerging techniques to enable distributed and seamless cooperative transmission techniques that include on-the-run relay selection to improve energy efficiency and reduce interference foot-print.

%%%%%%%%%%%%%%%%%%%%%%%%%%%%

\subsection{Massive Connectivity} 

The research on \gls{b5g} cellular IoT has already begun and \gls{6g} is currently slated to enable a connection density of $10^7$ devices/km$^2$. 
%Massive connectivity solutions are in the directions we discussed in Section~\ref{sec:NOMA}. 
\subsubsection{NOMA} In theory, downlink \gls{noma} can be straightforwardly applied in almost any existing OMA-based solutions. However, several important challenges need to be addressed before adopting \gls{noma} in IoT devices in practice. Two important issues are  obtaining global, perfect \gls{csi} at each node and implementing \gls{sic} with acceptable complexity and error.  Neither of the two is affordable in low-cost devices, so it is unlikely to have downlink \gls{noma} in massive IoT applications. In fact, such limited-capacity devices cannot perform complex operations like \gls{sic}. Novel encoding/decoding techniques are to be designed before applying \gls{noma} to \gls{mmtc} applications. \gls{noma} may also be used in broadband IoT and in general in high-capacity devices like cell phones. Even in this case, extensive experimental results are needed to prove the superiority of \gls{noma} in practice where \gls{csi} is imperfect and \gls{sic} is prone to error propagation.  Preliminary results on NOMA with imperfect CSI can be found in \cite{fang2017joint,do2018wireless,arzykulov2018outage}. Besides, new over-the-air experiments of NOMA using software-defined radio as well as a  list of existing literature and  remaining challenges are reported in \cite{qi2021over}.

Uplink \gls{noma} schemes are more likely to be adopted in the near future provided that the encoding complexity of code-domain \gls{noma} schemes is reduced. Therefore, it is critically important to investigate low-complexity encoding schemes with an acceptable decoding error. Novel \gls{ml}- and DL-based solutions can be developed to decrease the design complexity and increase performance. Applying such methods to massive access is
expected to decrease the complexity of receivers. 

\subsubsection{Random access}
In grant-free massive IoT, the \gls{bs} is receiving data from a large number of uncoordinated devices. This results in resulting a high computational complexity at the \gls{bs}.  Another important issue in grant-free random access is device-activity detection. Approximate message passing algorithms \cite{donoho2009message,liu2018massive,ke2020compressive} are proposed for this purpose. The application of DL is
expected to decrease complexity further. 
More importantly, DL-aided solutions may bring noticeable gain 
for joint device detection and channel estimation as well as 
joint device and data detection. Overall, complexity reduction is the most crucial criterion for the wide adoption of any algorithm for massive connectivity and still requires intensive research. Finally, unsourced random access is a new promising approach toward massive connectivity  \cite{polyanskiy2017perspective}, which deserves more investigation.

\subsection{Channel Coding and Decoding}

Future wireless systems should provide service to various applications with a diverse range of requirements. For IoT, these requirements vary from high-reliability and low-latency to extreme low-energy consumption and deep coverage. Channel coding techniques need to designed to meet these requirements. We believe that the main focus for future wireless systems should be as follows:
\subsubsection{Short packet communications} Most channel codes have been designed for long block lengths to meet Shannon's Limit. These codes are not working well under practical decoders at short block lengths. Although some designs have been proposed for polar and \gls{ldpc} codes at short block lengths, they are not capable of supporting bit-level granularity of codeword length and rate. Puncturing and extending which are being commonly used for generating rate-compatible codes are sub-optimal which results in performance degradation at short block lengths. The design should be focused on developing novel techniques to generate rate-compatible codes with bit-level granularity with an optimized performance at any desired rate and length. The code should also demonstrate a very low error floor. Rateless code and rate adaptation techniques can be potential candidates to solve this problem. \gls{ldpc} codes over higher-order fields can be also a potential approach, however effective and low-complexity decoders should be designed. Moreover, with the cost of increased complexity, concatenated codes can be used to reduce the error floor. Moreover, some applications, like industrial IoT, require power burst error correction capability. Effective interleaving and de-interleaving approached should be designed for short packet communications.

\subsubsection{Low-capacity and \gls{csi}-free communications}
Many IoT applications will operate at an extreme-low power, which means that the received power is very low. The operating rate of the channel code is accordingly very low. Current channel codes, such as \gls{ldpc} and polar codes, are unable to work efficiently at very low rates in the short block length regime. Therefore new channel codes that can operate at very low \glspl{snr} and be able to work efficiently with no/unreliable \gls{csi} are required. Moreover, these codes need the \gls{csi} to be available at the receiver to perform the decoding. At very low \glspl{snr} or when the \gls{csi} is not available at the receiver, the soft information is not accurate or reliable at the receiver; therefore, the decoding performance will dramatically degrade. Low-complexity \gls{osd} algorithms \cite{Chentao2021OSD} can be a universal decoding technique that can work effectively under these conditions. 

  The bottleneck of the decoding in BCH-DFT is to find the position of errors. This is confirmed by assuming a genie-aided error localization \cite{vaezi2014thsis} which improves the performance improves largely. Hence, any solution improving the decoding will be a valuable contribution and can make this class of codes suitable for low latency communications.
  Further, these codes can be used for joint source-channel coding to further reduce the delay  incurred in the communication system \cite{gabay2007joint,vaezi2014distributed}.  Another promising direction in this line is to seek methods that bypass the need for quantization of the codeword. If viable, such a solution will enjoy a much better decoding performance and can pave the road for widespread adoption of this class of codes in communication systems.

\subsubsection{Decoder design}
The end-to-end latency requirement of 0.5 ms of many low-latency applications, implies a
physical layer latency of 8 to 67 $\mu$s. The
channel decoder must share this latency budget with channel estimation and demodulation. Channel decoder should target processing latencies as low as 0.665 s \cite{shao2019survey}. This can be achieved by employing internal parallelism. 
The decoding techniques \gls{ldpc} codes gain benefit from the parallel structure of the belief propagation decoding, which can be implemented efficiently. This provides low decoding latency and complexity. However, these decoders are sub-optimal for short packet communications. The decoding of polar codes mostly involves successive operations which introduces significant delays. It also involves large list sizes to achieve near-optimal performance, which requires large memory. These decoders are not efficient for resource-constraint IoT devices. The focus should be on using near-optimal decoders, such as \gls{osd} with sufficient and necessary conditions to reduce the complexity. These decoders are universal and can be applied to any linear code \cite{Chentao2021OSD,Chentao2019SDOSD,Chentao2020PBOSD}. Further research is required to allow parallel computations to reduce the decoding run-time.

\subsection{Risk-aware Learning for IoT}
Introduction of mission-critical IoT into cellular networks mandates a
transition from average-utility network design into risk-aware
network design, where a rare event may incur a huge loss \cite{ragf}.
Instead of legacy \gls{ml} algorithms, which aim at maximizing
the average return of agents, or equivalently minimizing the
average regret of agents, risk-aware \gls{ml} aims at minimizing the regret of huge
loss due to potential performance degradation. Towards this end, extra emphasis must be put on the definition of a loss function. For example, consider risk-aware design of a traffic scheduler when serving solely one traffic type. In such a case, the scheduler design must  enforce maximizing the first moment of the return, i.e. reliability, while minimizing
the higher moments, e.g. the variance. Minimizing higher
moments of the return, i.e. optimizing the worst-case return,
could be achieved by defining the utility function as a nonlinear function of the return. A well-known example of a non-linear utility function is the exponential utility function \cite{ragf}, and works as follows. Considering latency $L$ as a KPI,  the risk-aware scheduler design minimizes the expected value of $e^L$  instead of minimizing the expected value of $L$. In \cite{ragf}, a risk-aware \gls{ml} solution for radio resource management of non-scheduled \gls{urllc} traffic has been proposed. This solution enables  safe  delivery of short update packets by leveraging hybrid orthogonal/non-orthogonal radio
resource slicing, and proactively regulating the spectrum needed
for satisfying the required \gls{qos}. 

A recent applications of risk-aware \gls{ml} has emerged in the \gls{aoi} research direction. Time-sensitive IoT applications,
such as real-time monitoring by drone and terrestrial sensors, must rely on a timely status information delivery to the control entity. Then, the 
freshness of the received status updates at the  control center is of crucial importance.
The state-of-the-art approaches in \cite{borzo,when2} follow a risk-neutral
approach, in which,  the expected value of AoI cost functions is minimized. In
order to capture the effect of rare events with a significant impact on the cost function, in \cite{riskb}, the notion of conditional value-at-risk for minimization
of AoI for real-time IoT reporting has been proposed.

\subsection{Security Issues}

While Blockchains and \gls{sdn} architectures can assist with some IoT security challenges, we are still struggling with effective key management. Keys and certificates are needed on a device to set up a secure channel to cloud services. Traditional ways of storing keys and certificates do not scale with the volume of IoT. An emerging solution would be to use a hardware-based authentication system that relies on the physical characteristics of the IoT device itself. These are inimitable properties unique to the hardware of each device that can not be modified, cannot be cloned and cannot be extracted. There are various types of \glspl{puf} that can assist in this domain, some examples are \gls{sram} \glspl{puf} \cite{9233424}, butterfly \glspl{puf} \cite{9233424}, Ring oscillator \glspl{puf} \cite{8552142}, etc. This hardware-based technology provides a unique identity to every node (a very similar concept to a private key), which the node can use to authenticate as well as encrypt all data to protect data from the moment it leaves the device, all the way to the cloud services, and eventually achieve a true end-to-end protection.

\section{Conclusions} 
\label{sec:con}
IoT is playing an increasingly important role in the world around us and IoT market continues to grow at a fast pace. IoT connected to cellular and wide area networks are the fastest growing sectors of IoT, and current and new cellular standards are among the main enablers of various IoT applications.  In this paper, we have
provided a comprehensive survey of recent literature that lies at
the intersection between IoT and \gls{5g}/\gls{6g} cellular networks.
We have listed and discussed the key performance indicators of \gls{5g} and \gls{6g} networks, indicated how IoT contributes to them and summarized recent advances maid along with those directions.
In particular, we have described to what extent current research supports massive access, energy efficiency, reliability, latency, and security required for IoT, and what steps need to be taken to bridge the gap in the 2030s. This survey covers several aspects of IoT overlooked by previous surveys such as energy efficiency and channel coding for reliability and air latency.  
In addition, we have reviewed basic concepts and principles
of various deep learning models and distributed and federated learning, emphasizing work specific
to the IoT across different application scenarios.  Further, we have
discussed how to tailor deep learning models to IoT applications. We have also reviewed the solution for the integration of satellites and drones to next-generation IoT networks.   
We have concluded the survey by pinpointing several remaining challenges, open research
issues, and promising directions, which may lead to valuable future research results. We hope that this survey becomes
definite guide to researchers and developers interested in the IoT and a valuable resource for the field.

\bibliographystyle{IEEEtran}
\bibliography{1Refs_amin,1Refs_Mahdi,1Refs_NOMA,1Refs_Mahyar,1Refs_SK,1Refs_danai}

\section{Appendix A: List of Abbreviation}		
\label{sec:appendix} 

A complete list of abbreviation can be found in the table below.

\printglossary[type=\acronymtype]

% \bibliographystyle{IEEEtran}
% %\bibliographystyle{apa}
% %\bibliographystyle{spphys}
% \bibliography{Refs_Mahdi}

\end{document}